%% file: grained_stellar_kinetics.tex
\documentclass[fleqn,usenatbib]{mnras}

\usepackage{newtxtext,newtxmath}
\usepackage{multicol}
\usepackage[T1]{fontenc}
\usepackage{ae,aecompl}

\usepackage{hyperref}
\usepackage{lineno}
\usepackage[utf8]{inputenc}
\usepackage{dutchcal}
\usepackage{pgfplots}
\usepackage{subfigure}
\usepackage{siunitx}
\usepackage[percent]{overpic}
\usepackage{tikz}
\usepackage{natbib}
\usepackage[mmddyyyy]{datetime}
\usepackage{booktabs}
\usepackage[titletoc,title]{appendix}
\usepackage{subfigure}
\usepackage{graphicx}	
\usepackage{amsmath}	
\usepackage{amssymb}	
\usepackage{amsfonts}
\usepackage{amssymb}
\usepackage{etoolbox}
\makeatletter
\patchcmd\@combinedblfloats{\box\@outputbox}{\unvbox\@outputbox}{}{%
	\errmessage{\noexpand\@combinedblfloats could not be patched}%
}%
\makeatother
\modulolinenumbers[5]
\numberwithin{equation}{section}
\usetikzlibrary{calc}
\usetikzlibrary{plotmarks}
\usetikzlibrary{positioning}
\usetikzlibrary{fit}
\usetikzlibrary{decorations.pathmorphing}
\usetikzlibrary{backgrounds}
\pgfplotsset{compat = newest}
\pgfplotsset{ legend style={font=\tiny} }
\definecolor{bgreen}{rgb}{0.0,0.5,0.0}
\definecolor{bblue}{rgb}{0.0,0.0,0.9}
\definecolor{bgold}{rgb}{0.7,0.5,0.0}
\definecolor{bred}{rgb}{0.9,0.0,0.0}

\newcommand\mathplus{+}

\title[Truncated stellar kinetics]{A convergent kinetic theory of collisional star clusters (i) a self-consistent 'truncated' mean-field acceleration of stars}
\author[Y. Ito]{
Yuta Ito$^{1,2,3}$\thanks{E-mail: yito@gradcenter.cuny.edu}
\\
$^{1}$Department of Physics, CUNY Graduate Center, 365 Fifth Avenue, New York, NY 10016, USA\\
$^{2}$Department of Engineering Science and Physics,
College of Staten Island, 2800 Victory Boulevard, Staten Island, NY 10314, USA\\
$^{3}$Department of Mathematics,
College of Staten Island, 2800 Victory Boulevard, Staten Island, NY 10314, USA
}
\date{Accepted XXX. Received YYY; in original form ZZZ}

\pubyear{2018}

\begin{document}
\label{firstpage}
\pagerange{\pageref{firstpage}--\pageref{lastpage}}
\maketitle

\begin{abstract}
Fundamental relaxation processes in the secular evolution of a collisional star cluster of $N$-'point' stars have been conventionally discussed based on either of collision kinetic theory (for strong two-body encounters) and wave one (for statistical acceleration and gravitational polarization). If combining the both theories together, one must introduce a self-consistent 'truncated' Newtonian mean-field (m.f.) acceleration of star at position $\bmath{r}$ and time $t$ due to a phase-space distribution function $f\left(\bmath{r}', \bmath{p}',t\right)$ for stars	
\begin{align}
&\bmath{A}^{\triangle}(\bmath{r},t)=-Gm\left(1-\frac{1}{N}\right)\int_{\mid\bmath{r}-\bmath{r}' \mid > \triangle}\frac{\bmath{r}-\bmath{r}'}{\mid \bmath{r}-\bmath{r}' \mid^{3}} f\left(\bmath{r}',\bmath{p}',t\right)\text{d}^{3}{\bmath{r}'}\text{d}^{3}{\bmath{p}'},\nonumber
\end{align}
where $G$ is the gravitational constant and $m$ the mass of stars. The lower limit $\triangle$ of the distance between two stars is order of the Landau distance. The truncated m.f. acceleration is a necessary consequence due to the strong encounters and m.f. acceleration being not able to 'coexist' at specific distance between stars. 

The present paper aims at initiating a star-cluster convergent kinetic theory to self-consistently derive kinetic equations of star clusters, mathematically non-divergent in distance- and wavenumber- spaces based on the truncated m.f. acceleration, correct at time scales of the secular evolution. This will be achieved by focusing on mathematical formulations of the Kandrup's generalised-Landau equation including the effect of the strong encounters and by extending the Grad's truncated distribution function and Klimontovich's theory of non-ideal systems. The formulations cover the following physical situations; (i) No star can approach another star closer than the distance $\triangle$ (ii) Only 'test' star can approach one of 'field' stars closer than the distance $\triangle$ at a time while the rest of field stars can not (iii) All stars can approach each other limitless.

\end{abstract}

\begin{keywords}
gravitation -- methods: analytical -- globular clusters: general--galaxies: general
\end{keywords}





\input{Section1_intro.tex}
\input{Section2_basic_kinetics_stellar_scalings.tex}
\input{Section3_truncatedBBGKY_nonideality}

\input{Section4_completely_WC_limit.tex}
\input{Section5_strong_interaction.tex}
\input{Section6_CKT.tex}
\input{Section7_conclusion.tex}
\section*{Acknowledgements}
I appreciate my adviser Carlo Lancellotti for allowing me to pursue this topic.




\bibliographystyle{mnras}
\bibliography{science}

\appendix
\input{Appendix.tex}


\bsp	
\label{lastpage}
\end{document}

%% file: Section1_intro.tex
\section{Introduction}\label{sec:intro}
A general point of view to understand statistical dynamics of dense star clusters is to introduce the effect of 'discreteness' of the clusters. The discreteness means the finiteness of total number $N$ of stars in a dense star cluster, say  $N\approx 10^{5}\sim 10^{7}$. In the present paper, the system of concern is collisional star clusters, e.g. globular clusters and collisional nuclear star clusters without super massive black holes. As a first approximation $(N\to\infty)$, the system can be assumed smooth and its evolution is dominated by a self-consistent mean field (m.f.) potential. The effect of m.f. potential is of significance on a few of dynamical-time scales and may freeze the system into a quasi-stationary state due to rapid fluctuations in m.f. field potential (i.e. violent relaxation). The evolutions of long-time lived star clusters might have been driven by less probable relaxation process (two-body close encounters ) and 'slow' many-body relaxations (statistical acceleration and gravitational polarization), in addition to the effect of m.f. potential\footnote{The present work focuses on systems modeled by kinetics of one-body distribution function of stars ('point particles' interacting via pair-wise Newtonian forces), neglecting the effect of triple encounters and some realistic effects (gas/dust/dark-mater dynamics, stellar evolution, inelastic direct collisions, formation of stars and binaries, stellar mass distribution, ...)}. 

The most fundamental relaxation process in the evolution of collisional star clusters may be arguably the statistical acceleration that stands for a non-collective relaxation, originated from the deviation of the actual force on 'test' star due to ($N-1$)-'field' stars from the smooth force due to the m.f. potential. The statistical acceleration is in association with the effect of stochastic many-body encounters. Conventionally, the effect of many-body encounters approximately gives place to that of cumulative \emph{two-body} encounters between stars. The relaxation processes described by the cumulative- and strong- two-body encounters may be termed \emph{collision kinetic theory} that is based on a heuristic employment of either of collision-Boltzmann, forward Komologouv-Feller (f.KF), master, Fokker-Planck (FP)/ Landau equations whose two-body encounters are local events in configuration space (section \ref{subsec:intro_collision}). On scales larger than the average distance of stars on which the two-body-encounter approximation breaks down, the effects of the statistical acceleration and gravitational polarization of the system are of significance and described by \emph{wave kinetic theory} that relies on inhomogeneous- Landau/ FP or Balescu-Lennard equation derived from first principles or the Holtsmark distribution (section \ref{subsec:intro_wave}).

While the cumulative two-body encounters become more probable on larger-space scales due to the long-range nature of Newtonian pair-wise potential, the statistical acceleration becomes greater in magnitude on smaller scales. This implies the relaxation effects on intermediate-space scales in the both theories are of significance in evolution of the system, and share the same model; FP/ Landau kinetic equation under the approximations of local encounter, homogeneous field stars and weak-coupling limit. The relaxation time $t_\text{r}$ is scaled as $t_\text{r}\sim t_\text{d}[\ln{ N}]/N$ 
for both collision kinetic theory \citep{Ambartsumian_1938} and wave one \citep{Severne_1976}. Fundamental statistical-dynamics approach to investigate the evolution of a collisional star cluster is to approximate the two-stage dynamics (due to the m.f. potential and two-body relaxation) into the relaxation evolution averaged over the dynamical scales. Especially one has numerically integrated the orbit-averaged FP equation, the FP equation averaged over unperturbed orbits of stars for a distribution function (DF) in terms of integrals of motion, for the evolution of the system on the secular-time scales, $t_\text{sec}\sim \text{tens of } t_\text{r}$, \citep[e.g.][]{Henon_1961, Henon_1965, Cohn_1979}. The successes of the approach are, however, acknowledged \citep{Heggie_2003, Binney_2011,Merritt_2013} only in sense that the approach can efficiently retrieve basic results of the $N$-body numerical simulations $(\text{still limited to } N\lesssim 10^6 )$ and is compatible with the other models (e.g. moment/gaseous models). Especially, the mathematical formulation of the secular evolution of stars clusters is far from its completion; there does not exist even a formulation for kinetic equations correct at order of time scales $t_\text{sec}\sim N$, to which the present paper contributes. After the collision- and wave- kinetic theories are reviewed in sections \ref{subsec:intro_collision} and \ref{subsec:intro_wave} respectively, the two motivations are explained for establishing mathematically non-divergent kinetic equations of star clusters in distance spaces in section \ref{subsec:intro_Identity} and for initiating \emph{convergent kinetic theory} (CKT) in section \ref{subsec:intro_convergent}.

\subsection{Collision kinetic theories of star clusters}\label{subsec:intro_collision}
The relaxation process described by collision kinetic theories of star clusters, in general, relies on stochastic two-body encounters. The description approximately models the incoherent behavior of the statistical acceleration as cumulative two-body encounters of stars and has well prevailed to explain 'two-body' relaxation process from the classical works \citep{Chandra_1942} to the recent ones \citep{Heggie_2003,Binney_2011,Merritt_2013}. Since the essential description relies on two-body interaction, the theories have been developed to understand the effect of strong two-body encounters in stellar dynamics\footnote{The effect of strong encounter is a long-standing problem in stellar dynamics since the early works \citep{Eddington_1914, Charlier_1917a,Ambartsumian_1938}. Refer to \citep{Kandrup_1980,Shoub_1992,Ashurov_2004} and the references there for the detail.} and have revealed the role of (ejected) energetic stars in formation of the structures of star clusters, especially the halo, and in dissipation of the systems \citep[e.g.][]{Lin_1980,Goodman_1983, Goodman_1984,Shoub_1992}. Up to date, one has focused on improving only \emph{collision terms} at kinetic-equation level since the early works \citep{Charlier_1917a, Jeans_1928, Chandra_1941_coll1,Williamson_1941,Chandra_1941_coll2}, to the f.KF model for a test star in homogeneous stationary Maxwellian background \citep{Agekyan_1959, Agekyan_1959a, Petrovskaya_1970, Petrovskaya_1970a, Kaliberda_1971, Retterer_1979} and to the f.KF- or collision-Boltzmann- models for more realistic (anisotropic, multi-mass, two-body relaxation...) systems \citep{Henon_1960, Henon_1965, Agekyan_1962, Goodman_1983, Ipser_1983, Shoub_1992, Ashurov_2004}. However, the FP/ Landau equation based on stochastic theory under the two-body approximation does not correctly describe the many-body effect, and is not clear as a correct mathematical expression to describe the long-time relaxation \citep{Alexandre_2004}. Also, stochastic theories can not self-consistently explain the relation between the collision term and m.f. potential; one has presumed a simple addition of a collision term to collisionless Boltzmann equation. The validity of the heuristic operation has never been carefully discussed. A correct incorporation of the effects of long-time relaxation and strong encounters into a kinetic equation must be made through first principles though, the discussion was made only for the former as explained in section \ref{subsec:intro_wave}. 

\subsection{Wave kinetic theories of star clusters}\label{subsec:intro_wave}
Wave kinetic theories of star clusters have aimed at an essential understanding of how the discreteness of a cluster and Newtonian man-body interaction affect the evolution, while the theories appears only as an introduction to relaxation processes in basic texts \citep{Saslaw_1985,Spitzer_1988} and less prevailed compared to collision ones. The onset of wave kinetic theory \citep{Chandra_1941} is an incorporation of the effect of many-body encounters into the relaxation time based on the Holtsmark distribution of force strengths\footnote{See e.g. \citep{Chavanis_2013b,Chavanis_2013a} and recent papers \citep{Sridhar_2016, Heyvaerts_2017} for thorough review of wave kinetic theories.}. \cite{Gasiorowicz_1956} derived a master equation including the Coulombian polarization and the statistical acceleration for plasmas based on the Vlasov equation and Holtsmark distribution. The \citep{Gasiorowicz_1956}'s idea was extended to stellar dynamics by employing the functional derivative method \citep{Gilbert_1968} and later BBGKY hierarchy \citep{Gilbert_1971}. In the Gilbert's works, the hierarchy was expanded up to the first order of smallness parameter, $1/N$, with weak-coupling limit, meaning the derived equation is presumed correct on the secular time scales while the effect of three-body encounters and strong-close encounters among stars were neglected. Application of the \citep{Gilbert_1968}'s equation has been limited to studies of the gravitational polarization effect in test-particle problem \citep{Gilbert_1970} and on homogeneous system \citep{Weinberg_1993} and studies of the effect of inhomogeneity without polarization effect \citep{Severne_1976, Haggerty_1976, Parisot_1979}. The \citep{Kandrup_1981}'s generalised-Landau (g-Landau) equation\footnote{Mathematically, the g-Landau equation is the same form as the \citep{Gilbert_1968}'s equation without gravitational polarization effect but physically different. The g-Landau equation is a \emph{self-consistent} 'point-particle' description, meaning only test star undergoes relaxation process with field stars while any field star does not undergo any relaxation process with the other field stars (See Appendix \ref{sec.many_to_two}).} conceptually well explains the relaxation process due to the statistical acceleration from perspectives of many-body interaction regarding with $N$-body DF description \citep{Kandrup_1981} and the discreteness of slowly-changing m.f. potential \citep{Kandrup_1988}. After the success of derivation of the explicit forms of the \citep{Gilbert_1968}'s type equation without polarization \citep{Polyachenko_1982} and with polarization \citep{Heyvaerts_2010, Chavanis_2012}, one has began to find its direct application to stellar discs; 'collisional'-relaxation dominant systems \citep{Fouvry_2015,Fouvry_2015_a} and resonant-relaxation dominant system \citep{Sridhar_2016, Sridhar_2016_a, Fouvry_2017a, Fouvry_2017}. Recently, \citep{Wren_2018} applied a linear-perturbation method to the \citep{Gilbert_1968}'s equation for a Maxwellian DF of stars, to revisit the gravithermal-instability problem \citep{Antonov_1985}. However, the previous works essentially neglect the effect of strong encounters by exploiting the weak-coupling limit which heuristically avoids mathematical divergences at kinetic-equation- and BBGKYH-hierarchy- level.

\subsection{Motivation for establishing non-divergent kinetic equation in distance space}\label{subsec:intro_Identity}
To avoid any mathematical divergence at kinetic-equation level one must combine the collision- and wave- theories of star clusters. The \citep{Kandrup_1981}'s g-Landau equation does not diverge on large scales (of distance between stars) if the statistical acceleration \citep{Kandrup_1981_a} or the effect of m.f. potential \citep{Gilbert_1971,Chavanis_2008,Chavanis_2010,Chavanis_2013a} on test star is taken into account, even without the polarisation effect. Also it is acknowledged that the Boltzmann (or f.KF) equation is essentially divergence-free on small scales (of impact parameter) \citep[e.g.][]{Landau_1987,Liboff_2003} for square-inverse forces. This implies that one can obtain a divergence-free star-cluster kinetic equation \emph{in distance space} by patching collision terms appearing in the g-Landau- and Boltzmann- equations at a specific distance between stars. 

As discussed in \citep{Chandra_1941,Takase_1950} the randomness in the Holtsmark distribution of force strength must have its lower limit in distance space, i.e. the Landau-distance scale; correspondingly, the statistical acceleration or the 'discreteness fluctuations \citep{Kandrup_1988}' in m.f. potential must do (See section \ref{sec:complete_WC}).  This encourages one to separate the collision- and wave- kinetic descriptions at the Landau-distance scale. Similar separations between strong two-body encounters and cumulative two-body ones has been discussed in collision kinetic theories \citep[e.g.][]{Ipser_1983,Shoub_1992}, while the separations complicate the formulation of kinetic equations due to the relative-velocity dependence of the impact parameter. Also, the collision kinetic theories lack the discussion of the effect of m.f. potential on test star that is important for finite star clusters. Hence, for simplicity, one may employ the 'conventional' Landau distance \citep{Montgomery_1964,Spitzer_1988} whose impact parameter does not depend on the relative velocity, corresponding with a spatial scale that is approximately smaller than the system size by a factor of $N$. Such simplification is possible since the strong encounter can occur locally in a star cluster on scales smaller than the 'conventional' Landau distance. (This will be discussed in section \ref{subsec:scalings}.). In this case, typical m.f. potential must be truncated on small scales on which the strong encounters dominate the motion of test star.

To truncate the m.f. potential on small distance scales, the present work resorts to the \citep{Grad_1958}'s \emph{truncated DFs} that were originally employed for DFs of particles in rarefied gases, truncated on the effective scales of short-range interaction force between the particles. However, such an artificial approximation on small scales must be carefully discussed since a typical scenario of the evolution of a star cluster (without binaries) predicts a gravothermal-instability, resulting in a 'core-halo' structure with a high density and inhomogeneity on the Landau-distance scale. Hence, one also needs another method to consider the small-scale strong inhomogeneity of m.f. potential at core/inner halo. One may resort to the \citep{Klimontovich_1992}'s theory of non-ideal gases and plasmas, where the treatment of m.f. potential was systematically discussed between collision- and wave- kinetic descriptions. Hence, the two methods (\citep{Grad_1958} and \citep{Klimontovich_1992}) will be employed to 'patch' the Boltzmann collision term to the g-Landau one at the Landau distance between stars in section \ref{sec:strong}.

\subsection{Motivation for initiating star-cluster Convergent kinetic theory}\label{subsec:intro_convergent}
Patching the collision kinetic equation and wave one at a distance between stars does not mean one can obtain a mathematically divergence-free kinetic equation \emph{in wavenumber space} too, since the patching method does not correctly account for the correlation-time dependence of the collision terms (section \ref{subsec:Log_col}). Hence, one must learn from the plasma CKT to find equations divergence-free even in wavenumber spaces too.

'Conventional' plasma CKTs \citep[e.g.][]{Aono_1968, Landau_1987, Liboff_2003} represents the methods to find non-divergent kinetic equations and the Coulomb logarithm for classical pure electron-ion plasmas, by taking into account the effects of polarisation and strong encounters between plasma constituent particles. For the temperature-equilibration problem of weakly-coupled plasmas \citep{Landau_1936}, the conventional plasma CKTs were established to improve the classical estimations \citep[e.g.][]{Cohen_1950, Spitzer_1962} of the value of factor $\gamma$ in Coulomb logarithm $\left(\ln\left[\gamma N \right]\right)$. Recently, the plasma CKTs have been revisited and extended to investigate moderately- and strongly- coupled plasmas \citep[e.g.][]{Dharma_wardana_1998, Dharma_wardana_2001, Gericke_2002, Brown_2005, Baalrud_2012,Baalrud_2013,Baalrud_2014}. It turns out, as long as limiting one's concern into the weak-coupling limit $\left(\ln[\gamma N]\gtrsim 8\right)$, the new CKTs, especially dimensional-continuation method \citep{Brown_2005}, can \emph{self-consistently} give the value of $\gamma$ in excellent agreements with molecular dynamics simulations \citep{Glosli_2008, Dimonte_2008, Daligault_2009, Grabowski_2013}. Even the (non-self-consistent) conventional CKTs, especially the matched-asymptote method at equation-level \citep[e.g.][]{Frieman_1963, Gould_1967}, can give a correct value of Coulomb logarithm at order of $\sim 1$.

On the other hand, in stellar dynamics, the Coulomb logarithm and star-cluster CKTs have not been the first concern \citep[e.g.][]{Shoub_1992}. One of the reasons is that typical orbit-averaged FP models do not necessitate a correct value of Coulomb Logarithm if the relaxation time is taken as CPU time unit \citep[e.g.][]{Cohn_1979,Takahashi_1995}. However, it is acknowledged that the correct value must be taken into account when considering some effects whose effective time scales are other than the relaxation time scale (e.g. binary heating, direct collision, triple encounter...). Typically, the value of factor $\gamma$ is determined based on the results of many $N$-body simulations \citep[e.g.][]{Farouki_1982,Smith_1992,Farouki_1994,Giersz_1994}. Also, stellar dynamics may, very likely, demand a time-dependent Coulomb logarithm \citep{Alexander_2012} in addition to the space-dependence \citep[e.g.][]{Spitzer_1988}.  Hence, it is desirable to initiate a star-cluster CKT to self-consistently find the correct relaxation time.

The present work aims at \emph{incorporating the effect of strong two-body encounters into the \citep{Kandrup_1981}'s g-Landau equation correct up to order of secular-time scales based on \citep{Grad_1958}'s truncated DF, then  initiating a star-cluster CKT based on the \citep{Klimontovich_1982}'s theory}. The present paper is organised as follows. In section \ref{sec:basic_kinetics}, fundamental concepts of kinetic theories are reviewed. In section \ref{sec:truncated_BBGKY}, the \citep{Grad_1958}'s method and \citep{Klimontovich_1992}'s theory are adjusted for stellar dynamics. In sections \ref{sec:complete_WC} and \ref{sec:strong}, kinetic equations are derived from the Liouville equation without- and with- the effect of strong encounters respectively, based on the \citep{Grad_1958}'s and \citep{Klimontovich_1982}'s method. In section \ref{sec:discussion}, by resorting to \citep{Klimontovich_1982}'s method, a star-cluster CKT is systematically constructed showing the relation between the equations derived by the \citep{Grad_1958}'s method and a conventional CKT \citep{Frieman_1963} respectively. Section \ref{sec:conclusion} is Conclusion. 

%% file: Section2_basic_kinetics_stellar_scalings.tex
\section{BBGKY hierarchy for distribution function and scalings of physical quantities in stellar dynamics}\label{sec:basic_kinetics}
In sections \ref{subsec_Liouville} and \ref{subsec:s-tuple} fundamental concepts of kinetic theory are reviewed and in section \ref{subsec:scalings}  a scaling of orders of the magnitudes (OoM) of physical quantities to describe a star cluster and encounters is explained. In section \ref{subsec:trajectory} the trajectory of test star in encounters is explained. In section \ref{subsec:Log_col} the logarithmic divergences in collision- and wave- kinetic theories are explained.
\subsection{The $N$-body Liouville equation}\label{subsec_Liouville}

Consider a star cluster of $N$-'point' stars of equal masses $m$ interacting each other purely via Newtonian gravitational potential
\begin{equation}
\phi(r_{ij})=-\frac{Gm}{r_{ij}} \qquad(1\lid i, j \lid N \quad \text{with} \quad i\neq j ),\label{Eq.phi}
\end{equation}
where $G$ is the gravitational constant and $r_{ij}\left(=\mid \bmath{r}_{i}-\bmath{r}_{j}\mid\right)$ is the distance between star $i$ at position $\bmath{r}_{i}$ and star $j$ at $\bmath{r}_{j}$. The Hamiltonian for the motions of stars in the system reads
\begin{equation}
H=\sum_{i=1}^{N}\left(\frac{\bmath{p}^{2}_{i}}{2m}+m\sum_{j>i}^{N}\phi(r_{ij})\right),\label{Eq.H}
\end{equation}
where $\bmath{p}_{i}(=m\bmath{\varv}_{i})$ is the momentum of star $i$ moving at velocity $\bmath{\varv}_{i}$. Assume the corresponding $6N$ Hamiltonian equations can be alternatively written in form of the $N$-body Liouville equation
\begin{align}
\frac{\text{d}F_{N}}{\text{d}t}=\left(\partial_{t}+\sum_{i=1}^{N}\left[\bmath{\varv}_{i}\cdot\nabla_{i}+\bmath{a}_{i}\cdot \bmath{\partial}_{i}\right]\right)F_{N}(1,\cdots,N,t)=0,\label{Eq.Liouville}
\end{align}
where the symbols for the operators are abbreviated by $\partial_{t}=\frac{\partial}{\partial t}$, $\nabla_{i}=\frac{\partial}{\partial \bmath{r}_{i}}$, and $\bmath{\partial}_{i}=\frac{\partial}{\partial \bmath{\varv}_{i}}$. The acceleration $\bmath{a}_{i}$ of star $i$ due to the pair-wise Newtonian forces from the rest of stars is defined as
\begin{equation}
\bmath{a}_{i}\equiv\sum_{j=1(\neq i)}^{N}\bmath{a}_{ij}\equiv-\sum_{j=1(\neq i)}^{N}\nabla_{i}\phi(r_{ij}).
\end{equation}
The arguments $\{1,\cdots,N\}$ of the $N$-body (joint-probability) DF $F_{N}$ are the Eulerian position coordinates and momenta $\{\bmath{r}_{1},\bmath{p}_{1},\cdots,\bmath{r}_{N},\bmath{p}_{N}\}$ of stars in the system at time $t$. The $N$-body DF $F_{N}$ is interpreted as the phase-space probability density of finding stars $1, 2, \cdots, N$ at phase-space points $(\bmath{r}_{1},\bmath{p}_{1})$, $(\bmath{r}_{2},\bmath{p}_{2})$, $\cdots$ and $(\bmath{r}_{N},\bmath{p}_{N})$ respectively at time $t$, and is normalized as
\begin{equation}
\int F_N(1,\cdots,N, t)\text{d}_{1}\cdots\text{d}_{N}=1,\label{Eq.normalizedDF}
\end{equation}
where an abbreviated notation is employed for the phase-space volume elements, $\text{d}_{1}\cdots\text{d}_{N}(=\text{d}\bmath{r}_{1}\text{d}\bmath{p}_{1}\cdots\text{d}\bmath{r}_{N}\text{d}\bmath{p}_{N})$. In addition, the function $F_{N}$ is assumed symmetric about a permutation between any two phase-space states of stars \citep{Balescu_1997,Liboff_2003}, and the Hamiltonian equation \eqref{Eq.H} in phase space (obviously) holds the same symmetry, meaning stars $1$, $\cdots$, $N$ are assumed identical and indistinguishable respectively.

\subsection{The $s$-tuple distribution function and correlation function}\label{subsec:s-tuple}
A reduced DF of stars in a star cluster is, in general, introduced in form of $s$-body (joint-probability) DF
\begin{equation}
F_{s}(1\cdots s, t)=\int F_N(1,\cdots,N, t)\quad\text{d}_{s+1}\cdots\text{d}_{N},\label{Eq.SbodyJointDF}
\end{equation} 
or in form of $s$-tuple DF:
\begin{equation}
f_{s}(1\cdots s, t)=\frac{N!}{(N-s)!}F_{s}(1\cdots s, t).\label{Eq.StupleDF}
\end{equation} 
The $s$-tuple DF describes the probable number (phase-space) density of finding stars $1,2,\cdots,s$ at phase-space points $1,2,\cdots,s$ respectively. The $s$-tuple DF simplifies the relation of macroscopic quantities with irreducible $s$-body dynamical functions. For example, the total energy of the system at time $t$ may read
\begin{subequations}
	\begin{align}
	E(t)&=\int\cdots\int\sum_{i=1}^{N}\left(\frac{\bmath{p}^{2}_{i}}{2m}+m\sum_{j>i}^{N}\phi(r_{ij})\right)F_{N}(1\cdots N,t)\text{d}_1\cdots\text{d}_N,\\
	&=N\int \frac{\bmath{p}_{1}^{2}}{2m}F_{1}(1,t)\text{d}_{1}+m\frac{N(N-1)}{2}\int\phi(r_{12})F_{2}(1,2,t)\text{d}_1\text{d}_2,\\
	&=\int \frac{\bmath{p}_{1}^{2}}{2m}f_{1}(1,t)\text{d}_{1}+\frac{m}{2}\int\phi(r_{12})f_{2}(1,2,t)\text{d}_1\text{d}_2.\label{totE}
	\end{align}
\end{subequations}
where the symmetry of permutation between two phase-space points for both the Hamiltonian and the $s$-body DF are applied. The total energy $E(t)$ can turn into a more physically meaningful form by introducing typical $s$-ary DFs to understand the effect of correlation between stars, as follows.
\begin{align*}
\quad	f(1,t) \quad&:\text{(unary) DF}  \\
\quad	f(1,2,t) \quad &:\text{binary DF} \\
\quad	g(1,2,t) \quad &:\text{(binary) correlation function} \\
\quad	f(1,2,3,t) \quad &:\text{ternary DF}\\
\quad	T(1,2,3,t) \quad &:\text{ternary correlation function}
\end{align*}
Ignoring the effect of ternary correlation function $T(1,2,3,t)$ (i.e. the effect of three-body interactions, e.g. triple encounters of stars), the single-, double- and triple- DFs may be, in general, rewritten as following Mayer cluster expansion \citep[e.g.][]{Mayer_1940,Green_1956}\footnote{The DFs and correlation functions for stars, in general, may depend on the number $N$ as
\begin{align}
&f(1,t), f(2,t), f(3,t)\propto N,\nonumber\\
&g(1,2,t), g(2,3,t), g(3,1,t)\propto N(N-1),\label{Eq.scale_g(1,2,t)}
\end{align}
where the normalisation condition for DFs and correlation functions follows \citep{Liboff_1966}.}
\begin{subequations}
\begin{align}
&f_{1}(1,t)\equiv f(1,t),\label{Eq.singleDF}\\
&f_{2}(1,2,t)\equiv f(1,2,t)= f(1,t)f(2,t)+\left[g(1,2,t)-\frac{f(1,t)f(2,t)}{N}\right],\label{Eq.doubleDF}\\
&f_{3}(1,2,3,t)=f(1,t)f(2,t)f(3,t)+ g(1,2,t)f(3,t)\nonumber\\
&\qquad\qquad\quad  +g(2,3,t)f(1,t)+ g(3,1,t)f(2,t)\nonumber\\
&\qquad\qquad\quad-\frac{1}{N}\left(2g(1,2,t)+f(1,t)f(2,t)\right)f(3,t)\nonumber\\
&\qquad\qquad\quad-\frac{1}{N}\left(2g(2,3,t)+f(2,t)f(3,t)\right)f(1,t)\nonumber\\
&\qquad\qquad\qquad-\frac{1}{N}\left(2g(3,1,t)+f(3,t)f(1,t)\right)f(2,t),\label{Eq.tripleDF}
\end{align}
\end{subequations}
and under the weak-coupling approximation
\begin{align}
&f_{3}(1,2,3,t)=f(1,t)f(2,t)f(3,t)\nonumber\\
&\qquad\qquad\quad+ \left(g(1,2,t)-\frac{f(1,t)f(2,t)}{N}\right)f(3,t)\nonumber\\
&\qquad\qquad\qquad+\left(g(2,3,t)-\frac{f(2,t)f(3,t)}{N}\right)f(1,t)\nonumber\\
&\qquad\qquad\qquad\quad+\left(g(3,1,t)-\frac{f(3,t)f(1,t)}{N}\right)f(2,t).\label{Eq.tripleDF_WC}
\end{align}
The important difference of star clusters from classical plasmas and ordinary neutral gases can be characterised by the effect of smallness parameter, $1/N$, in equations \eqref{Eq.doubleDF} and \eqref{Eq.tripleDF_WC}; the parameter is not ignorable for dense star clusters $\left(10^{5}\lesssim N \lesssim 10^{7}\right)$. As proved under the weak-coupling approximation by \cite{Liboff_1965,Liboff_1966} and employed by \cite{Gilbert_1968,Gilbert_1971}, the correlation function $g(i,j,t)$ has the anti$-$normalization property for self-gravitating systems
\begin{equation}
\int g(i,j,t)\text{d}_{i}=\int g(i,j,t)\text{d}_{j}=0. \qquad(i,j=1, 2, \text{ or } 3\quad \text{with } i\neq j).\label{Eq.Anti-norm}
\end{equation}
Employing the correlation function $g(1,2,t)$, the total energy, equation \eqref{totE}, may be rewritten as
\begin{align}
&E(t)=\int \frac{\bmath{p}_{1}^{2}}{2m}f_{1}(1,t)\text{d}_{1}+U_{\text{m.f.}}(t)+U_{\text{cor}}(t),\label{Eq.E}
\end{align}
where
\begin{subequations}
\begin{align}
&U_{\text{m.f.}}(t)=\frac{m}{2}\int\Phi(\bmath{r}_{1},t)f(1,t)\text{d}_{1},\label{Eq.U_id}\\
&U_{\text{cor}}(t)=\frac{m}{2}\int\phi(r_{12})g(1,2,t)\text{d}_{1}\text{d}_{2},\label{Eq.U_cor}
\end{align}
\end{subequations}
and the self-consistent gravitational m.f. potential is defined as
\begin{align}
&\Phi(\bmath{r}_{1},t)=\left(1-\frac{1}{N}\right)\int \phi(r_{12})f(2,t)\text{d}_{2},\label{Eq.Phi}
\end{align}
where the factor $\left(1-\frac{1}{N}\right)$ is also the effect of discreteness; the m.f. potential on a star is due to $(N-1)$-field stars \citep[e.g.][]{Kandrup_1986}. Also, the corresponding self-consistent gravitational m.f. acceleration of star 1 reads
\begin{align}
&\bmath{A}(\bmath{r}_{1},t)=-\left(1-\frac{1}{N}\right)\int \nabla_{1}\phi(r_{12})f(2,t)\text{d}_{2}.\label{Eq.A}
\end{align}

\subsection{Scaling of the order of magnitudes of physical quantities}\label{subsec:scalings}
Section \ref{subsec:basic_scaling} explains the basic scalings of physical quantities employed in the present work and in section \ref{subsection:scaling_strong} the scaling associated with strong two-body encounters.

\subsubsection{basic scalings}\label{subsec:basic_scaling}
One needs two scaling parameters for non-divergent kinetic theory; the discreteness parameter, $1/N$, and the distance $r_{12}$ between two stars (say, star 1 is test star at $\bmath{r}_{1}$ and star 2 is one of field stars at $\bmath{r}_{2}$.). The fundamental scaling of physical quantities associated with the discreteness parameter follows the scaling employed in \citep[][Appendix A]{Chavanis_2013a} except for the correlation function $g(1,2,t)$ (equation \eqref{Eq.scale_g(1,2,t)}). For $r_{12}$, following the scaling of the order of magnitudes (OoM) of physical quantities for classical electron-ion plasmas \citep[][pg 22]{Montgomery_1964}, one may classify the effective distance of two-body Newtonian interaction and m.f. acceleration into the following four ranges of distance between two stars depending on the magnitude of forces due to the accelerations on test star in a star cluster system;
\begin{enumerate}
	\item m.f.(many-body) interaction \qquad\quad \quad $d$$<r_{12}<$ $R$
	\item weak m.f.(many-body) interaction \quad $ a_\text{BG}$ $<r_{12}<$ $d$
	\item weak two-body interaction \qquad \quad \quad $r_\text{o}$ $<r_{12}<$ $a_\text{BG}$
	\item strong two-body interaction \qquad  \quad\quad$ 0<r_{12}<$ $r_\text{o}$
\end{enumerate}
where $R$ is the characteristic size of a finite star cluster (e.g. the Jeans length and tidal radius), $d$ the average distance of stars in the system, $r_\text{o}$ the 'conventional' Landau radius (to be explained in section \ref{subsection:scaling_strong}) and $a_\text{BG}$ the Boltzmann-Grad(BG) radius. The BG radius separates the distance $r_{12}$ at which two-body encounters are dominant from those at which the effect of m.f. acceleration (many-body encounters) is dominant; $a_\text{BG}$ corresponds with the scaling of Boltzmann-Grad limit \citep{Grad_1958}\footnote{It is to be noted that the BG radius is in essence the same as the '\emph{encounter radius} \citep{Ogorodnikov_1965}' to separate the encounter and passage of stars.}. For relaxation processes in plasmas \citep{Montgomery_1964}, the BG radius $a_\text{BG}$ is of no essence since the fundamental mathematical formulation assumes homogeneous plasmas and the Thermodynamic limit,
\begin{align}
n=N/V\to\mathcal{O}(1)\qquad (\text{with}\quad V \to \infty\quad \text{and}\quad N \to \infty),
\end{align} 
where $V$ is the system volume of plasmas.

In the present work, the 'Landau radius' $r_{90}$ is newly defined as the closest spatial separation of two stars when the impact parameter of test star is equal to the Landau distance $b_{90}$ (the impact parameter to deflect test star thorough an encounter by $90^\text{o}$ from the original direction of motion);
\begin{subequations}
	\begin{align}
	&r_{90}(\varv_{12}(-\infty))=\frac{b_{90}(\varv_{12}(-\infty))}{1+\sqrt{2}},\label{Eq.r_90}\\
	&b_{90}(\varv_{12}(-\infty))=\frac{2Gm}{\varv^{2}_{12}(-\infty)}\label{Eq.b_90},
	\end{align}
\end{subequations}
where $\varv_{12}(-\infty)$ is the relative speed between star 1 and star 2 before encounter.
 Refer to Tables \ref{table:scale} and \ref{table:scale_fig} for the scalings of basic physical quantities and Appendix \ref{Appendix:interaction} for how some of the scalings, especially the ranges of distances, could be determined. The characteristic time scales of the relevant evolution of DFs and correlation function are defined as 
\begin{subequations}
	\begin{align}
	&\frac{1}{t_\text{dyn}}\simeq \left|\frac{\bmath{\varv}_{1}\cdot\nabla_{1} f(1,t)}{f(1,t)}\right|,\\
	&\frac{1}{t_\text{sec}}\simeq \left|\frac{1}{f(1,t)}\left(\frac{\partial f(1,t)}{\partial  t}\right)\right|,\\
	&\frac{1}{t_\text{cor}}\simeq \left|\frac{1}{g(1,2,t)}\left(\frac{\partial  g(1,2,t)}{\partial  t}\right)\right|.
	\end{align} 
\end{subequations}

\begin{table}\centering
	\caption{A scaling of the order of magnitudes of physical quantities associated with the evolution of a star cluster that has not gone through a core-collapse. The scaling will be especially employed for a completely weakly-coupled- and weakly-inhomogeneous- star clusters in sections \ref{sec:complete_WC} and \ref{sec.WC_strong} respectively, whose density contrast is much less than the order of $N$. The OoM are scaled by $N$ and $r_{12}$ except for the correlation time $t_\text{cor}$, which needs the change in velocity, $\delta\bmath{\varv}_{a}\left(=\int\bmath{a}_{12}\text{d}t_\text{cor}\right)$, due to Newtonian two-body interaction.}
	\begin{tabular}{l l}
		\hline
		quantities & order of magnitude\\
		\hline
		$t_\text{r}$ & $\sim N/\ln[N],$\\
		$f(1,t), t_\text{sec}$ & $\sim N,$\\
		$\bmath{A}_{1},R, m,\bmath{\varv}_{1},\varv_{12},\bmath{r}_{1},t_\text{dyn}$&$\sim 1,$\\
		$d$ &$\sim 1/N^{1/3},$\\
		$a_\text{BG}$ & $\sim1/N^{1/2}$\\
		$G, r_\text{o}, K_{n}$&$\sim 1/N,$\\
		\hline
		$g(1,2,t)$&$ \sim N/r_{12},\quad$ for $r_\text{o}<r_{12}<R$ \\
		          &$\sim N^{2},\qquad$ for $r_{12}<r_\text{o}$\\
		$\bmath{a}_{12}$&$\sim 1/(r_{12}^{2}N) $\\
		$t_\text{cor}$, $\delta\bmath{\varv}_{A}\left(=\int\bmath{A}_{1}\text{d}t_\text{cor}\right)$&$\sim r_{12}$ $\quad$ for  $r_\text{o}<r_{12}<R$\\
		                                                                                            &$\sim\delta \bmath{\varv}_{12} r_{12}^{2}N$ $\quad$ for $r_{12}<r_\text{o}$\\
		\hline
	\end{tabular}\label{table:scale}
\end{table}

\begin{table}
	\caption{A scaling of physical quantities according to the effective interaction range of Newtonian interaction accelerations and close encounter}
	\begin{tikzpicture}
	\filldraw[thin, fill=pink!20!white, opacity=0.1, shading=ball] (0,1)--(8,1) arc(0:30:8);
	\filldraw[thin, fill=pink!30!white, opacity=0.2, shading=ball] (0,1)--(6,1) arc(0:30:6);
	\filldraw[thin, fill=pink!40!white, opacity=0.3, shading=ball] (0,1)--(4,1) arc(0:30:4);
	\filldraw[thin, fill=pink!50!white, opacity=0.4, shading=ball] (0,1)--(2,1) arc(0:30:2);
	\draw(6,1) arc(0:30:6);
	\draw(4,1) arc(0:30:4);
	\draw(2,1) arc(0:30:2);
	\draw[densely dashed, thick] (8,0.9)--(8,-0.4);
	\draw[densely dashed, thick] (6,0.9)--(6,-0.4);
	\draw[densely dashed, thick] (4,0.9)--(4,-0.4);
	\draw[densely dashed, thick] (2,0.9)--(2,-0.4);
	\node[below] at (8,-0.6) {$R$};
	\node[below] at (6,-0.6) {$d$};
	\node[below] at (4,-0.6) {$a_\text{BG}$};
	\node[below] at (2,-0.6) {$r_\text{o}$};
	\node at (0,-0.7) {${r}_{12}$};
	\node at (1,0.5) {strong 2-body};\node at (1,0.2) {(Boltzmann)};
	\node at (3,0.5) {weak 2-body};\node at (3,0.2) {(Landau)};
	\node at (5,0.5) {weak m.f};\node at (5,0.2) {(g-Landau)};
	\node at (7,0.5) {m.f.(many-body)};\node at (7,0.2) {(g-Landau)};
	
	\node[below] at (8,-1.3) {$\sim 1$};
	\node[below] at (6,-1.2) {$\sim 1$};
	\node[below] at (4,-1.2) {$\sim 1$};
	\node[below] at (2,-1.2) {$\sim 1$};
	\node[below] at (0,-1.3) {$\bmath{A}_{1}$};
	
	\node[below] at (8,-2.3) {$\sim N$};
	\node[below] at (6,-2.2) {$\sim N^{4/3}$};
	\node[below] at (4,-2.2) {$\sim N^{3/2}$};
	\node[below] at (2,-2.2) {$\sim N^{2}$};
	\node[below] at (0,-2.3) {$g(1,2,t)$};
	
	\node[below] at (8,-3.0) {$\sim 1/N$};
	\node[below] at (6,-2.9) {$\sim N^{-1/3}$};
	\node[below] at (4,-3) {$\sim 1$};
	\node[below] at (2,-3) {$\sim N $};
	\node[below] at (0,-3) {$\bmath{a}_{12}$};
	
	\node[below] at (8,-3.7) {$\sim 1/N$};
	\node[below] at (6,-3.6) {$\sim N^{-2/3}$};
	\node[below] at (4,-3.6) {$\sim N^{-1/2}$};
	\node[below] at (2,-3.6) {$\sim 1 $};
	\node[below] at (0,-3.6) {$\delta\varv_{\bmath{a}}$};
	
	\node[below] at (8,-4.4) {$\sim 1$};
	\node[below] at (6,-4.3) {$\sim N^{-1/3}$};
	\node[below] at (4,-4.3) {$\sim N^{-1/2}$};
	\node[below] at (2,-4.3) {$\sim N^{-1}$};
	\node[below] at (0,-4.3) {$\delta\varv_{\bmath{A}}$, $t_\text{cor}$};
	
	\shadedraw[inner color=orange, outer color=yellow, draw=black] (0,1) circle (0.1cm);
	\node [above] at (0,1.1) {star 1};
	\shadedraw[inner color=orange, outer color=yellow, draw=black] (6.2,1.2) circle (0.1cm);
	\node [right] at (6.3,1.4) {star 2};
	\end{tikzpicture}\label{table:scale_fig}
\end{table}
\subsubsection{Close encounter and encounters with large-deflection angle and large-speed change }\label{subsection:scaling_strong}
A special focus of the scaling is the Landau radius $r_\text{o}$, equation \eqref{Eq.r_90}, since it especially depends on the relative speed between two stars. A mathematically correct treatment on the Landau distance has been discussed for Newtonian interaction \citep{Retterer_1979,Ipser_1983,Shoub_1992} and Coulombian one \citep{Chang_1992}, until then one had simplified the Landau distance by approximating the relative speed $\varv_{12}$ to the velocity dispersion $<\varv>$ of the system; the 'conventional' Landau- distance, $b_\text{o}$, and and radius, $r_\text{o}$, are defined as
\begin{align}
&b_{90}\simeq\frac{2Gm}{<\varv>^{2}}\equiv b_\text{o},\label{Eq.thermo_approx_b}\\
&r_\text{o}\equiv\frac{b_\text{o}}{1+\sqrt{2}}.\label{Eq.thermo_approx_r}
\end{align}
Assume the dispersion speed may be determined by the Virial theorem for a finite spherical star cluster of radius of $R$ as follows
\begin{align}
<\varv>\equiv c\sqrt{\frac{GmN}{R}},\label{Eq.<v>}
\end{align}
where $c$ is a constant and the radius $R$ may be the Jeans length or tidal radius to hold the finiteness of the system size. Simple examples for the value of the constant $c$ are; $c=\sqrt{3/5}$ if the system is finite and spatially homogeneous and $c$ is order of unity if the system follows the King model \citep{King_1966}. In the present paper, the dispersion approximation is still employed since it simplifies the scaling of the Landau distance without losing the essential property of strong encounters. Employing equations \eqref{Eq.thermo_approx_b}, \eqref{Eq.thermo_approx_r} and \eqref{Eq.<v>},  one finds the relation between the system radius and the Landau distance as follows
\begin{align}
\frac{r_\text{o}}{R}=\frac{2}{1+\sqrt{2}}\frac{1}{c^{2}N}. \label{Eq.ratio_R_ro}
\end{align}

\begin{figure}
	\centering
	\begin{tikzpicture}[scale=1]
	\begin{loglogaxis}[ xlabel=\Large{$\tilde{\varv}_{12}=\varv_{12}(-\infty)/<\varv>$},ylabel=\Large{$b/R$},xmin=0.0008/sqrt(2),xmax=5,ymin=2e-7,ymax=100, legend pos=north east, ytick={1,1e-3,1e-6}, minor ytick={3.33e-6}, ymajorgrids, yminorgrids, minor grid style={orange,dashed} ]
	\addplot [color = red ,mark=no,thick, solid] table[x index=0, y index=1]{b90_velocity_dep_3.txt}; 
	\addlegendentry{\large{$\Delta\varv=<\varv>,c=\sqrt{3/5}$}};
	\addplot [color = blue ,mark=no,thick, densely dashdotted ] table[x index=0, y index=4]{b90_velocity_dep.txt};
	\addlegendentry{\large{$\Delta\varv=\frac{<\varv>}{\sqrt{2N}}$,$c=\sqrt{2}$}}; 
	\addplot [color = green ,mark=no,thick, densely dotted] table[x index=0, y index=5]{b90_velocity_dep.txt}; 
	\addlegendentry{\large{$b=b_{90}(\varv_{12}),c=\sqrt{3/5}$}};
	\addplot [color = blue ,mark=no,thick, densely dashdotted ] table[x index=0, y index=4]{b90_velocity_dep_2.txt};
	\addplot [color = green ,mark=no,thick, densely dotted ] table[x index=0, y index=5]{b90_velocity_dep_2.txt}; 
	\node at (9e-3,10e-6){$b_\text{o}(N\sim10^{6},c=\sqrt{3/5})$}; 
	\end{loglogaxis}
	\end{tikzpicture}
	\caption{The normalised-speed-dependence of the impact parameter $b$  for different changes of speed $\Delta \varv$ of test star, where $b(\tilde{\varv}_{12})=2R[\left({\varv}_{12}(-\infty)/\Delta\varv\right)^{2}-1]^{1/2}/[\sqrt{N}c\tilde{\varv}_{12}]^{2}$ (, which can be derived from equations \eqref{Eq_close_encounter2} and \eqref{Eq_close_encounter3} \emph{without} the dispersion approximation, or see equation 13 of \citep{Shoub_1992}). The solid curve ($\Delta\varv=<\varv>$) separates encounter (scattering) events; the encounter  events described by the region below the solid curve are strong encounters, while those above the curve is weak one. The dotted line separates large-angle- and small-angle deflection of star. In the present paper considering the two-body encounters are spatially-local events due to m.f. acceleration being dominant on larger scales than $a_\text{BG}$, the local weak- and strong- encounters are defined only on the region below the horizontal-grid line of $a_\text{BG}$. Especially, the encounters defined above the Landau distance $r_\text{o}$ are to be called \emph{distant} two-body encounter and encounters below $r_\text{o}$ are \emph{close} one. }
	\label{fig:Landau_distance}
\end{figure}
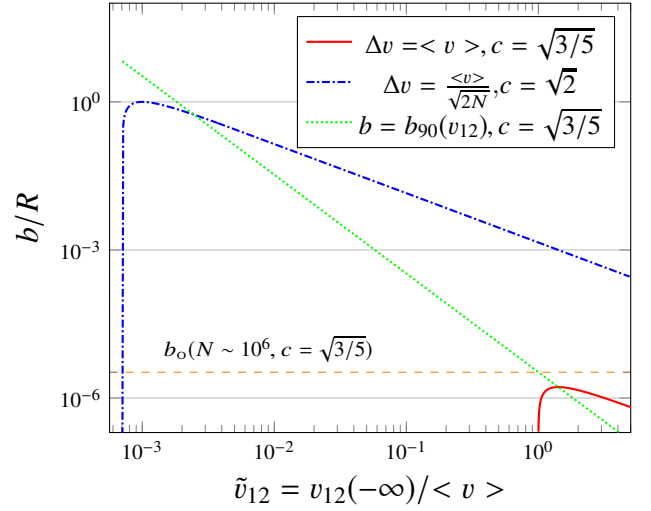

As discussed in \citep{Shoub_1992}, one may separate the impact parameter $b$ of encounter into weak- and strong- deflections following the change in speed of test star through an two-body encounter (Figure \ref{fig:Landau_distance}). In general, kinds of 'strong' two-body encounter is either of large-angle $\left(\gtrsim 90^\text{o}\right)$ deflection and large-speed $\left(\gtrsim <\varv>\right)$ change of test star. In figure \ref{fig:Landau_distance}, the former is described by the region below the dotted curve and the latter is described by the region below the solid curve.  For mathematical convenience, \citep{Shoub_1992} chose the speed change of $ <\varv>/\sqrt{2N}$ (the dashdotted curve on figure \ref{fig:Landau_distance}) to delimit the strong- and weak- encounters at which the change in speed of test star is the same order of the speed change caused by a distant field star on the system-size scale via Newtonian pair-wise acceleration. (Of course one does not have to delimit the encounters since even weak deflections can be described by the Boltzmann-collision description.). However, in more realistic systems, the upper limit of impact parameter for two-body encounter is approximately the BG radius, $a_\text{BG}$, up to which the Boltzmann-collision (collision kinetic) description may be defined. Also actual strong encounters occur only on scales smaller than $b_\text{o}$ (at most $R/(c^{2}N)$)and the slowest relative speed $\varv_{12}(-\infty)$ that causes a large change in speed is equal to the speed dispersion $<\varv>$. Correspondingly, the maximum impact parameter that includes both of strong encounter and large-angle-deflection encounter is the conventional Landau distance, equation \eqref{Eq.thermo_approx_b}. Hence, one may scale the maximum impact parameter as the conventional Landau distance;
\begin{align}
b_\text{max}= b_\text{90}\approx b_\text{o}\sim\mathcal{O}(1/N).\label{Eq.thermo_approx_2}
\end{align}
Equation \eqref{Eq.thermo_approx_2} can be reasonable under the following condition. If one neglects the contribution from energetic stars faster than the escape speed of the system ($\approx 2 <\varv>$), the strong-encounter is 'localized' around the velocity dispersion in relative-speed spaces. Only in this sense, one may employ a dispersion approximation for the relative speed
\begin{align}
\varv_{12}\approx <\varv>\sim\mathcal{O}(1)\label{Eq.thermo_approx_3}
\end{align}

Following the discussion above, one may understand that choosing the conventional Landau distance $b_\text{o}$ as the maximum impact parameter of encounters and employing the dispersion approximation are to focus on each \emph{close-strong} encounter that includes the effects of large-angle-deflection- and strong- encounters on scales smaller than the distance $b_\text{o}$.\footnote{Technically speaking, the relative speed $\tilde{\varv}_{12}$ is a function of the speed $\Delta \varv$ and impact parameter $b$ in finding the explicit form of Boltzmann equation as done in \citep{Shoub_1992} hence the domain of $\tilde{\varv}_{12}$ for strong encounter is to be determined by the relation between the change in speed $\Delta \varv$ and the dispersion $<\varv>$, while one does not need to resort to the serious discussion for the scaling purpose.}
  
Since the relative-speed dependence of the impact parameter may be loosely neglected, one can define the 'Knudsen number' for each close strong encounter by
\begin{align}
K_{n}= R\tilde{n} \sigma^{(p)},\label{Eq.Kn}
\end{align}
where the momentum-transfer cross section \citep[e.g.][]{McQuarrie_2000,Bittencourt_2004,Shevelko_2012}, $\sigma^{(p)}$, due to the close-strong encounters and the corresponding Newtonian-'scattering' relation may be characterised respectively by
\begin{subequations}
	\begin{align}
	&\sigma^{(p)}\approx 2\pi\int^{b_\text{o}}_{0} \left(1-\cos\theta\right)b\text{d}b, \label{Eq_close_encounter1}\\
	&\tan\frac{\theta}{2}\approx\frac{b_\text{o}}{b},\label{Eq_close_encounter2}\\
	&\Delta\varv_{1}\approx<\varv>\sin\frac{\theta}{2},\label{Eq_close_encounter3}
	\end{align}
\end{subequations}
where  $\theta$ is the deflection angle of the unperturbed trajectory of a test star due to each close strong encounter and  the impact parameter $b$ reaches the conventional Landau distance $b_\text{o}$ at $\theta=\pi/2$. It is to be noted that $b_\text{o}$ and $b$ in equations \eqref{Eq_close_encounter1}, \eqref{Eq_close_encounter2} and \eqref{Eq_close_encounter3} do not explicitly or even implicitly depend on relative speed $\varv_{12}$ since the equations are a direct consequence of equations \eqref{Eq.thermo_approx_2} and \eqref{Eq.thermo_approx_3} (i.e. the dispersion approximation).  Also, the mean density $\tilde{n}$ in equation \eqref{Eq.Kn} may be still the order of $N$ since the ejection- and evaporation- rates may be longer than the time scale of secular evolution are less significant except for the core-collapse stage \citep[e.g][]{Spitzer_1988,Binney_2011}. The order of the cross section $\sigma^{(p)}$ is
\begin{align}
&\sigma^{(p)}=2\ln[2]\pi b_\text{o}^{2}\sim\mathcal{O}(1/N^{2}),
\end{align}
correspondingly
\begin{align}
K_{n}= Rn \sigma^{(p)}\sim\mathcal{O}(1/N).
\end{align}
Hence, one may consider the close-strong two-body encounter is also characterized by the discreteness parameter, $1/N$. The Knudsen number $K_{n}$ may be understood as approximately the possibility of finding test star experiencing a close-strong encounter in the 'Landau sphere' (the sphere of radius $r_\text{o}$ around a field star) on dynamical-time scale, $t_\text{dyn}$. In more actual situation, the Landau sphere, of course, does not correctly isolate strong encounters from weak ones; weak encounters may occur even in the Landau sphere due to the relative-speed dependence of the impact parameter. Exactly speaking, the Landau sphere must be exploited to separate close two-body encounters from distant ones, or collision kinetic description (two-body encounters) from wave one (many-body encounters). The latter helps one to understand the importance of the truncated m.f. acceleration $\bmath{A}^{\blacktriangle}(\bmath{r}_{1},t)$ due to the insignificance of the m.f. acceleration on small spatial scales in the secular evolution of a finite system;
\begin{align}
\bmath{A}^{\blacktriangledown}(\bmath{r}_{1},t)&=-\left(1-\frac{1}{N}\right)\int_{{r}_{12}<r_\text{o}} \nabla_{1}\phi(r_{12})f(2,t)\text{d}_{2},\label{Eq.A_down}\\
&\sim \mathcal{O}(1/N).\nonumber
\end{align}
where the the scaling $\partial_{t}f(1,t)\sim\mathcal{O}(1)$ is to be recalled. 

\subsection{Trajectories of a test star}\label{subsec:trajectory}

The complete (Lagrangian) trajectory of star $i$ can be discussed by taking the sum of the m.f. acceleration of star $i$ due to smooth m.f. potential force and the Newtonian pair-wise acceleration via interaction with star $j$;
\begin{subequations}
	\begin{align}
	&\bmath{r}_{i}(t)=\bmath{r}_{i}(t-\tau)+\int_{t-\tau}^{t}\bmath{\varv}_{i}\left(t'\right)\text{d}t',\hspace{15pt} (i\neq j =1,2)\\
	&\bmath{\varv}_{i}(t)=\bmath{\varv}_{i}(t-\tau)+\int_{t-\tau}^{t}\left[\bmath{a}_{ij}\left(t'\right)+\bmath{A}_{i}\left(t'\right)\right]\text{d}t'.
	\end{align}\label{Eq.characteristics_general}
\end{subequations}
One can approximate the complete trajectory to a simpler form in each range of distance between stars $i$ and $j$, following the scaling of section \ref{subsec:scalings}. At distances $r_{ij}<a_\text{BG}$ where two-body Newtonian interaction dominates the other effects, the trajectory perfectly follows a pure Newtonian two-body problem
\begin{subequations}
	\begin{align}
	&\bmath{r}_{i}(t)=\bmath{r}_{i}(t-\tau)+\int_{t-\tau}^{t}\bmath{\varv}_{i}\left(t'\right)\text{d}t',\\
	&\bmath{\varv}_{i}(t)=\bmath{\varv}_{i}(t-\tau)+\int_{t-\tau}^{t}\bmath{a}_{ij}\left(t'\right)\text{d}t'.
	\end{align}\label{Eq.characteristics_two_body}
\end{subequations}
At relatively short distances $(r_{ij}\lesssim r_\text{o})$, the trajectory due to a strong-close encounter may be considered as local Newtonian interaction between two stars (i.e. the Boltzmann two-body collision description if one includes the Markovian approximation)
\begin{subequations}
	\begin{align}
	&\bmath{r}_{12}(t)=\bmath{r}_{12}(t-\tau)+\int_{t-\tau}^{t}\bmath{\varv}_{12}\left(t'\right)\text{d}t',\\
	&\bmath{R}=\frac{\bmath{r}_{1}+\bmath{r}_{2}}{2}\approx \bmath{r}_{1},\\
	&\bmath{r}_{1}=\bmath{r}_{1}(t-\tau)+\int_{t-\tau}^{t}\frac{\bmath{\varv}_{1}\left(t'\right)+\bmath{\varv}_{2}\left(t'\right)}{2}\text{d}t',\\
	&\bmath{\varv}_{i}(t)=\bmath{\varv}_{i}(t-\tau)+\int_{t-\tau}^{t}\bmath{a}_{ij}\left(t'\right)\text{d}t'.
	\end{align}\label{Eq.characteristics_v_jump}
\end{subequations}
At intermediate distances $(r_\text{o}<< r_{ij}\lesssim a_\text{BG})$, the trajectory due to two-body weak-distant encounters may take rectilinear motion local in space with weak-coupling limit
\begin{subequations}
	\begin{align}
	&\bmath{r}_{12}(t)=\bmath{r}_{12}(t-\tau)+\bmath{\varv}_{12}\tau,\\
	&\bmath{R}\approx \bmath{r}_{1},\\
	&\bmath{\varv}_{i}(t)=\bmath{\varv}_{i}(t-\tau).
	\end{align}\label{Eq.characteristics_rectilinear}
\end{subequations}
Lastly at long distances $(a_\text{BG}<<r_{ij}<R)$, the trajectory due to many-body weak-distant encounter may purely follows the motion of star under the effect of m.f. acceleration with weak-coupling limit
\begin{subequations}
	\begin{align}
	&\bmath{r}_{i}(t)=\bmath{r}_{i}(t-\tau)+\int_{t-\tau}^{t}\bmath{\varv}_{i}\left(t'\right)\text{d}t',\\
	&\bmath{\varv}_{i}(t)=\bmath{\varv}_{i}(t-\tau)+\int_{t-\tau}^{t}\bmath{A}_{i}\left(t'\right)\text{d}t'.
	\end{align}\label{Eq.characteristics_m.f.}
\end{subequations}

\subsection{Logarithmic divergences in collision terms}\label{subsec:Log_col}
The mathematical origin of the Coulomb logarithm can be simply found by employing the scaling of physical quantities (section \ref{subsec:scalings}) and the trajectories of stars (section \ref{subsec:trajectory}) for each of collision- and wave- kinetic theories. 

\subsubsection{Coulomb logarithm in collision kinetic theory}
The Boltzmann collision term has the following form (See section \ref{sec.WC_strong} and Appendix \ref{Appendix:Bogorigouv})
\begin{align}
I_\text{Bol}^{\text{(loc)}}=&\int\text{d}^{3}\bmath{p}_{2}\int b\text{d}b \int_{0}^{2\pi}\text{d}\psi \varv_{1,2}\nonumber\\
&\times[-f(1,t)f(\bmath{r}_{1},\bmath{p}_{2},t)+f(\bmath{r}_{1},\bmath{p}_{1}(-\infty),t)f(\bmath{r}_{1},\bmath{p}_{2}(-\infty),t)],\label{Eq.Bol_coll}
\end{align}
where $b$ is the impact parameter of encounter and $\psi$ the azimuth on encounter (scattering) cross section. In equation \eqref{Eq.Bol_coll} the encounter is assumed to be ideal (i.e. local in space and time) and follows the trajectory, equation \eqref{Eq.characteristics_v_jump}. Inside the Landau sphere, the scaling associated with the distance $r_{12}$ is only the impact parameter, hence the collision term, equation \eqref{Eq.Bol_coll}, may be scaled as follows
\begin{align}
I_\text{Bol}^{\text{(loc)}}\sim\int b\text{d}b\propto b^{2}. \label{Eq.Bol_coll_scale_in}
\end{align}

On the other hand, outside the Landau sphere the trajectories of star may approximately follow the rectilinear motion, equation \eqref{Eq.characteristics_rectilinear}, and weak-coupling limit can be applied. The collision term may be expanded as follows
\begin{align}
I_\text{Bol}^{\text{(loc)}}=&\int\text{d}^{3}\bmath{p}_{2}\int b\text{d}b \int_{0}^{2\pi}\text{d}\psi \varv_{1,2} \nonumber\\
&\qquad\times\left[-\Delta\bmath{\varv} \cdot\bmath{\partial}_{12}+\frac{1}{2}\Delta\bmath{\varv} \cdot\bmath{\partial}_{12}\Delta\bmath{\varv} \cdot\bmath{\partial}_{12}\right]f(1,t)f(\bmath{r}_{1},\bmath{p}_{2},t).\label{Eq.Bol_coll_scale_WC}
\end{align}
where the change in the relative velocity $\bmath{\varv}_{12}$ is defined as follows
\begin{align}
\Delta\bmath{\varv}=\int^{\infty}_{0}\bmath{a}_{12}(t-\tau')\text{d}\tau'.
\end{align}
To find the logarithmic divergence one must consider the gap of coordinates between the cylindrical coordinate associated with the impact parameter and the spherical coordinate of the weak-coupling approximation (refer to section \ref{sec:strong} for discussion of the coordinate-space gaps ). Projecting the velocity change $\Delta\bmath{\varv} $ onto the components parallel- and perpendicular- to the cross section, the OoM of the Boltzmann collision term associated with the impact parameter and the deflection (polar) angle $\theta$ is as follows
\begin{subequations}
	\begin{align}
	I_\text{Bol}^{\text{(loc)}}\sim \int b\text{d}b \left\{-\left[\Delta\bmath{\varv} \cdot\bmath{\partial}_{12}\right]_{\perp} \cos\theta+\frac{\sin^{2}\theta}{2}\left[\Delta\bmath{\varv} \cdot\bmath{\partial}_{12}\Delta\bmath{\varv} \cdot\bmath{\partial}_{12}\right]_{\parallel}\right\},
	\end{align}
\end{subequations}
Since a large impact parameter (larger than the Landau distance) is a reciprocal of $\sin[\theta/2]$ for small $\theta$ (See equation \eqref{Eq_close_encounter2}), one obtains 
\begin{align}
I_\text{Bol}^{\text{(loc)}}\sim \ln[b]\left\{\left[\Delta\bmath{\varv} \cdot\bmath{\partial}_{12}\right]_{\perp}+\frac{1}{2}\left[\Delta\bmath{\varv} \cdot\bmath{\partial}_{12}\Delta\bmath{\varv} \cdot\bmath{\partial}_{12}\right]_{\parallel} \right\}.  \label{Eq.Bol_coll_scale_out}
\end{align}

To avoid the logarithmic divergence in equation \eqref{Eq.Bol_coll_scale_out}, one needs to take an upper limit on the impact parameter while the logarithmic term itself is in general problematic. The Taylor expansion of the Boltzmann collision term is not a correct treatment since the higher order terms can be greater than the logarithmic term for stars moving at the high speeds $\varv_{1}>\sqrt{\ln[N]}<\varv>$ \citep{Shoub_1987} and large-angle deflections can occur even for large impact parameters as shown in Figure \ref{fig:Landau_distance}. However, the value of $\sqrt{\ln[N]}$ is approximately $3.4\sim 4.0$ for $N=10^{5}\sim10^{7}$, implying if one assumes that stars with velocities higher than the typical escape speed $(\sim2<\varv>)$ are less likely to exist in star clusters, the logarithmic term may be dominant yet. Also, the logarithmic divergence may be a  correct scaling for the Boltzmann collision term under the dispersion approximation of the Landau distance and in limit of $b\to\infty$ since the conditions can avoid the large-angle deflection for impact parameters much larger than the Landau radius, which is the case of the CKT (section \ref{sec:discussion}). For more serious discussion of logarithmic terms and cut-off problems, refer to \citep{Montgomery_1964,Liboff_2003} and \citep{Chang_1992,Shoub_1992} without- and with- thorough discussion of the velocity dependence of the impact parameter respectively.

\subsubsection{Coulomb logarithm in wave kinetic theory}\label{subsec:Coulomb_log}
Consider the general form of the Landau-collision term (See e.g. sections \ref{subsec:nonideal} and \ref{subsec:grainess_effect_tr} and Appendix \ref{Appendix.Ext.non_ideality}), the most often-used in stellar-dynamics study, 
\begin{align}
&I=\bmath{\partial}_{1}\cdot\int\text{d}_{2} \bmath{a}_{12}\int_{0}^{\tau}\text{d}\tau'\nonumber\\
&\quad\times \left[\bmath{a}_{12}\cdot\bmath{\partial}_{12}\right]_{t-\tau'} f(1(t-\tau'),t-\tau')f(2(t-\tau'),t-\tau'),\label{Eq.Landau}
\end{align}
where the trajectory of stars is assumed either of equations \eqref{Eq.characteristics_v_jump}, \eqref{Eq.characteristics_rectilinear} and \eqref{Eq.characteristics_m.f.}. As typically done to derive the explicit form of the Landau collision term, one may employ the Fourier transform of the accelerations $\bmath{a}_{12}$ and factors related to the distance $r_{12}$. However, to grasp the dependence of the collision term on wavenumber $k$, one does not need an extensive mathematics since all the factors in the integrand of the collision term can be described in the same spherical coordinate unlike the collision kinetic theory; one may simply take the following dimensional relation
\begin{align}
k\sim \frac{1}{r_{12}}.\label{Eq.recip}
\end{align}
Then, the collision term, equation \eqref{Eq.Landau}, associated with the wavenumber through the distance $r_{12}$ and correlation time, $\tau$, may approximately follow
\begin{align}
I\sim\int \tau\text{d}k.\label{Eq.g_Landau_approx}
\end{align}
The divergence problems for the wave kinetic theory relies on the correlation time integral. The wavenumber dependences of the collision term for the three trajectories, equations \eqref{Eq.characteristics_v_jump}, \eqref{Eq.characteristics_rectilinear} and \eqref{Eq.characteristics_m.f.}, are as follows
\begin{align}
I\sim\quad
\begin{cases}
k, & \text{if $\quad a_\text{BG}<< r_{12}\lesssim R$}\\
\ln[k], & \text{if $\quad r_\text{o}<<r_{12}<a_\text{BG}$}\\
1/k. & \text{if $\quad0<r_{12}<r_\text{o}$}
\end{cases}       
\end{align}
where the scalings of correlation time in section \ref{subsec:scalings} are employed. 

It, of course, is not meaningful to take a cutoff on a specific distance to avoid the mathematical divergence of the wavenumbers in the collision term as will be shown in section \ref{subsec:grainess_effect_tr} since the reciprocal relation, equation \eqref{Eq.recip}, itself is not a correct mathematical treatment; one must employ the complete trajectory, equation \eqref{Eq.characteristics_general}, to correctly manage the divergence problem for the collision term or the CKT as will be explained in section \ref{sec:discussion}. For more serious discussion of the explicit forms and divergence problems of the collision terms in wave kinetic theories, refer to \citep{Polyachenko_1982,Kandrup_1981_a,Heyvaerts_2010,Chavanis_2013a}.

%% file: Section3_truncatedBBGKY_nonideality.tex
\section{BBGKY hierarchy for truncated distribution function and non-ideal theory}\label{sec:truncated_BBGKY}
In section \ref{subsec:DF} non-standard DFs are explained, and section \ref{subsec:BBGKY} shows the BBGKY hierarchies for the DFs. The \citep{Klimontovich_1982}'s theory of non-ideal systems is explained in section \ref{subsec:nonideal}.

\subsection{Non-standard distribution functions}\label{subsec:DF}
Since the scaling of OoM of physical quantities to describe a star cluster is naive to the effective interaction ranges of Newtonian pairwise- and m.f.- potentials (section \ref{subsec:scalings}), one may resort to variant forms of two-body and higher-order-body DFs according to the OoM of the potentials. In the present section, the Grad's truncated DF, 'weakly-coupled' DF and 'domain-patched' DF are explained. 

\subsubsection{Truncated distribution function}\label{subsec:trunc_DF}
The truncated DF was originally introduced by \cite{Grad_1958} to derive the collisional Boltzmann equation for rarefied gases of particles interacting each other via short-range interaction of an effective potential distance $\triangle$. In the outside of sphere of radius $\triangle$ around test particle, one assumes no two-body interaction with a field particle occurs, or the pair-wise potential is much weaker than the inside of the sphere. The deficiency of the truncated DF, being not symmetric about permutation between the states of two stars, was improved in \citep{Cercignani_1972,Cercignani_1988} where the BBGKY hierarchies for the truncated DF of hard spheres and particles interacting via short-range pair potential were derived. The advantage of exploiting the truncated DF is three fold; (i) Among various derivations of the Boltzmann collision term, only the \citep{Grad_1958}'s method has a mathematically strict limit (Boltzmann-Grad limit); the ratio of particle size to the total particle numbers as proved in \citep{Lanford_1981} and can avoid the mathematical divergence problem at the $N$-body Liouville-equation level (ii) The \citep{Grad_1958}'s method allows one to derive a kinetic equation in spherical coordinates; one can discuss the effect of two-body encounters and the statistical acceleration in the same coordinates\footnote{The wave kinetic theories are in general discussed in spherical coordinates in terms of relative displacement between two stars, while collision ones typically assumes cylindrical coordinates (Appendix \ref{Appendix:Bogorigouv}).} (iii) Statistical dynamics of two-body encounter can be separated at $r_{ij}=\triangle$ from the deterministic Newtonian mechanics inside the Landau sphere \citep[e.g.][]{Cercignani_2008}.

An $s$-tuple truncated DF of stars may be defined as\footnote{\cite{Cercignani_1972,Cercignani_1988} used the $s$-body (symmetric) joint-probability DF and the Boltzmann-Grad limit ($N\triangle^{2}\rightarrow\mathcal{O}(1)$ as $N\to\infty$), meaning the small number $s$ in the factor $\frac{N!}{(N-s)!}$ is less important, while stellar dynamics necessitate the small $s$ to discuss the granularity. Accordingly, the formulas shown in the present work are slightly different from the Cercignanni's work due to the definition for DF.}
\begin{equation}
f^{\triangle}_{s}(1, \cdots, s, t)=\frac{N!}{(N-s)!}\int_{\Omega_{s+1,N}}F_N(1,\cdots,N, t)\text{d}_{s+1}\cdots\text{d}_N,\label{Eq.SbodytruncDF}
\end{equation} 
where  $1\lid s \lid N-1$. The effective interaction range $\triangle$ throughout the present paper is considered as the Landau radius $r_\text{o}$. Hence, the symbol $\triangle$ means a transition scales of the separation between wave- and collision kinetic- descriptions. Equation \eqref{Eq.SbodytruncDF} is in essence the same as the definition for the truncated DF used in \citep{Cercignani_1972} though, it can be rewritten in a reduced form due to the $s$-body DFs being symmetric in permutation between two phase-space states of stars; the domain of integration in equation \eqref{Eq.SbodytruncDF} must be taken over the limited phase-space volumes $\Omega_{s+1,N}$ defined by
\begin{equation}
	\Omega_{s+1,N}=\left(\{\bmath{r}_{s+1},\bmath{p}_{s+1}\cdots\bmath{r}_{N},\bmath{p}_{N}\}\Bigg\vert \prod_{i=s+1}^{N}\prod_{j=1}^{i-1}\left\{{\mid \bmath{r}_{i}-\bmath{r}_{j} \mid > \triangle}\right\}\right).\label{Int_Omega}
\end{equation}
For example,
\begin{subequations}
\begin{align}
&\Omega_{2,2}=\left(\{\bmath{r}_{2},\bmath{p}_{2}\} \Big\vert\{\mid \bmath{r}_{1}-\bmath{r}_{2} \mid > \triangle\}\right),\\
&\Omega_{3,3}=\left(\{\bmath{r}_{3},\bmath{p}_{3}\}\Big\vert\{\mid \bmath{r}_{1}-\bmath{r}_{3} \mid > \triangle\}\times\{\mid \bmath{r}_{2}-\bmath{r}_{3} \mid > \triangle\}\right),
\end{align}
\end{subequations}
and refer to Appendix \ref{Appendix_BBGHY} for more detail discussion. 
\begin{table}
	\caption{A schematic description of the truncated DF. Kinetic description for stars $1,2,\cdots,k,\cdots, N$ follows the wave kinetic description unless one of stars enters the Landau sphere of another star while the collision kinetic description must be employed if any two stars approaches closer than the Landau radius.}
	\begin{tikzpicture}
	\shade[ball color = gray!40, opacity = 0.4] (0,0) circle (1.41421cm);
	\draw (0,0) circle (1.41421cm);
	\draw (-1.41421,0) arc (180:360:1.41421 and 0.6);
	\draw[dashed] (1.41421,0) arc (0:180:1.41421 and 0.6);
	\draw[dashed] (0,0 ) -- node[above]{$\triangle$} (1.41421,0);
	\shadedraw[inner color=orange, outer color=yellow, draw=black] (0,0) circle (0.1cm);
	\node [below] at (0,0.0) {star 1};	
	
	\shade[ball color = gray!40, opacity = 0.4] (4.5,-3) circle (1.41421cm);
	\draw (4.5,-3) circle (1.41421cm);
	\draw (3.11421,-3) arc (180:360:1.41421 and 0.6);
	\draw[dashed] (5.91421,-3) arc (0:180:1.41421 and 0.6);
	\draw[dashed] (4.5,-3 ) -- node[above]{$\triangle$} (5.91421,-3);
	\shadedraw[inner color=orange, outer color=yellow, draw=black] (4.5,-3.0) circle (0.1cm);
	\node [below] at (4.5,-3.0) {star 2};
	
	\shade[ball color = gray!40, opacity = 0.4] (4,1) circle (1.41421cm);
	\draw (4,1) circle (1.41421cm);
	\draw (2.61421,1) arc (180:360:1.41421 and 0.6);
	\draw[dashed] (5.41421,1) arc (0:180:1.41421 and 0.6);
	\draw[dashed] (4,1) -- node[above]{$\triangle$} (5.41421,1);
	\shadedraw[inner color=orange, outer color=yellow, draw=black] (4,1) circle (0.1cm);
	\node [below] at (4.0,1) {star $k$};
	
	\end{tikzpicture}
\end{table}
The truncated $s$-tuple DF $f^{\triangle}_{s}(1, \cdots, s, t)$ is also assumed symmetric about a permutation between two states. The truncated single- and double- DFs explicitly read
\begin{subequations}
\begin{align}
&f^{\triangle}_{1}(1,t)\nonumber\\
&=f_{1}(1,t)-\frac{1}{2}\int_{r_{23} < \triangle}f_{3}(1,2,3,t)\text{d}_{2}\text{d}_3-\int_{r_{12} < \triangle}f_{2}(1,2,t)\text{d}_{2}\nonumber\\
&\quad+\left[i\int_{\substack{r_{12} < \triangle\\\times r_{13}< \triangle}}+\iint_{\substack{r_{12}  < \triangle\\\times r_{23}< \triangle}}-\iint_{\substack{r_{12}  < \triangle\\ \times r_{13}  < \triangle\\\times r_{23}< \triangle}}\right]f_{3}(1,2,3,t)\text{d}_{2}\text{d}_{3}\nonumber\\
&\qquad-...\label{Eq.single_truncated_DF}\\
&f^{\triangle}_{2}(1,2,t)\nonumber\\
&=f_{2}(1,2,t)-\frac{1}{2}\int_{r_{34} < \triangle}f_{4}(1,2,3,4,t)\text{d}_{3}\text{d}_4\nonumber\\
&\quad-\left[\int_{\mid r_{13} \mid < \triangle}+\int_{\mid r_{23} \mid < \triangle}-\int_{\substack{\mid r_{13} \mid < \triangle\\\times\mid r_{23} \mid < \triangle}}\right]f_{3}(1,2,3,t)\text{d}_{3}\nonumber\\
&\quad+\iint\text{d}_{3}\text{d}_{4} f_{4}(1,\cdots, 4,t)\nonumber\\
&\qquad\times\{\Theta(\triangle-r_{13})[\Theta(\triangle-r_{14})+\Theta(\triangle-r_{23})+\Theta(\triangle-r_{34})]\nonumber\\
&\qquad\quad+\Theta(\triangle-r_{23})[\Theta(\triangle-r_{24})+\Theta(\triangle-r_{34})]\nonumber\\
&\qquad\quad-2\Theta(\triangle-r_{13})\Theta(\triangle-r_{23})\nonumber\\
&\qquad\qquad\times[\Theta(\triangle-r_{14})+\Theta(\triangle-r_{23})+\Theta(\triangle-r_{34})]\}\nonumber\\
&\qquad-...\label{Eq.double_truncated_DF}
\end{align}
\end{subequations}
where $\Theta(\cdot)$ describes a Heaviside step function. Hence the truncated single (double) DF describes the probability $not$ to find star 1 (star 1 or 2) around star 2 (star 3) within the region inside a sphere of radius $\triangle$ (the Landau sphere) at time $t$. Despite of the mathematically strict definition for the truncated DFs, it does not have a straightforward physical meaning; one may resort to a simplification of the truncated DF\footnote{In \citep{Grad_1958,Cercignani_1972}, the interaction range essentially goes to zero due to the Boltzmann-Grad limit $\triangle\sim\frac{1}{\sqrt{N}} \to 0$ and the truncated DF is considered as a standard DF.}. Due to the shortness of the interaction range of  $\triangle$ between two stars
\begin{align}
\triangle\equiv r_\text{o}\sim\mathcal{O}\left(\frac{1}{N}\right),\label{Eq.Landau_distance}
\end{align}
the truncated single- and double- DFs can be approximated to
\begin{subequations}
	\begin{align}
	&f^{\triangle}_{1}(1,t)=f_{1}(1,t)-\frac{1}{2}\iint_{r_{23} < \triangle}f_{3}(1,2,3,t)\text{d}_{2}\text{d}_{3}+\mathcal{O}(1/N),\label{Eq.trunc_DF1_N}\\
	&f^{\triangle}_{2}(1,2,t)=f_{2}(1,2,t)-\frac{1}{2}\iint_{r_{34} < \triangle}f_{4}(1,2,3,4,t)\text{d}_{3}\text{d}_{4}+\mathcal{O}(1).\label{Eq.trunc_DF2_N}
	\end{align}\label{Eq.approx_truncated_DFs}
\end{subequations}
The second terms on the R.H.S of equations \eqref{Eq.trunc_DF1_N} and \eqref{Eq.trunc_DF2_N} show the effect of discreteness on the DFs. One should be aware of the effect of discreteness on the truncated DF being associated with the randomness (fluctuation in the m.f. potential \citep{Chandra_1943a,Takase_1950}) rather than that one generally discusses\footnote{It is obvious in stellar dynamics that a strict definition for typical DF itself is difficult to achieve due to the 'discreteness' or granularity of the system in phase space $(\bmath{r},\bmath{p})$. The 'discreteness' stands for 'sparse' physical infinitesimal elements of phase space \citep[][pg. 9]{Spitzer_1988}; what one can do is to take the DF in terms of integrals of motion and orbit-averaging it.}. The obvious complication of the DFs, equations \eqref{Eq.trunc_DF1_N} and \eqref{Eq.trunc_DF2_N}, may be comforted by excluding the possibility of triple encounter. In the Landau sphere of radius $r_{23}=\triangle$ or $r_{34}=\triangle$,  any stars other than the stars of concern (stars 2 and 3 or stars 3 and 4 respectively) can not exist in the Landau sphere under the two-body encounter approximation. Hence, equation \eqref{Eq.approx_truncated_DFs} can be reduced to
\begin{subequations}
	\begin{align}
	&f^{\triangle}_{1}(1,t)=f_{1}(1,t)\left(1-\frac{1}{2}\iint_{r_{23} < \triangle}f_{2}(2,3,t)\text{d}_{2}\text{d}_{3}\right),\label{Eq.trunc_DF1_N_2}\\
	&f^{\triangle}_{2}(1,2,t)=f_{2}(1,2,t)\left(1-\frac{1}{2}\iint_{r_{34} < \triangle}f_{2}(3,4t)\text{d}_{3}\text{d}_{4}\right).\label{Eq.trunc_DF2_N_2}
	\end{align}\label{Eq.approx_truncated_DFs_2}
\end{subequations}
The fundamental idea of truncated DF is that the truncation of phase-space volume makes the system 'open' on small scales. This may be clearly understood if one takes the integral $\int\cdot\text{d}_{1}$ over equation \eqref{Eq.trunc_DF1_N_2} and $\iint\cdot\text{d}_{1}\text{d}_{2}$ over equation \eqref{Eq.trunc_DF2_N_2};
\begin{subequations}
	\begin{align}
	&\int f^{\triangle}_{1}(1,t)\text{d}_{1}=N\left(1-\frac{1}{2}\iint_{r_{23} < \triangle}f_{2}(2,3,t)\text{d}_{2}\text{d}_{3}\right),\label{Eq.trunc_N1}\\
	&\iint f^{\triangle}_{2}(1,2,t)\text{d}_{1}\text{d}_{2}\nonumber\\
	&\qquad=N(N-1)\left(1-\frac{1}{2}\iint_{r_{34} < \triangle}f_{2}(3,4t)\text{d}_{3}\text{d}_{4}\right).\label{Eq.trunc_N2}
	\end{align}\label{Eq.approx_truncated_N}
\end{subequations}
The total number of stars described by the truncated DFs does not conserve since the DFs 'overlook' counting the probable number of stars in the Landau spheres(, which is useful only for binary formation and disruption/coalescence.). This obvious complication may be avoided by assuming two different assumptions. First, one may assume no star can approach another star than the Landau radius. Such stars will be termed \emph{weakly-coupled} (WC) stars in the present paper. The WC stars are mathematically defined in section \ref{sec:WC_star} and applied to a star cluster in section \ref{sec:complete_WC}. Second, one may also apply the 'test-particle' method of \citep{Kaufman_1960,Kandrup_1981} explained in Appendix \ref{sec.many_to_two}. In the present paper, 'test-particle' method means that only test star (star 1) can approach one of field stars closer than the Landau radius but none of the other field stars can, meaning one does not find any stars in the Landau sphere of radius $r_{23}=\triangle$ or $r_{34}=\triangle$ (See section \ref{sec.WC_strong}). (It is to be noted whether star 1 is in the Landau sphere of star 2 or not is not a crucial discussion since it comes into a play at order of $1/N^{2}$ as seen in the third term on the R.H.S of equation \eqref{Eq.single_truncated_DF}.). The both of assumptions (WC-stars approximation or the 'test-particle' method) can avoid the non-conservation of total number of stars;
\begin{subequations}
	\begin{align}
	&f^{\triangle}_{1}(1,t)=f_{1}(1,t)+\mathcal{O}(1/N^{2}),\label{Eq.trunc_DF1_test}\\
	&f^{\triangle}_{2}(1,2,t)=f_{2}(1,2,t)+\mathcal{O}(1/N).\label{Eq.trunc_DF2_test}
	\end{align}
\end{subequations}
Hence, the truncated $s$-tuple DFs of stars may be treated as the standard DFs, equations \eqref{Eq.singleDF} and \eqref{Eq.doubleDF}.

The total energy of stars in a star cluster has the following forms in terms of the truncated DFs
\begin{align}
&E(t)^{\triangle}=\int \frac{\bmath{p}_{1}^{2}}{2m}f^{\triangle}(1,t)\text{d}_{1}+U^{\triangle}(t),\label{Eq.E_trun}
\end{align}
where
\begin{align}
	U^{\triangle}(t)=\frac{m}{2}\int_{r_{12} > \triangle}\phi(r_{12})f^{\triangle}(1,2,t)\text{d}_{1}\text{d}_{2},\label{Eq.U_trun}
\end{align}
In the same way as the non-conservation of total number of stars, the truncated DFs do not conserve the total energy. If one does not resort to any approximation, equation \eqref{Eq.E_grain} states even the total energy of a finite star cluster must be conserved only up to order of $\mathcal{O}(1)$. Hence, employing the WC-star approximation or 'test-particle' method, one obtains the total energy of stars outside the Landau spheres
\begin{align}
&E(t)^{\triangle}=\int \frac{\bmath{p}_{1}^{2}}{2m}f(1,t)\text{d}_{1}+U_{\text{m.f.}}^{\triangle}(t)+U_{\text{cor}}^{\triangle}(t),\label{Eq.E_grain}
\end{align}
where
\begin{subequations}
	\begin{align}
	&U_{\text{m.f.}}^{\triangle}(t)=\frac{m}{2}\int\Phi^{\triangle}(\bmath{r}_{1},t)f(1,t)\text{d}_{1},\label{Eq.U_id_grain}\\
	&U_{\text{cor}}^{\triangle}(t)=\frac{m}{2}\int_{r_{12} > \triangle}\phi(r_{12})g(1,2,t)\text{d}_{1}\text{d}_{2},\label{Eq.U_cor_grain}
	\end{align}
\end{subequations}
and the self-consistent \emph{truncated} m.f. potential is defined as 
\begin{align}
&\Phi^{\triangle}(\bmath{r}_{1},t)=\left(1-\frac{1}{N}\right)\int_{r_{12} > \triangle} \phi(r_{12})f(2,t)\text{d}_{2}.\label{Eq.Phi_grain}
\end{align}
The corresponding \emph{truncated} m.f. acceleration reads
\begin{align}
&\bmath{A}^{\triangle}(\bmath{r}_{1},t)=-\left(1-\frac{1}{N}\right)\int_{r_{12} > \triangle} \nabla_{1} \phi(r_{12})f(2,t)\text{d}_{2}.\label{Eq.A_grain}
\end{align}
One must recall that the DFs, equations \eqref{Eq.trunc_DF1_test} and \eqref{Eq.trunc_DF2_test}, inside the Landau sphere do not have a statistically strict meaning. The truncated m.f. potential, equation \eqref{Eq.Phi_grain}, and acceleration, equation \eqref{Eq.A_grain}, seem an artificial concept though, it gives a clear physical meaning. The truncated DF assigns a \emph{geometrical} constraint on a standard double DF (both of the product of uncorrelated DFs and correlation function) that the dynamics of stars (e.g. Newtonian two-body interaction, formation of binaries, coalescence and disruption) inside the Landau sphere does not 'coincide' with the statistical quantity at the same distance to describe the system, which corresponds with the 'rough approximation \citep{Takase_1950}' of randomness in Holtsmark DF. Hence, fluctuations in m.f. acceleration can be excited only outside the sphere. The truncated m.f. acceleration, equation \eqref{Eq.A_grain}, also stands for a case in which the m.f. acceleration of a star due to stars traveling in a Landau sphere does not contribute to the stellar dynamics. (Hence, the polarisation across the surface of the Landau sphere must be ignored.). In section \ref{subsec:grainess_effect_Poisson}, the Poisson equation for the truncated DF of stars will be explained.

\subsubsection{'Weakly-coupled' Distribution Function}\label{sec:WC_star}
To avoid the non-conservation of total- number and energy of stars described by the truncated DF, in the present section, the hard-sphere DF \citep{Cercignani_1972} will be extended to the weakly-coupled DF of stars. \cite{Cercignani_1972} extended the Grad' truncated DF into the hard-sphere DF to derive the collisional Boltzmann equation for rarefied gases of hard-sphere particles. The hard-sphere model does not allow any particles of radii $\triangle$ exist inside the other particles of radii $\triangle$ in a rarefied gas; it is defined as
\begin{align}
&f_{N}^{\blacktriangle}(1,\cdots, N,t)=
\begin{cases}
f_{N}(1,\cdots,N,t),     & \text{if}\quad r_{ij}\geq \triangle \text{with}\quad i\neq j \\
0,              & \text{otherwise} \label{Eq.truncDf_HS}
\end{cases}
\end{align}
Following the definition of single- and double- truncated DFs, equations \eqref{Eq.single_truncated_DF} and \eqref{Eq.double_truncated_DF}, the first two $s$-tuple hard-sphere DFs explicitly read
\begin{subequations}
\begin{align}
&f_{1}^{\blacktriangle}(1,t)=f_{1}(1,t),\\
&f_{2}^{\blacktriangle}(1,2,t)=
\begin{cases}
f_{2}(1,2,t),     & \text{if}\quad r_{12}\geq \triangle\\
0,              & \text{otherwise} \label{Eq.truncDf_HS_12}
\end{cases}
\end{align}
\end{subequations}
In equation \eqref{Eq.truncDf_HS_12}, the hard-sphere double DF is smooth and continuous, well-defined as limit of $r_{12}\rightarrow\triangle^{+}$, while it can be discontinuous  as limit of $r_{12}\rightarrow\triangle^{-}$. Hence, the value of the double DF at the radius $r_{12}=\triangle$ is defined as the limit value 
\begin{align}
\left[f_{2}(1,2,t)\right]_{r_{12}=\triangle} = \lim_{r_{12}\to \triangle+}f_{2}^{\blacktriangle}(1,2,t). \label{Eq.truncDf_HS_lim}
\end{align}
This mathematical definition gives the thresh point $(r_{12}=\triangle)$ a physical causality in space, i.e. a direct collision between two spheres occurs only from the outside of each sphere. One must be careful to deal with the explicit form of double or higher order of $s$-tuple hard-sphere DF. The double DF may be explicitly defined as
\begin{subequations}
	\begin{align}
	f_{2}^{\blacktriangle}(1,2,t)&=
	\begin{cases}
	\left(1-\frac{1}{N}\right)f(1,t)f(2,t)+g(1,2,t),     & \text{if}\quad r_{12}\geq \triangle\\
	0,              & \text{otherwise} \label{Eq.truncDf_HS_expl}
	\end{cases}\\
	                             &\equiv \left(1-\frac{1}{N}\right)\left[f(1,t)f(2,t)\right]_{r_{12}\geq \triangle}+g(1,2,t)_{r_{12}\geq \triangle},\label{Eq.hard_f(1,2,t)}
	\end{align}
\end{subequations}
where the DFs $f(1,t)$ and $f(2,t)$ are not exactly statistically uncorrelated since the geometrical condition assigned on the interaction range, $r_{12}>\triangle$, must be considered; only the DFs $f^{\blacktriangle}(1,t)$ and $f^{\blacktriangle}(2,t)$ are statistically independent each other. Hence,
\begin{align}
 \left[f(1,t)f(2,t)\right]_{r_{12}\geq \triangle}\neq f^{\blacktriangle}(1,t)f^{\blacktriangle}(2,t). \label{Eq.neq_DFs}
\end{align}
Also, the hard-sphere DF is different from the DF, equation \eqref{Eq.SbodytruncDF}, in sense that the phase-space domain of truncated DF is limited always through that of integration, while hard-sphere does not have domain itself in the Landau sphere.

On a star cluster if one assumes a strong constraint that any star can not approach any other stars closer than the Landau distance, equation \eqref{Eq.Landau_distance} (while the maximum separation between stars is bounded by the system size), the weak-coupling approximation may be actually embodied:
\begin{align}
&r_\text{o}\qquad  < \quad r_{12}\quad \leq\qquad  R, \label{Eq.weak_coupling}\\
&\mathcal{O}\left(\frac{1}{N}\right) \qquad\qquad\qquad\quad \mathcal{O}(1)\nonumber
\end{align}
This ideal mathematical condition is interpreted as an extreme case of the hard-sphere DF, equation\eqref{Eq.truncDf_HS}, with the limit value of zero for $f(1,2,t)$ at $r_{12}=\triangle$;
\begin{align}
\left[f_{2}(1,2,t)\right]_{r_{12}=\triangle} = 0.\label{Eq.WC_r12}
\end{align}
The hard-sphere DF with the condition, equation \eqref{Eq.WC_r12}, is termed a weakly-coupled DF in the present work to isolate itself from hard-sphere DF. The weakly-coupled DF in essence corresponds with the 'Rough approximation \citep{Takase_1950}' of the random factor for the Holtsmanrk distribution of Newtonian force strength, meaning the relative velocity dependence between test- and a field- star will be neglected when the test star entering the Landau sphere in the present work for simplicity.

\subsubsection{Domain-patched DF}
Since the truncated- and weakly-coupled- DFs are meaningful only outside the Landau sphere, one needs another DF to discuss the statistical dynamics inside the Landau sphere for a late stage of evolution of core collapse of a star cluster. In the present section, the hard-sphere $s$-body DF is modified into a 'domain-patched' $s$-body DF:
\begin{align}
f_{N}^{p}(1,\cdots, N,t)\equiv f_{N}^{\blacktriangle}(1,\cdots,N,t)+f_{N}^{\blacktriangledown}(1,\cdots, N,t),
\end{align}
where
\begin{align}
&f_{N}^{\blacktriangledown}(1,\cdots, N,t)=
\begin{cases}
0,     & \text{if}\quad r_{ij}>\triangle\quad (i\neq j)  \\
f_{N}(1,\cdots, N,t),     & \text{otherwise} \label{Eq.truncDf_patched}
\end{cases}
\end{align}
where the patched DF $f_{N}^{p}(1,\cdots, N,t)$ is assumed mathematically satisfactory smooth, continuous and differentiable at any points in phase space (even at the transition distance $r_{ij}=\triangle$). The corresponding $s$-body DF may be assumed computed in the same way as standard DFs, hence the first two patched DFs explicitly read
\begin{subequations}
\begin{align}
&f_{1}^{p}(1,t)=f_{1}(1,t),\\
&f_{2}^{p}(1,2,t)=f_{2}^{\blacktriangle}(1,2,t)+f_{2}^{\blacktriangledown}(1,2,t),\\
&\hspace{35pt}=
	\begin{cases}
	f(1,t)f(2,t)+g(1,2,t),     & \text{if}\quad r_{12}\geq \triangle\\
	f(1,t)f(2,t)+g(1,2,t),              & \text{otherwise} \label{Eq.truncDf_HS_patched}
	\end{cases}\\
&\hspace{35pt}= \left[f(1,t)f(2,t)\right]_{r_{12}\geq \triangle}+g(1,2,t)_{r_{12}\geq \triangle}\nonumber\\
&\hspace{35pt}\qquad+\left[f(1,t)f(2,t)\right]_{r_{12}\leq \triangle}+g(1,2,t)_{r_{12}\leq \triangle}.
	\end{align}
\end{subequations}

\subsection{BBGKY hierarchies for standard, truncated and hard-sphere DFs}\label{subsec:BBGKY}
In a very similar way to the derivation of standard BBGKY hierarchy, the BBGKY hierarchy for the truncated $s$-body function can be found (refer to Appendix \ref{Appendix_BBGHY}, or see \cite{Cercignani_1972,Cercignani_1988}) as 
\begin{align}
&\partial_{t}f_{s}+\sum_{i=1}^{s}\left[\bmath{\varv}_{i}\cdot\nabla_{i}+\sum_{\quad{j=1(\neq i)}}^{s}\bmath{a}_{ij}\cdot\bmath{\partial}_{i}\right]f_{s}\nonumber\\
&\qquad +\sum_{i=1}^{s}\left[\bmath{\partial}_{i}\cdot\int_{\Omega_{s+1,s+1}}f_{s+1}\bmath{a}_{i,s+1}\text{d}_{s+1}\right]\nonumber\\
&=\sum_{i=1}^{s}\left[\int\text{d}^{3}\varv_{s+1}\oiint f_{s+1} \bmath{\varv}_{i,s+1}\cdot\text{d}\bmath{\sigma}_{i,s+1}\right]\nonumber\\
&\qquad+\frac{1}{2}\int\text{d}^{3}\varv_{s+2}\int\text{d}_{s+1}\oiint f_{s+2}  \bmath{\varv}_{s+1,s+2}\cdot\text{d}\bmath{\sigma}_{s+1,s+2},\label{Eq.BBGKY_truncated}
\end{align}
where
\begin{align}
\bmath{\varv}_{ij}=\bmath{\varv}_{i}-\bmath{\varv}_{j},
\end{align}
and $\bmath{\sigma}_{ij}$ is the normal surface vector perpendicular to the surface of the Landau sphere spanned by the radial vector $\triangle(\bmath{r}_i-\bmath{r}_j)/r_{ij}$ around the position $\bmath{r}_j$ and the surface integral $\oiint$ is taken over the surface components $\text{d}\bmath{\sigma}_{ij}$. The L.H.S of equation \eqref{Eq.BBGKY_truncated} is the same as a standard BBGKY hierarchy except for the truncated DF, while the two terms on the R.H.S appears due to the effects of stars entering or leaving the surface of the Landau sphere; those two extra terms may turn into collisional terms as explained in section \ref{sec.WC_strong}. 

For the weakly-coupled DF, the contributions from the surface integral vanish; the two terms on the R.H.S of equation \eqref{Eq.BBGKY_truncated} vanish since any star does no exist inside the Landau sphere, i.e. equations \eqref{Eq.weak_coupling} and \eqref{Eq.WC_r12} are valid. Hence, the BBGKY hierarchy for the weakly-coupled DF is
\begin{align}
&\partial_{t}f_{s}+\sum_{i=1}^{s}\left[\bmath{\varv}_{i}\cdot\nabla_{i}+\sum_{\quad{j=1(\neq i)}}^{s}\bmath{a}_{ij}\cdot\bmath{\partial}_{i}\right]f_{s}\nonumber\\
&\qquad +\sum_{i=1}^{s}\left[\bmath{\partial}_{i}\cdot\int_{\Omega_{s+1,s+1}}f_{s+1}\bmath{a}_{i,s+1}\text{d}_{s+1}\right]=0.\label{Eq.BBGKY_WC}
\end{align}

Lastly, in limit of $\triangle\rightarrow 0$, one can retrieve a standard BBGKY hierarchy for standard DF from both equations  \eqref{Eq.BBGKY_truncated} and \eqref{Eq.BBGKY_WC}
\begin{align}
&\partial_{t}f_{s}+\sum_{i=1}^{s}\left[\bmath{\varv}_{i}\cdot\nabla_{i}+\sum_{\quad{j=1(\neq i)}}^{s}\bmath{a}_{ij}\cdot\bmath{\partial}_{i}\right]f_{s}\nonumber\\
&\qquad +\sum_{i=1}^{s}\left[\bmath{\partial}_{i}\cdot\int f_{s+1}\bmath{a}_{i,s+1}\text{d}_{s+1}\right]=0.\label{Eq.BBGKY_orth}
\end{align}
The domain-patched DFs also obey the hierarchy, equation \eqref{Eq.BBGKY_orth}, since the domain-patched DFs are still symmetric about interchange of the states of stars.

\subsection{The Klimontovich's kinetic theory of non-ideal systems}\label{subsec:nonideal}
The \citep{Klimontovich_1982}'s theory of non-ideal systems was introduced to study the effect of non-ideality,\footnote{In basic kinetic theoretical description \citep[e.g.][]{Klimontovich_1982,Schram_2012}, the collision terms in kinetic equations for ideal systems are local in time and space (meaning collisions of concern occur in small spatial scales and short time scale compared to hydrodynamic space- and time- scales of the system of concern.), while the corresponding collision terms for non-ideal systems are non-local. In non-ideal systems, the total energy of test particle and a field particle during the two-body interaction can conserve \citep{Klimontovich_1982,Snider_1995} and the hydrodynamic and thermodynamic quantities (even the Boltzmann entropy) may be affected \citep[e.g.][]{Belyi_2002,Belyi_2003}. For example, fundamental kinetic equations, such as Boltzmann equation \citep{Boltzmann_1964}, FP/Landau equation \citep{Landau_1936}, and Balescu-Lenard \citep{Balescu_1960,Lenard_1960} can describe only \emph{ideal} systems, hence, the stellar collision kinetic equations \citep[e.g.][]{Retterer_1979,Goodman_1983,Ipser_1983} are to be used for ideal systems. On one hand, stellar wave kinetic equations \citep[e.g.][]{Gilbert_1968,Kandrup_1981,Heyvaerts_2010,Chavanis_2012} can describe \emph{non-ideal} systems.} i.e. the effect of non-locality of test-particle trajectory in time and space that typical collisional Boltzmann equation does not count due to its ideal trajectory, equation \eqref{Eq.characteristics_v_jump}. To the best of my reading, the \citep{Klimontovich_1982}'s theory can be interpreted as a conversion relation between collision- and wave- kinetic descriptions. Assume test star (star 1) undergoing purely two-body Newtonian interaction with a field star (star 2) and the stars follow the trajectories, equation \eqref{Eq.characteristics_two_body}. Also, following the \citep{Klimontovich_1982}'s theory of non-ideal gaseous system, neglect the effects of discreteness, polarisation and three-body encounter. The second equation of standard BBGKY hierarchy from equation \eqref{Eq.BBGKY_orth} is
\begin{align}
\frac{\text{d}}{\text{d}t}f(1,2,t)&\equiv\left(\partial_{t}+\bmath{\varv}_{1}\cdot\nabla_{1}+\bmath{\varv}_{2}\cdot\nabla_{2}+\bmath{a}_{12}\cdot\bmath{\partial}_{12}\right)f\left(1,2,t\right),\nonumber\\
                                &=\left(\partial_{t}+\bmath{\varv}_{1}\cdot\nabla_{1}+\bmath{\varv}_{2}\cdot\nabla_{2}\right)f(1,t)f(2,t).\label{Eq.BBGKY_2nd_two_body}
\end{align} 
To find the formal solution, solve equation \eqref{Eq.BBGKY_2nd_two_body} for $f(1,2,t)$ employing the method of characteristics
	\begin{align}
	&f(1,2,t)\nonumber\\
	&=f(1(t-\tau),t-\tau)(2(t-\tau),t-\tau)\nonumber\\
	&\quad+\int^{t}_{t-\tau} \left(\partial_{t}+\bmath{\varv}_{1}\cdot\nabla_{1}+\bmath{\varv}_{2}\cdot\nabla_{2}\right)_{t=t'}f\left(1\left(t'\right),t'\right)f\left(2\left(t'\right),t'\right)\text{d}t'.
	\label{Eq.twoDF_strong_chact}
	\end{align}
The basic idea of \citep{Klimontovich_1982}'s theory is simple; to exploit the integral-by-parts along the trajectory of the star to the R.H.S of equation \eqref{Eq.twoDF_strong_chact}. Then, one obtains
	\begin{align}
	&f(1(t-\tau),t-\tau)(2(t-\tau),t-\tau)\nonumber\\
	&+\int^{t}_{t-\tau} \left(\partial_{t}+\bmath{\varv}_{1}\cdot\nabla_{1}+\bmath{\varv}_{2}\cdot\nabla_{2}\right)_{t=t'}f\left(1\left(t'\right),t'\right)f\left(2\left(t'\right),t'\right)\text{d}t'\nonumber\\
	&\quad=-\int^{t}_{t-\tau}\left[\bmath{a}_{12}\cdot\bmath{\partial}_{12}\right]_{t=t'}f\left(1\left(t'\right),t'\right)f\left(2\right(t'\left),t'\right)\text{d}t'\nonumber\\
	&\qquad+f(1,t)f(2,t). \label{Eq.twoDF_strong_weak}
	\end{align}
Hence, one can find the conversion relation between collision- and wave- kinetic description in form of correlation function
\begin{subequations}
	\begin{align}
	&g(1,2,t)\nonumber\\
	&=f(1(t-\tau),t-\tau)(2(t-\tau),t-\tau)-f(1,t)f(2,t)\nonumber\\
	&\quad+\int^{t}_{t-\tau} \left(\partial_{t}+\bmath{\varv}_{1}\cdot\nabla_{1}+\bmath{\varv}_{2}\cdot\nabla_{2}\right)_{t=t'}f\left(1\left(t'\right),t'\right)f\left(2\left(t'\right),t'\right)\text{d}t',\label{Eq.g(1,2,t)_collision}\\
	&=-\int^{t}_{t-\tau}\left[\bmath{a}_{12}\cdot\bmath{\partial}_{12}\right]_{t=t'}f\left(1\left(t'\right),t'\right)f\left(2\right(t'\left),t'\right)\text{d}t'.\label{Eq.g(1,2,t)_wave}
	\end{align}
\end{subequations} 
The correlation function, equation \eqref{Eq.g(1,2,t)_collision}, was intensively studied in \citep{Klimontovich_1982} and represents a collision kinetic description. Following \citep{Klimontovich_1982}, the correlation function, equation \eqref{Eq.g(1,2,t)_collision}, is composed of correlation functions associated with a Boltzmann-type collision (the first term), m.f. potential (the second terms) and the retardation effect (the third term)\footnote{It is to be noted that one must realise the relation of the correlation function with binary DF $f(1,2,t)$
	\begin{align}
	&f(1,2,t)\nonumber\\
	&=f(1(t-\tau),t-\tau)(2(t-\tau),t-\tau)\nonumber\\
	&\quad+\int^{t}_{t-\tau} \left(\partial_{t}+\bmath{\varv}_{1}\cdot\nabla_{1}+\bmath{\varv}_{2}\cdot\nabla_{2}\right)_{t=t'}f\left(1\left(t'\right),t'\right)f\left(2\left(t'\right),t'\right)\text{d}t'.
	\end{align}.
The form above was one of the main equations employed from the beginning of the theory in \citep{Klimontovich_1982}.}. On the other hand, the correlation function, equation \eqref{Eq.g(1,2,t)_wave}, represents a wave kinetic description; it is, of course, the fundamental form of the Landau-type collision term having been used in stellar dynamics. The difference between the two descriptions, if the system is a star cluster, is that the correlation function, equation \eqref{Eq.g(1,2,t)_wave}, describes a less probable two-body encounter and the retardation effects while equation \eqref{Eq.g(1,2,t)_wave} describes 'slow' weak relaxation effect due to many-body interaction at two-body-DF level. I believe one may go back and forth between collision kinetic description, equation \eqref{Eq.g(1,2,t)_collision}, and wave one, equation \eqref{Eq.g(1,2,t)_wave}\footnote{It is to be noted that the explicit forms of DFs, equations \eqref{Eq.g(1,2,t)_collision} and \eqref{Eq.g(1,2,t)_wave}, are \emph{special} solutions of the second equation of the standard BBGKY hierarchy, not the general solution. This is since one does not need to assume the form of correlations functions and DFs based on the Mayer expansion for DFs. See discussion in \citep{Snider_1995} where the non-locality of Boltzmann collision terms is discussed without time integral terms.}. The conversion relation can be simplified to some convenient forms by employing proper approximations (See Appendix \ref{Appendix.Ext.non_ideality}) and will be employed in section \ref{sec:discussion} to find convergent kinetic equations of star clusters. The important property in the convergent relation, equation \eqref{Eq.twoDF_strong_weak}, is that the m.f. acceleration of star 1 (originated from the second term on the R.H.S of the equation) and the Boltzmann collision term (originated from the first term on the L.H.S of the equation) does not self-consistently coexist at specific distance between stars. This will be throughly discussed in section \ref{sec:strong}.
 
In addition to the conversion relation, the non-ideal theory is of importance in discussion of the effect of strong-interaction potential in dense systems of particles interacting with long-range interaction (that was the actual purpose for the studies in \citep{Klimontovich_1982}). To improve the 
Boltzmann-collision description so that the conservation of total energy between two stars in encounter holds, one needs to take into account the effect of finiteness of correlation- time $t_\text{cor}$ and distance $r_\text{cor}$. Following  discussions of non-ideal gases \citep{Green_1952,Snider_1995} for pure two-body encounter between two stars, one can expand the two-body DF $f(1(t-\tau),t-\tau)f(2(t-\tau),t-\tau)$, associated with a non-local Boltzmann collision term, in series of the following ratios $t_\text{cor}/t_\text{sec}$ and $r_\text{cor}/R$;
\begin{align}
&f(1(t-\tau),t-\tau)f(2(t-\tau),t-\tau)\nonumber\\
&\approx  \left(1-\tau\left[\partial_{t}+\bmath{V}\cdot\nabla_{1}\right]-\bmath{r}_{12}\cdot\nabla_{1}\right) f(\bmath{r}_{1},\bmath{p}_{1},t)f(\bmath{r}_{1},\bmath{p}_{2},t).\label{Eq.f(1,2,t)_two_non-ideal}
\end{align}
Equation \eqref{Eq.f(1,2,t)_two_non-ideal} is a kind of density expansion of the DFs without the effect of three-body DF, while \citep{Klimontovich_1982}'s theory can correctly discuss the effect of existence of another field star (star 3) at the second-equation level of the standard BBGKY hierarchy; the effect corresponds with the time-integral term on the R.H.S of equation \eqref{Eq.g(1,2,t)_collision}. Star 3 at three-body-DF level represents the effect of the existence of field stars on the motion of test star through the rapid-temporal- and significant-spatial- change in the DF (the density), roughly speaking, in a similar way to star 3 being associated with the effects of gravitational polarization and statistical acceleration in \citep{Gilbert_1968}'s equation.  Also, the non-ideal theory is suitable in discussion of the relation between the strong-close encounter and m.f. acceleration since the effect of encounters appears only through the correlation term $g(1,2,t)$ in both of collision- and wave- kinetic descriptions; even for the collision kinetic theory, the m.f. acceleration can be separated from the collision term. The \citep{Klimontovich_1982}'s theory will be discussed in detail for stellar dynamics inside the Landau sphere in section \ref{sec.strong_Klimon}.

%% file: Section4_completely_WC_limit.tex
\section{The generalised Landau equation for the 'weakly-coupled' distribution function of stars}\label{sec:complete_WC}
In the present section, the weakly-coupled DFs (section \ref{subsec:DF}) is employed to derive a kinetic equation to model evolutions of a 'completely weakly-coupled' star cluster in which no star can approach the other stars closer than the Landau radius $r_\text{o}$. In section \ref{sec:grainess_effect} the effects of truncation of phase-space volume on the collision term (relaxation time of the system) and on the m.f. acceleration (Poisson equation) are discussed.

\subsection{Completely weakly-coupled stellar systems}\label{subsec.compl_WC}
Assume that a star cluster at the early stage of evolution may be modeled by weakly-coupled DFs for stars. The first two equations of the hierarchy, equation \eqref{Eq.BBGKY_WC}, for DF $f_{1}(1,t)$ and the first equation of the hierarchy for DF $f_{1}(2,t)$ respectively read
\begin{subequations}
	\begin{align}
	&(\partial_{t}+\bmath{\varv}_{1}\cdot\nabla_{1})f_{1}(1,t)=-\bmath{\partial}_{1}\cdot\int_{\Omega_{2,2}}f_{2}(1,2,t)\bmath{a}_{12}\text{d}_{2},\label{Eq.1stBBGKY_WC}\\
	&\left(\partial_{t}+\bmath{\varv}_{1}\cdot\nabla_{1}+\bmath{\varv}_{2}\cdot\nabla_{2}+\bmath{a}_{12}\cdot\bmath{\partial}_{12}\right)f_{2}(1,2,t)\nonumber\\
	&\quad\qquad=-\int_{\Omega_{3,3}}\left[\bmath{a}_{1,3}\cdot\bmath{\partial}_{1}+\bmath{a}_{2,3}\cdot\bmath{\partial}_{2}\right]f_{3}(1,2,3,t)\text{d}_{3},\label{Eq.2nd_BBGKY_WC}\\
	&(\partial_{t}+\bmath{\varv}_{2}\cdot\nabla_{2})f_{1}(2,t)=-\bmath{\partial}_{2}\cdot\int_{r_{23}>\triangle}f_{2}(2,3,t)\bmath{a}_{23}\text{d}_{3},\label{Eq.1stBBGKY_WC_m.f}
	\end{align}
\end{subequations}
where $\bmath{\partial}_{12}=\bmath{\partial}_{1}-\bmath{\partial}_{2}$ and the domains of DFs and the accelerations are defined only at distances $r_{ij}>\triangle$. 

To simplify equations \eqref{Eq.1stBBGKY_WC} and \eqref{Eq.2nd_BBGKY_WC}, assume the system is not gravitaitonally-polarizable and employ the definitions for \emph{s}-ary DFs, equations \eqref{Eq.singleDF}, \eqref{Eq.doubleDF} and \eqref{Eq.tripleDF_WC}. One obtains
\begin{subequations}
\begin{align}
&\left(\partial_{t}+\bmath{\varv}_{1}\cdot\nabla_{1}+\bmath{A}_{1}^{(2,2)}\cdot\bmath{\partial}\right)f(1,t)=-\bmath{\partial}_{1}\cdot\int_{\Omega_{2,2}}g(1,2,t)\bmath{a}_{12}\text{d}_{2},\label{Eq.1stBBGKY_Kadrup}\\
&\left(\partial_{t}+\bmath{\varv}_{1}\cdot\nabla_{1}+\bmath{\varv}_{2}\cdot\nabla_{2}+\bmath{A}_{1}^{(3,3)}\cdot\bmath{\partial}_{1}+\bmath{A}_{2}^{(3,3)}\cdot\bmath{\partial}_{2}\right)g(1,2,t)\nonumber\\
&=\left(-\bmath{a}_{12}\cdot\bmath{\partial}_{12}+\frac{1}{N}\bmath{A}_{1}^{(3,3)}\cdot\bmath{\partial}_{1}+\frac{1}{N}\bmath{A}_{2}^{(3,3)}\cdot\bmath{\partial}_{2}\right)f(1,t)f(2,t)\nonumber\\
&\quad-\left(1-\frac{1}{N}\right)\left[\bmath{A}_{1}^{(2,2)}-\bmath{A}_{1}^{(3,3)}\right]\cdot\bmath{\partial}_{1}f(1,t)f(2,t)\nonumber\\
&\quad-\left(1-\frac{1}{N}\right)\left[\bmath{A}_{2}^{(2,2)}-\bmath{A}_{2}^{(3,3)}\right]\cdot\bmath{\partial}_{2}f(1,t)f(2,t)\nonumber\\
&\quad-\partial_{1}\cdot\left(\int_{\Omega_{3,3}}\bmath{a}_{13}g(1,3,t)\text{d}_{3}-\int_{\Omega_{2,2}}\bmath{a}_{13}g(1,3,t)\text{d}_{3}\right)f(2,t)\nonumber\\
&\quad-\partial_{2}\cdot\left(\int_{\Omega_{3,3}}\bmath{a}_{23}g(2,3,t)\text{d}_{3}-\int_{\Omega_{2,2}}\bmath{a}_{23}g(2,3,t)\text{d}_{3}\right)f(1,t),\label{Eq.2nd_BBGKY_grained_many}
\end{align}
\end{subequations}
where the lowest OoM of the terms are left with $\mathcal{O}(1)$\footnote{One may realise that the lowest order at equation level is $\sim\mathcal{O}(1/N^{2})$ due to the truncated acceleration $\bmath{A}^{(2,2)}/N$ to hold the self-consistency of the kinetic equation.} and the truncated m.f. accelerations are defined as
\begin{subequations}
	\begin{align}
	&\bmath{A}_{i}^{(2,2)}(\bmath{r}_{i},t)=\left[1-\frac{1}{N}\right]\int_{r_{i3}>\triangle}f(3,t)\bmath{a}_{i3}\text{d}_{3},\quad(i,j=1,2)\\
	&\bmath{A}_{i}^{(3,3)}(\bmath{r}_{i},\bmath{r}_{j},t)=\left[1-\frac{1}{N}\right]\int_{\Omega_{3,3}}f(3,t)\bmath{a}_{i3}\text{d}_{3}.
	\end{align}
\end{subequations}
The last six terms on the R.H.S of equation \eqref{Eq.2nd_BBGKY_grained_many} may be simplified, by neglecting the existence of the third star in  two-body encounter between two stars of concern;
\begin{align}
 \Omega_{2,2}\approx\Omega_{3,3}.
\end{align}
This is possible since the truncated phase-space volume of the truncated DF, equation \eqref{Eq.single_truncated_DF}, for the third star contributes to equation \eqref{Eq.2nd_BBGKY_grained_many} only as a margin of error with order of $\mathcal{O}(1/N^{2})$; corresponding to
\begin{align}
\int_{\Omega_{2,2}}\cdot\quad\text{d}_{3}\approx\int_{\Omega_{3,3}}\cdot\quad\text{d}_{3}+\mathcal{O}\left(1/N^{2}\right).
\end{align}
Hence, equation \eqref{Eq.2nd_BBGKY_grained_many} simply reduces to
\begin{align}
&\left(\partial_{t}+\bmath{\varv}_{1}\cdot\nabla_{1}+\bmath{\varv}_{2}\cdot\nabla_{2}+\bmath{A}_{1}^{(2,2)}\cdot\bmath{\partial}_{1}+\bmath{A}_{2}^{(2,2)}\cdot\bmath{\partial}_{2}\right)g(1,2,t)\nonumber\\
&=-\left[\tilde{\bmath{a}}^{\triangle}_{12}\cdot\bmath{\partial}_{1}+\tilde{\bmath{a}}^{\triangle}_{21}\cdot\bmath{\partial}_{2}\right]f(1,t)f(2,t),\label{Eq.2nd_BBGKY_grained_Kandrup}
\end{align}
where 
\begin{subequations}
	\begin{align}
	\tilde{\bmath{a}}^{\triangle}_{12}=\bmath{a}_{12}-\frac{1}{N}\bmath{A}_{1}^{(2,2)},\\
	\tilde{\bmath{a}}^{\triangle}_{21}=\bmath{a}_{21}-\frac{1}{N}\bmath{A}_{2}^{(2,2)}.
	\end{align}
\end{subequations}
Employing the method of characteristics, one obtains the correlation function from equation \eqref{Eq.2nd_BBGKY_grained_Kandrup}
\begin{align}
&g(1,2,t)\nonumber\\
&=g(1(t-\tau),2(t-\tau),t-\tau)\nonumber\\
&\quad-\int^{t}_{t-\tau}\left[\tilde{\bmath{a}}^{\triangle}_{12}\cdot\bmath{\partial}_{1}+\tilde{\bmath{a}}^{\triangle}_{21}\cdot\bmath{\partial}_{2}\right]_{t=t'}f\left(1(t'),t'\right)f\left(2(t'),t'\right)\text{d}t'.\label{Eq.Generalized_Landau_truncated_g}
\end{align}

In the scenario for the g-Landau equation in \citep{Kandrup_1981}, all the stars in a star cluster are perfectly uncorrelated at the beginning of correlation time $t-\tau$, implying that the destructive term $g(1(t-\tau),2(t-\tau),t-\tau)$ vanishes at two-body DF level. To apply the same simplification for a secular evolution of the system of concern, one must necessarily consider the memory effect, that is of importance if the time duration between encounters is comparable to the correlation-time scale. The memory effect, however, may be of less significance in stellar dynamics due to the violent relaxation, short-range two-body encounters, spatial inhomogeneities and anisotropy \citep[e.g.][pg. 34]{Saslaw_1985}. Hence, the destructive term on the R.H.S of equation \eqref{Eq.Generalized_Landau_truncated_g} may vanish. One obtains the g-Landau equation with the effect of discreteness from equations \eqref{Eq.1stBBGKY_Kadrup} and \eqref{Eq.Generalized_Landau_truncated_g}
\begin{align}
&\left(\partial_{t}+\bmath{\varv}_{1}\cdot\nabla_{1}+\bmath{A}_{1}^{(2,2)}\cdot\bmath{\partial}_{1}\right)f(1,t)\nonumber\\
&=\bmath{\partial}_{1}\cdot\int_{\Omega_{2,2}}\text{d}_{2} \bmath{a}_{12}\int_{0}^{\tau}\text{d}\tau'\left[\tilde{\bmath{a}}_{12}^{\triangle}\cdot\bmath{\partial}_{1}+\tilde{\bmath{a}}_{21}^{\triangle}\cdot\bmath{\partial}_{2}\right]_{t-\tau'}\nonumber\\
&\quad\times f(1(t-\tau'),t-\tau')f(2(t-\tau'),t-\tau').\label{Eq.Generalized_Landau_truncated}
\end{align}

The effect of retardation in the collision term of equation \eqref{Eq.Generalized_Landau_truncated} may be discussed. Since the trajectory of test star is chracterised by equation \eqref{Eq.characteristics_m.f.}, the correlation time would be at most the free-fall time of test star under the effect of the m.f. acceleration while the shortest correlation time scale is longer than the time scale for test star to travel across a Landau sphere to hold the weak-coupling approximation;
\begin{align}
\mathcal{O}(1/N)<t_\text{cor}\lesssim\mathcal{O}(1),
\end{align} 
meaning the non-Markovian effect on the relaxation process is less significant;
\begin{align}
\mathcal{O}(1/N^{2})<\frac{t_\text{cor}}{t_\text{rel}}\lesssim\mathcal{O}(1/N).
\end{align} 
Hence, one may assume the Markovian limit\footnote{For the Markovian limit, one should not change the other arguments of the DF in the collision term since the changes in momentum and position of test star in encounter is not ignorable due to the effect of m.f. acceleration.} for the collision term for the correlation time $0<\tau'<t_\text{cor}$
\begin{align}
f(1(t-\tau'),t-\tau')f(2(t-\tau'),t-\tau')\approx f(1(t-\tau'),t)f(2(t-\tau'),t),
\end{align}
Taking the limit of $\tau\to\infty$, one obtains
\begin{align}
&\left(\partial_{t}+\bmath{\varv}_{1}\cdot\nabla_{1}+\bmath{A}_{1}^{(2,2)}\cdot\bmath{\partial}_{1}\right)f(1,t)\nonumber\\
&=\bmath{\partial}_{1}\cdot\int_{\Omega_{2,2}}\text{d}_{2} \bmath{a}_{12}\int_{0}^{\infty}\text{d}\tau'\nonumber\\
&\quad\times\left[\tilde{\bmath{a}}_{12}^{\triangle}\cdot\bmath{\partial}_{1}+\tilde{\bmath{a}}_{21}^{\triangle}\cdot\bmath{\partial}_{2}\right]_{t-\tau'} f(1(t-\tau'),t)f(2(t-\tau'),t).\label{Eq.Generalized_Landau_truncated_Marko}
\end{align}
Employing the anti-normalization condition, equation \eqref{Eq.Anti-norm}, for the correlation function and taking a limit of $\triangle\to 0$, one may retrieve the \citep{Kandrup_1981}'s g-Landau equation;
\begin{align}
&\left(\partial_{t}+\bmath{\varv}_{1}\cdot\nabla_{1}+\bmath{A}_{1}\cdot\bmath{\partial}_{1}\right)f(1,t)=\bmath{\partial}_{1}\cdot\int\text{d}_{2} \tilde{\bmath{a}}_{12}\int_{0}^{\infty}\text{d}\tau'\nonumber\\
&\quad\times\left[\tilde{\bmath{a}}_{12}\cdot\bmath{\partial}_{1}+\tilde{\bmath{a}}_{21}\cdot\bmath{\partial}_{2}\right]_{t-\tau'} f(1(t-\tau'),t)f(2(t-\tau'),t),\label{Eq.Generalized_Landau}
\end{align}
where the statistical acceleration can be found in the forms
\begin{subequations}
	\begin{align}
	\tilde{\bmath{a}}_{12}=\bmath{a}_{12}-\frac{1}{N}\bmath{A}_{1},\\
	\tilde{\bmath{a}}_{21}=\bmath{a}_{21}-\frac{1}{N}\bmath{A}_{2}.
	\end{align}
\end{subequations}
The truncated g-Landau equation \eqref{Eq.Generalized_Landau_truncated} is different from the g-Landau equation \eqref{Eq.Generalized_Landau}, not only in the domain of interaction range, but also in the form of physical quantities; the truncated- acceleration and collision term. The truncation of the phase-space volume in integrals is termed as 'the effect of discreteness' in the present work and discussed in section \ref{sec:grainess_effect}.

\subsection{The effect of discreteness on the relaxation time and mean-field acceleration}\label{sec:grainess_effect}
In the present section, the effects of 'weakly-coupled' DF on the relaxation time of a system modeled by equation \eqref{Eq.Generalized_Landau_truncated_Marko} and on the corresponding Poisson equation are discussed.
\subsubsection{The relaxation time}\label{subsec:grainess_effect_tr}
To evaluate the effect of 'discreteness' on the relaxation time, assume test star follows the rectilinear motion, equation \eqref{Eq.characteristics_rectilinear}, and the encounter is local for the truncated g-Landau collision term in equation \eqref{Eq.Generalized_Landau_truncated_Marko}, meaning the truncated Landau collision term is examined;
\begin{align}
I_\text{L}^{\blacktriangle}=&\bmath{\partial}_{1}\cdot\int_{\Omega_{2,2}} \bmath{a}_{12}\int_{0}^{\infty}\left[\bmath{a}_{12}\left(t-\tau'\right)\right]_{r_{12}\left(t-\tau'\right)>\triangle}\text{d}\tau'\text{d}^{3}\bmath{r}_{12}\nonumber\\
                         &\qquad\cdot\bmath{\partial}_{12} f(1,t)f(\bmath{r}_{1},\bmath{p}_{2},t)\text{d}^{3}\bmath{p}_{2},\label{Eq._Landau_truncated_Marko}
\end{align}
where the effect of non-ideality (retardation and spatial non-locality) for the Landau collision term was neglected for simplicity. The Fourier-transform of the acceleration of star 1 due to star 2 at distances $r_{12}>\triangle$ is as follows\footnote{It is to be noted that the Fourier transform of the potential $\phi_{12}$ typically done to find the explicit form of the Landau collision term necessitates a 'convergent factor',$e^{-\lambda r_{12}}$, where $\lambda$ is a vanishing low number to be taken as zero after the Fourier transform. The factor can remove singularities of (generalised) functions on complex planes and slow decays of potentials in three dimensional spaces \citep[e.g][]{Adkins_2013}. One, however, does not need to employ the factor in the Fourier transform of the truncated acceleration, $\bmath{a}_{12}\Theta(r_{12}>\triangle)$, and even in the corresponding inverse Fourier transform, $\mathcal{F^{-1}}\left[\mathcal{F}\left[\bmath{a}_{12}\Theta(r_{12}>\triangle)\right]\right]$. Rendering the transform, $\mathcal{F^{-1}}\left[\mathcal{F}\left[\bmath{a}_{12}\Theta(r_{12}>\triangle)\right]\right]$, is a simple task, hence it will be left for readers; one will need the following identity to find the step function
\begin{align}
\int^{\infty}_{0}\frac{\sin k}{k}\text{d}k=\frac{\pi}{2}.
\end{align}
}
\begin{subequations}
	\begin{align}
	\mathcal{F}\left[\bmath{a}_{12}(r_{12}>\triangle)\right]&=-Gm\int_{\Omega_{2,2}}\exp(-\text{i}\bmath{k}\cdot \bmath{r}_{12})\frac{\bmath{r}_{12}}{r_{12}^{3}}\text{d}^{3}\bmath{r}_{12},\\
	&=-\frac{Gm\sin[k\triangle]}{2i\pi^{2}k^{2}\triangle}\hat{k}. \label{Eq.Fourier_a}
	\end{align}
\end{subequations}
The same transform must be employed for $\bmath{a}_{12}(t-\tau')$ in equation \eqref{Eq._Landau_truncated_Marko} but the time of $\bmath{r}$ is fixed to $t-\tau'$ and the corresponding wavenumber must be exploited. It is to be noted that equation \eqref{Eq.Fourier_a} is in essence the same as the Fourier transform of the truncated acceleration $\bmath{a}_{12}\Theta(r_{12}-\triangle)$, meaning the corresponding acceleration of star 1 is null within the volume of the Landau sphere. This is since the existence of stars in the Landau sphere is not of concern due to the spatial locality and the effect of truncation on DF must be controlled through truncation of acceleration. In the limit of $\triangle\to 0$, equation \eqref{Eq.Fourier_a} results in a well-known Fourier transform of acceleration or pair-wise Newtonian potential in wave kinetic theory \citep[e.g.][Appendix C]{Chavanis_2012}
\begin{subequations}
	\begin{align}
	\lim_{\triangle\to 0}\mathcal{F}\left[\bmath{a}_{12}\right]&=-\frac{Gm}{2i\pi^{2}k}\hat{k},\\
	&\xrightarrow[\times\frac{1}{-\text{i}\bmath{k}}]{}\mathcal{F}\left[\phi_{12}\right]\hat{k}.
	\end{align}
\end{subequations}
After a proper calculation following \citep[][Appendix C]{Chavanis_2012}, the collisional term results in
\begin{subequations}
	\begin{align}
	&I_\text{L}^{\blacktriangle}=\bmath{\partial}_{1}\cdot \int \overleftrightarrow{T}(\bmath{p}_{12},\bmath{p}_{2})\cdot\bmath{\partial}_{12} f(\bmath{r}_{1},\bmath{p}_{1},t)f(\bmath{r}_{1},\bmath{p}_{2},t) \text{d}^{3}\bmath{p}_{2}, \\
	&\overleftrightarrow{T} \equiv -B\frac{\bmath{p}^{2}_{12}\overleftrightarrow{I}-\bmath{p}_{12}\bmath{p}_{12}}{p^{3}},\qquad (\bmath{p}_{12}\equiv\bmath{p}_{1}-\bmath{p}_{2})\\
	&B\equiv 2\pi G m^{2}\int^{\infty}_{0}\frac{\sin^{2}[k\triangle]}{k^{3}\triangle^{2}}\text{d}k. \label{Eq.g_Logarithm}
	\end{align}
\end{subequations}
where for expressions of the tensor $\overleftrightarrow{T}$, typical dyadics are exploited. Following the works \citep{Severne_1976,Kandrup_1981,Chavanis_2013a} if one assumes the cut-offs $k\in[2\pi/R,2\pi/r_\text{o}]$ on each limit of the integral domain of the collision term, the factor $B$, equation \eqref{Eq.g_Logarithm}, explicitly reads
\begin{align}
B=\ln[N]+1.5+\sum_{m=1}^{\infty}\frac{(-16\pi^{2})^{m}}{2m(2m)!}+\mathcal{O}(1/N^{2}),\label{Eq.B}
\end{align}
where the following indefinite integral formula \citep[e.g.][]{Zeidler_2004} was employed\footnote{A Similar calculation for a weakly-nonideal self-gravitating system appears in \citep{Bose_2012}, in which the upper limit is also assigned on the domain of the integration.} 
\begin{align}
	\int\frac{\cos[\alpha k]}{k}\text{d}k=\ln[\alpha k]+\sum_{m=1}^{\infty}\frac{\left(-[\alpha k]^{2}\right)^{m}}{2m(2m)!}.\label{Eq.math_formula}
\end{align}
It would be obvious that the lower limit of the distance $r_{12}$, the Landau radius, can not remove the logarithmic singularity in the collision term as shown in equation \eqref{Eq.B}. This is of course since an application of the weak-coupling approximation to the g-Landau collision term is inconsistent especially at $r_{12}\to r_\text{o}$;to avoid the singularity associated with high wavenumbers, one needs all the higher orders of weak-coupling approximation as correction to the rectilinear-motion approximation, or the trajectory  of test star must follow pure Newtonian two-body problem, equation \eqref{Eq.characteristics_two_body}. Since the value of the parameter $c$ in equation \eqref{Eq.ratio_R_ro} is in essence a user-choice parameter, the following ideal (often-employed in wave kinetic theory) relation is assumed for simplicity
\begin{align}
R=Nr_\text{o}.\label{Eq.simple_R_ro}
\end{align}

The result of numerical integration of the factor $B$, equation \eqref{Eq.g_Logarithm}, is as follows
\begin{align}
B-\ln[N]=1.5+ \sum_{m=1}^{\infty}\frac{\left(-[4\pi]^{2}\right)^{m}}{2m(2m)!}\approx-3.11435,
\end{align}
and the decrease rate of the Coulomb logarithm for different $N$ is shown in Table \ref{tabel:trunc_Coulomb_log}. It turns out, the effect of discreteness \emph{decreases} the coulomb logarithm $\ln[N]$ for relatively low-number star cluster $\left(N=10^{5}\right)$ by 14.0 $\%$, for high-number cluster $\left(N=10^{7}\right)$ by 10.0 $\%$, and (as a reference) for large galaxies $\left(N=10^{11}\right)$ by 6.37$\%$; accordingly, the corresponding relaxation times \emph{increase} from typical one (that has the same physical condition but effect of discreteness) by the same factors.
\begin{table}
	\caption{The decrease rate of the Coulomb logarithm, $100(1-B/\ln[N])$, due to the effect of discreteness.}
	\label{tabel:trunc_Coulomb_log}
	\begin{tabular}{|l|c|l|c|}
    \hline
	    $N$                          & decrease rate [\%]   & $N$                          & decrease rate [\%] \\ 
	    \hline
	    $10^{5}$                     & 14.02                   & $10^{9}$                     & 7.790 \\
	    $10^{6}$                     & 11.69                   & $10^{10}$                    & 7.011 \\
	    $10^{7}$                     & 10.02                   & $10^{11}$                    & 6.373 \\
	    $10^{8}$                     & 8.764                   &                              &        \\
	\hline
	\end{tabular}\\
\end{table}

\subsubsection{Truncated Poisson equation}\label{subsec:grainess_effect_Poisson}
Utilizing the identity
\begin{align}
\nabla^{2}_{1}\left(\frac{1}{r_{12}}\right)=0,
\end{align}
one can derive the Poisson equation for the truncated m.f. acceleration (or the 'truncated Poisson equation')
\begin{align}
\nabla_{1}\cdot\bmath{A}_{1}^{(2,2)}=-Gm\left(1-\frac{1}{N}\right) \int n_{1}(\bmath{r}_{1}-\triangle\hat{r})\text{d}\Omega,\label{Eq.grained_poisson_A}
\end{align}
where $\text{d}\Omega$ is the element of solid angle spanned by a unit vector $\hat{r}$ in radial direction. Typical observations for star clusters are done at radii 0.01 $\sim$ 1 pc from the center of the clusters even for possibly collapsed clusters \citep[e.g.][]{King_1985,Lugger_1995}; this corresponds with $r_{1}>> r_\text{o}$ if the system dimension reaches $\sim$ tens of parsec. Hence, as a practical application, one may approximate the truncated Poisson equation \eqref{Eq.grained_poisson_A} to
\begin{align}
\nabla_{1}\cdot\bmath{A}_{1}^{(2,2)}\approx& -4\pi Gm\left(1-\frac{1}{N}\right) n_{1}(\bmath{r}_{1})+\triangle Gm\int\hat{r}\cdot\nabla_{1}n_{1}(\bmath{r}_{1})\text{d}\Omega\nonumber\\
&\qquad+\mathcal{O}(1/N^{2}),\\
=&-4\pi Gm \left(1-\frac{1}{N}\right)n_{1}(\bmath{r}_{1})+\mathcal{O}(1/N^{2}).
\end{align}
The relation of the truncated m.f. acceleration with the truncated potential may be written as
\begin{align}
\nabla^{2}_{1}\Phi^{(2,2)}=-\nabla_{1}\cdot\bmath{A}^{(2,2)}_{1}-Gm\left(1-\frac{1}{N}\right) \int\triangle\hat{r}\cdot\nabla_{1}n_{1}(\bmath{r}_{1}-\triangle\hat{r})\text{d}\Omega.
\end{align}
Employing equation \eqref{Eq.grained_poisson_A}, the Poisson equation for the truncated potential reads
\begin{align}
\nabla^{2}_{1}\Phi^{(2,2)}=Gm \left(1-\frac{1}{N}\right)\int (1-\triangle\hat{r}\cdot\nabla_{1}) n_{1}(\bmath{r}_{1}-\triangle\hat{r})\text{d}\Omega,\label{Eq.grained_Poisson_Phi}
\end{align}
and in a limit of $\triangle\to 0$
\begin{align}
\nabla^{2}_{1}\Phi^{(2,2)}\approx 4\pi Gm \left(1-\frac{1}{N}\right)n_{1}(\bmath{r}_{1})+Gm \int\left[\triangle\hat{r}\cdot\nabla_{1}\right]^{2}n_{1}(\bmath{r}_{1})\text{d}\Omega.\label{Eq.grained_Poisson_Phi_approx}
\end{align}
The standard Poisson equation for the m.f. potential of star clusters is also applicable to any star cluster at radii $r_{1}>>r_\text{o}$ since the second term on the R.H.S in equation \eqref{Eq.grained_Poisson_Phi_approx} is order of $\mathcal{O}\left(1/N^{2}\right)$. It is to be noted that the truncated Poisson equation \eqref{Eq.grained_poisson_A} or \eqref{Eq.grained_Poisson_Phi} itself shows a kind of coarse-graining on the surface of the Landau sphere through istropising the density of the system of concern at radius of $\triangle$, while the truncated Poisson equation does not function to model the dynamics inside the Landau sphere. 
 
For theoretical/numerical studies of stellar dynamics, the dynamics inside the Landau sphere may be of importance since the core size of the system of concern can mathematically reach the size of the Landau radius and the halo may have a strong inhomogeneity in density as a result of gravothemal-instability \citep[e.g][]{Cohn_1979,Takahashi_1995}. In this case, one can no longer employ typical Poisson equation, hence one must hold the form of the truncated Poisson equation \eqref{Eq.grained_poisson_A} or \eqref{Eq.grained_Poisson_Phi}. The small-scale dynamics inside the Landau sphere will be discussed in section \ref{sec:strong}. 

\subsection{Remarks on the generalised Landau equation for the Weakly-Coupled DF of stars}
Equation \eqref{Eq.Generalized_Landau_truncated} represents a kinetic formulation of the discussion for the minimum scale of fluctuation in m.f. acceleration (randomness of the Holtsmark distribution) done in \citep{Chandra_1943a, Takase_1950}, especially for the 'Rough approximation \citep{Takase_1950}'. The physical importance of equation \eqref{Eq.Generalized_Landau_truncated} is an isolation of the wave kinetic description from the deterministic dynamics inside the Landau sphere. 

In collision kinetic description at two-body-DF level without strong encounters, stars modeled by the uncorrelated DFs can enter the Landau sphere (or the m.f. acceleration can exist there) while stars modeled by the two-body DF  can not enter the sphere (or the stochastic collision term has lower limit at the conventional Landau radius). 
\begin{align}
f(1,2,t)=f(1,t)f(2,t)+\left[g(1,2,t)\right]_{r_{12}>\triangle}.
\end{align}
This inconsistency is the result of a discretion in collision kinetic theory; a simple addition of a stochastic (FP) collision term and the collisionless Boltzmann equation. 

In a similar way, in wave kinetic theory, the weak-coupling approximation is correct only outside the Landau sphere while stars modeled by the uncorrelated DFs and correlation functions can enter the Landau spheres;
\begin{align}
f(1,2,t)=f(1,t)f(2,t)+g(1,2,t).
\end{align}
However, to hold the weak-coupling limit, one needs to employ the weakly-coupled DFs, equations \eqref{Eq.hard_f(1,2,t)} and \eqref{Eq.neq_DFs}, resulting in the truncated m.f. acceleration (section \ref{subsec:grainess_effect_tr}) and the truncated collision term (section \ref{subsec:grainess_effect_Poisson}) of concern. Equation \eqref{Eq.Generalized_Landau_truncated} for the weakly-coupled DF can avoid the inconsistencies among the uncorrelated DF, correlation function and weak-coupling approximation by correctly limiting the distance between stars at the BBGKY hierarchy level. The relation between strong encounters and the m.f. acceleration will be discussed in section \ref{sec:strong}.

%% file: Section5_strong_interaction.tex
\section{The generalised Landau equation for the truncated DF of stars}\label{sec:strong}
In section \ref{sec.WC_strong}, the effect of close, strong two-body encounters is incorporated in the g-Landau equation by employing the truncated DF of stars at relatively early stage of the evolution of a weakly-inhomogeneous star cluster. In section \ref{sec.strong_Klimon}, based on the domain-patched DF and the \citep{Klimontovich_1982}'s theory of non-ideal systems, the validity of the kinetic equation derived in section \ref{sec.WC_strong} is examined on small space scales. Especially, section \ref{sec.strong_Klimon} considers the core or inner halo of a star cluster at the late stage of core-collapse whose density reaches up to order of $N^{2}n_\text{o}$ in the core or whose density of the halo follows $\sim 1/r_{1}^{2}$, where $n_\text{o}$ is the mean density of the system. 
 
\subsection{Strong encounter in a weakly inhomogeneous system based on truncated DF with 'test-particle' approximation}\label{sec.WC_strong}
Consider test star, being immersed in a weakly-coupled star cluster of field stars, may encounter one of the field stars in the Landau sphere of test star while none of field stars can approach the other field stars closer than the Landau radius. This kind of 'test-particle method \citep{Kaufman_1960, Kandrup_1981}' may not be exactly correct though, it is convenient to find a self-consistent equation in kinetic-theoretical formulation and to avoid the non-conservation of total- energy and number of star clusters modeled by the truncated DF of stars (section \ref{subsec:trunc_DF}). One possible excuse to employ the method is that one may not observe the event of close two-body encounters in more than one Landau sphere \emph{at the same time} (at least on average of a few of dynamical time scales) if the system is not at the late stage of core collapse. This is since the close encounters may be likely to occur locally in the core of the system rather than the halo due to the inhomogeneity in density; the close encounters do not occur uniformly in space at a time. Hence, in the present section, any simultaneous strong encounters will be neglected just for simplicity. Also in order to employ the truncated DF, the system must not be strongly inhomogeneous to avoid the non-ideality of encounters inside the Landau sphere; one may assume that the OoM of physical quantities follow the scaling for weakly-inhomogeneous systems (section \ref{subsec:scalings}).
 
Under the 'test-particle' approximation, one may approximate the BBGKY hierarchy, equation \eqref{Eq.BBGKY_truncated}, for the truncated DFs according to equations \eqref{Eq.trunc_DF1_test} and \eqref{Eq.trunc_DF2_test}. The term associated with $(s+2)$-body DF in the hierarchy can vanish even if the non-ideality of encounters holds (Appendix \ref{Appendix:extra}). Then, one can take into account the strong two-body encounter between stars, without resorting to higher orders of the smallness parameter, $1/N$, by keeping the surface-integral terms in the hierarchy, equation \eqref{Eq.BBGKY_truncated}. Recalling the OoM of the Knudsen number (section \ref{subsection:scaling_strong}), the OoM of the surface-integral term is at least $ N K_{n}\sim \mathcal{O}(1)$. Hence, the first equation of BBGKY hierarchy for the truncated DF reads
\begin{align}
&\partial_{t}f_{1}+\bmath{\varv}_{1}\cdot\nabla_{1}f_{1}\nonumber\\
&\quad=-\bmath{\partial}_{1}\cdot\int_{\Omega_{2,2}}f_{2}\bmath{a}_{12}\text{d}2+\int\text{d}^{3}\varv_{2}\oiint f_{2} \bmath{\varv}_{1,2}\cdot\text{d}\bmath{\sigma}_{1,2},\label{Eq.1stBBGKY_truncated_strong}
\end{align}
and the second equation
\begin{align}
&\left(\partial_{t}+\bmath{\varv}_{1}\cdot\nabla_{1}+\bmath{\varv}_{2}\cdot\nabla_{2}+\bmath{A}_{1}^{(2,2)}\cdot\bmath{\partial}_{1}+\bmath{A}_{2}^{(2,2)}\cdot\bmath{\partial}_{2}+\bmath{a}_{12}\cdot\bmath{\partial}_{12}\right)f_{2}\nonumber\\
&=-\int_{\Omega_{3,3}}\left[\bmath{a}_{1,3}\cdot\bmath{\partial}_{1}+\bmath{a}_{2,3}\cdot\bmath{\partial}_{2}\right]f_{3}\text{d}3\nonumber\\
&\quad+\int\text{d}^{3}\varv_{3}\left(\oiint f_{3} \bmath{\varv}_{1,3}\cdot\text{d}\bmath{\sigma}_{1,3}+\oiint f_{3} \bmath{\varv}_{2,3}\cdot\text{d}\bmath{\sigma}_{2,3}\right).\label{Eq.2ndBBGKY_truncated_strong}
\end{align}
As mentioned in section \ref{subsec:DF}, equations \eqref{Eq.1stBBGKY_truncated_strong} and \eqref{Eq.2ndBBGKY_truncated_strong} are meaningful only at distances $r_{12}\gtrsim r_\text{o}$ and on the spatial scales $r_{1}\gtrsim r_\text{o}$.

Recalling that the surface integrals in equation \eqref{Eq.2ndBBGKY_truncated_strong} are less significant compared to the dynamical terms by order of $\mathcal{O}(1/N)$, assuming that the outside of the Landau sphere can be described by weak-coupling limit, and employing the definition for correlation functions, equations \eqref{Eq.doubleDF} and \eqref{Eq.tripleDF_WC}, then equation \eqref{Eq.1stBBGKY_truncated_strong} results in
\begin{align}
&\left(\partial_{t}+\bmath{\varv}_{1}\cdot\nabla_{1}+\bmath{A}_{1}^{(2,2)}\cdot\bmath{\partial}_{1}\right)f(1,t)=-\bmath{\partial}_{1}\cdot\int_{\Omega_{2,2}} g(1,2,t)\bmath{a}_{12}\text{d}2\nonumber\\
&\quad +\int\text{d}^{3}\varv_{2}\oiint [f(1,t)f(2,t)+g(1,2,t)] \bmath{\varv}_{1,2}\cdot\text{d}\bmath{\sigma}_{1,2},\label{Eq.1stBBGKY_wc_strong}
\end{align}
and equation \eqref{Eq.2ndBBGKY_truncated_strong} reduces to  
\begin{align}
&\left(\partial_{t}+\bmath{\varv}_{1}\cdot\nabla_{1}+\bmath{\varv}_{2}\cdot\nabla_{2}+\bmath{A}_{1}^{(2,2)}\cdot\bmath{\partial}_{1}+\bmath{A}_{2}^{(2,2)}\cdot\bmath{\partial}_{2}+\bmath{a}^{\triangle}_{12}\right)g(1,2,t)\nonumber\\
&=-\left[\tilde{\bmath{a}}^{\triangle}_{12}\cdot\bmath{\partial}_{1}+\tilde{\bmath{a}}^{\triangle}_{21}\cdot\bmath{\partial}_{2}\right]f(1,t)f(2,t).\label{Eq.2nd_BBGKY_grained_Kandrup_stro}
\end{align}

One's concern is to find an approximated form of equation \eqref{Eq.2nd_BBGKY_grained_Kandrup_stro} on the Landau-radius scales. One, of course, would like to derive a Boltzmann-collision-type equation to model the close encounters. A moderate assumption can be made for test star entering the Landau sphere around a field star to hold famous '\emph{Stosszahlansatz} (molecular chaos)' or to have the destructive term $g(1,2,t\to -\infty)$ vanished. In the same way as gaseous systems \citep{Cercignani_2008}, the Stosszahlansatz must be in association with the randomness of test star \emph{entering} the Landau sphere, not that of star leaving. For the star cluster of concern, since fluctuations in the m.f. acceleration are assumed to exist at least on the scale of the Landau radius in a similar way to the randomness of Holtsmark distribution discussed in \citep{Takase_1950}, it would be 'natural' to consider that the way test star enters a field star's Landau sphere may be random yet. Even aside from the assumption taken for the randomness in the present work, one may consider that the \emph{actual beginning} of strong two-body encounter of test star with a field star is at most the mean distance of stars away, following gaseous systems\footnote{In kinetic theory of ordinary gaseous systems \citep[e.g.][]{Landau_1987}, the negative of infinity of time $t$ in the destructive term corresponds with the time duration of a particle travels the average distance of particles before collision being much greater than the correlation time. If one assumes two-body encounters begins at a distance of the average distance of stars, regardless of how the test particle interact with field particles during the flight time (retardation effect), the destructive term may vanish due to the memory loss via pre-random collision with another particles (Markovian effect).}. This is since the memory loss that test star undergoes may be efficient due to the effect of m.f. potential that dominates the motion of test star down to the BG radius $a_\text{BG}$, in addition to the violent relaxation, inhomogeneity of the system, short-range-encounter relaxation... In addition, the effect of retardation is much weaker than the product of DFs, $f(\bmath{r}_{1},\bmath{p}_{1},t)f(\bmath{r}_{1},\bmath{p}_{2},t)$, since the meaningful correlation time before entering the Landau sphere is at most the life-time of fluctuation. Hence, one may assume that test star reaches the Landau sphere of a field star in a purely random manner.

One may resort to the methods in \citep{Grad_1958,Cercignani_1972,Cercignani_1988} to find the collision-Boltzmann equation based on the Stosszahlansatz. The second equation of BBGKY hierarchy, however, must be solved for two-body DF on the surface of the Landau sphere for strictness as discussed in \citep{Cercignani_2008}, Hence, the present work reproduces the same result as those works by directly solving the second equation. This is especially since the kinetic descriptions are different for stars entering and leaving Landau spheres, which may not be suitable to resorting to the Stosszahlansatz. Neglecting the effect of m.f. acceleration due to its weakness on the scale of the Landau radius, equation \eqref{Eq.2nd_BBGKY_grained_Kandrup_stro} is simplified to 
\begin{align}
&\left(\partial_{t}+\bmath{\varv}_{1}\cdot\nabla_{1}+\bmath{\varv}_{2}\cdot\nabla_{2}+\bmath{a}_{12}\cdot\bmath{\partial}_{12}\right)g(1,2,t)\nonumber\\
&\qquad\qquad =-\bmath{a}_{12}\cdot\bmath{\partial}_{12}f(1,t)f(2,t). \qquad (r_{12}=r_\text{o})\label{Eq.2ndBBGKY_truncated_strong_surface_g}
\end{align}
For test star entering the Landau sphere around a field star, one must employ the wave kinetic description. Assuming the memory loss of test star is efficient enough so that the destructive term may not be of significance, equation \eqref{Eq.2ndBBGKY_truncated_strong_surface_g} reduces to
\begin{align}
g(1,2,t)=-\int^{t}_{t-\tau}\left[\bmath{a}_{12}\cdot\bmath{\partial}_{12}f(1,t)f(2,t)\right]_{t=t'}\text{d}t',\qquad (r_{12}=r_\text{o})
\end{align}
If the initial time of correlation is assumed to be at most the time of fluctuation, the corresponding two-body DF follows the Stosszahlanstz;
\begin{align}
f(1,2,t)=f(1,t)f(\bmath{r}_{1},\bmath{p}_{2},t)+\mathcal{O}\left(N^{4/3}\right),\qquad (r_{12}=r_\text{o})
\end{align}
where the order of correlation function is assumed to be $\sim N^{4/3}$ while $f(1,2,t)$ is order of $N^{2}$ following the scaling of section \ref{subsec:scalings}.

On the other hand, test star leaving the Landau sphere follows the collision kinetic description since the memory of the star originates from the motion of the star inside the Landau sphere. Following the theory of non-ideal system (section \ref{subsec:nonideal}), the corresponding collision kinetic description is equation \eqref{Eq.BBGKY_2nd_two_body}. Since the effect of non-idelity in encounter may be neglected, exploiting the method of characteristics, the solution, equation \eqref{Eq.twoDF_strong_chact}, may be approximated to
\begin{align}
f(1,2,t)=f(\bmath{r}_{1}, \bmath{p}_{1}(t-\tau), t)f(\bmath{r}_{1}, \bmath{p}_{2}(t-\tau), t)+\mathcal{O}\left(N^{3/2}\right),\label{Eq.Stoss}
\end{align}
where stars 1 and 2 follow the trajectories, equation \eqref{Eq.characteristics_v_jump}, for local encounters of stars. Equation \eqref{Eq.Stoss} still holds even if the DF is not well defined inside the Landau sphere since the ideal encounter essentially assumes the spatial locality $(\bmath{r}_{1}=\bmath{r}_{2})$ and the Newtonian interaction can be realized only through the momenta $\bmath{p}_{1}(t-\tau)$ and $\bmath{p}_{2}(t-\tau)$ in the same way as the Bogoliubov's derivation (Appendix \ref{Appendix:Bogorigouv}) for the Boltzmann equation.

\begin{figure}
	\caption{A schematic trajectory of test star around a field star. (a) test star entering the Landau sphere, which corresponds with the loss of test star from states $f(\bmath{r}_{1},\bmath{p}_{1},t)f(\bmath{r}_{1},\bmath{p}_{2},t)$ to be described in wave kinetic description; since stars approaches each other, only the surface of top hemisphere contributes to the collision term. (b) test star leaving the sphere, which corresponds with the gain of stars to state $f(\bmath{r}_{1},\bmath{p}_{1}(t-\tau),t)f(\bmath{r}_{1},\bmath{p}_{2}(t-\tau),t)$ to be described in collision kinetic description: since the stars reseeds each other, only the surface of right hemisphere contributes to the collision term.}
	\label{tabel:strong_encounter}
\begin{tikzpicture}
\shade[ball color = gray!20, opacity = 0.4] (0,0) circle (1.41421cm);
\draw (0,0) circle (1.41421cm);
\draw (-1.41421,0) arc (180:360:1.41421 and 0.6);
\draw[densely dashed] (1.41421,0) arc (0:180:1.41421 and 0.6);
\draw[densely dashed] (0,0 ) -- node[above]{$\triangle$} (1.41421,0);
\shadedraw[inner color=orange, outer color=yellow, draw=black] (-1,2.6) circle (0.1cm);
\node [right] at (-1,2.3){$\bmath{p}_{1}$};
\node [right] at (-0.9,2.6) {test star};
\shadedraw[inner color=orange, outer color=yellow, draw=black] (0,0) circle (0.1cm);
\node at (0.1,-0.2) {field star};
\node[above] at (-.1,0){$\bmath{r}_{2}$};

\draw[red,thick,solid](-1,2.5)--(-1,1);
\node[right] at (-1.,0.9){$\bmath{r}_{1}$};
\draw[thick,red,->](-1,2.5)--(-1,2.0);
\draw[thick,blue,->,>=stealth](-1,1)--(-1.5,1.5);
\node[above] at (-1.4,1.5) {$\text{d}\bmath{\sigma}_{1,2}$};
\draw[red!100,thick,dotted](-1,1)..controls (-1,-0.5) and(0.5,-1).. (1,-1);
\draw[red,thick,solid](1,-1)--(2.0,-1);
\draw[red,thick,->](1.5,-1)--(2.0,-1);
\node[below]  at (1.5,-1.2){$\bmath{p}_{1}'$};
\node at (0,-2.3){(a) Loss from  $(\bmath{p}_{1},\bmath{p}_{2})$};
\end{tikzpicture}
\begin{tikzpicture}
\draw[thick] (-2,3.0)--(-2,-2.5);
\shade[ball color = gray!10, opacity = 0.4] (0,0) circle (1.41421cm);
\draw (0,0) circle (1.41421cm);
\draw (0,1.41421) arc (90:270:0.6 and 1.41421);
\draw[densely dashed] (0,1.41421) arc (90:-90: 0.6 and 1.41421);
\draw[densely dashed] (0,0 ) -- node[left]{$\triangle$} (0,1.41421);
\shadedraw[inner color=orange, outer color=yellow, draw=black] (0,0) circle (0.1cm);
\node at (0.1,-0.2) {field star};

\draw[red,thick,solid](-1,2.5)--(-1,1);
\node [right] at (-1,2.5){$\bmath{p}_{1}(t-\tau)$};
\draw[thick,red,->](-1,2.5)--(-1,2.0);
\draw[thick,blue,->,>=stealth](1,-1)--(1.5,-1.5);
\node[below] at (1.5,-1.6) {$\text{d}\bmath{\sigma}_{1,2}$};
\draw[red!100,thick,dotted](-1,1)..controls (-1,-0.5) and(0.5,-1).. (1,-1);
\draw[red,thick,solid](1,-1)--(2.0,-1);
\draw[red,thick,->](1.5,-1)--(2.0,-1);
\shadedraw[inner color=orange, outer color=yellow, draw=black] (2.1,-1) circle (0.1cm);
\node [below] at (1.9,-1.1) {test star};
\node  at (2,-0.7){$\bmath{p}_{1}$};
\node at (0.5,-2.3){(b) Gain from $(\bmath{p}_{1}(t-\tau),\bmath{p}_{2}(t-\tau))$};
\end{tikzpicture}
\end{figure}
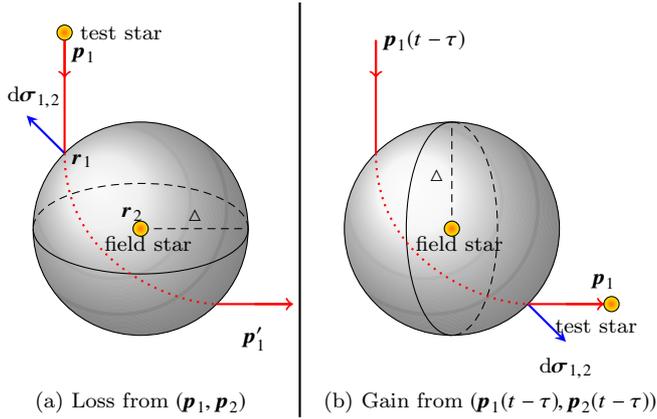

As typically discussed for the collision Boltzmann equation, the two encounters may be separated into two events (i) the loss from state $f(\bmath{r}_{1},\bmath{p}_{1},t)f(\bmath{r}_{1},\bmath{p}_{2},t)$  and (ii) the gain from the other state $ f(\bmath{r}_{1},\bmath{p}_{1}(t-\tau),t)f(\bmath{r}_{1},\bmath{p}_{2}(t-\tau),t)$ (See Figure \ref{tabel:strong_encounter}). Due to the symmetry of the trajectory, only the surface of hemisphere associated with the state $(\bmath{r}_{1},\bmath{p}_{2},t)$ is needed for the surface integral of the collision term in equation \eqref{Eq.1stBBGKY_wc_strong} for both of the events. 
Following \citep{Grad_1958,Cercignani_1972}, the surface integral can be projected onto a domain of thin disk with radius $\triangle$
\begin{align}
&\bmath{\varv}_{1,2}\cdot\text{d}\bmath{\sigma}_{1,2}=
\begin{cases}
-\varv_{12}\text{d}\sigma_{1,2},     & \text{(if stars approach each other)} \\
\varv_{12}\text{d}\sigma_{1,2}.      & \text{(if stars reseeds each other)} \label{Eq.projection}
\end{cases}
\end{align} 
The surface-integral term in equation \eqref{Eq.1stBBGKY_wc_strong} reduces to the collisional-Boltzmann term for ideal close encounters (i.e. encounters are local in space and time)
	\begin{align}
	&I_\text{Bol}^{\blacktriangledown\text{(loc)}}=\int\text{d}^{3}\bmath{p}_{2}\int_{0}^{\triangle}b\text{d}b \int_{0}^{2\pi}\text{d}\psi \varv_{1,2}\delta f(\bmath{r}_{1},\bmath{r}_{12},\bmath{p}_{1},\bmath{p}_{2},t),\nonumber\\
	&\delta f(\bmath{r}_{1},\bmath{r}_{12},\bmath{p}_{1},\bmath{p}_{2},t)\nonumber\\
	&\quad\equiv[-f(1,t)f(\bmath{r}_{1},\bmath{p}_{2},t)+f(\bmath{r}_{1},\bmath{p}_{1}(t-\tau),t)f(\bmath{r}_{1},\bmath{p}_{2}(t-\tau),t)],
	\end{align}
where $b$ is the impact parameter (the radial coordinate on the disc) and $\psi$ is the azimuth.
Hence, the g-Landau equation including the effect of close encounters based on the 'point-particle method' is
\begin{align}
&\left(\partial_{t}+\bmath{\varv}_{1}\cdot\nabla_{1}+\bmath{A}_{1}^{(2,2)}\cdot\bmath{\partial}_{1}\right)f(1,t)=I_\text{g-L}^{\blacktriangle}+I^{\blacktriangledown\text{(loc)}}_\text{Bol}.\label{Eq.correct_kinetic_eqn}
\end{align}
Equation \eqref{Eq.correct_kinetic_eqn} embodies a mathematically non-divergent kinetic equation at distances $r-\text{o}<r_{12}<R$. The method above is important as the first approximation of discreteness since it employs only terms up to $\mathcal{O}(1/N)$ without including any mathematical divergence in the process of formulation. As one realises, equation \eqref{Eq.correct_kinetic_eqn} includes the truncated m.f. acceleration, of course since one can not define the statistical quantity inside the Landau sphere due to the definition for the truncated DF. It, however, is important to consider the statistical dynamics inside the Landau sphere when one considers the dense core or halo with a strong inhomogeneity of a star cluster at the late stage of core-collapse. Since \citep{Grad_1958}'s method does not explicitly includes any information inside the Landau sphere, one may resort to \citep{Klimontovich_1982}'s theory (See section \ref{sec.strong_Klimon}). Also, the mathematical divergence problem for equation \eqref{Eq.correct_kinetic_eqn} will be discussed in section \ref{sec:convergent}.

\subsection{Strong encounter in a strongly-inhomogeneous system based on the 'domain-patched' DF}\label{sec.strong_Klimon}
In the present section, the statistical dynamics inside the Landau sphere is discussed, which was neglected in the truncated-DF description. The relevant important case would be dynamics in the core or inner halo of a strongly inhomogeneous weakly-coupled star cluster which is undergoing a gravithermal-instability. Assume the density of the core may reach order of $N^{2}\tilde{n}$ at the late stage of the core collapse as reference for the maximum density that the truncated DF may possibly achieve. If the Landau sphere of concern is situated in the core of the system, where the field stars are possibly considered to be a (quasi-)Maxwellian and the m.f. acceleration is less significant compared to that in the inner halo, one can focus on two-body problem. In the core, The pair-wise interaction may be considered only due to short-range Newtonian forces and the randomness of test star entering the Landau sphere may be assumed without a strong conflict (See section \ref{sec.ret}). On the other hand, if the Landau sphere of concern is situated in the inner halo of the system that reaches a strong inhomogeneity in density modeled by a density of $\sim 1/r_{1}^{2}$ (assuming a kind of the isothermal gas sphere), one may still assume that the effects of inhomogeneity and violent relaxation may hold the randomness of test star entering the sphere (See section \ref{sec.inh}).

Since the two cases above (high density of the core and strong inhomogeneity of the halo density) necessitate one to consider the non-locality of two-body encounters in time and space inside the Landau sphere, one may resort to the domain-patched $s$-body DFs (section \ref{subsec:DF}) to discuss the small-scale dynamics. The first equation of the BBGKY hierarchy for the domain-patched DF is
\begin{align}
&\left(\partial_{t}+\bmath{\varv}_{1}\cdot\nabla_{1}+\bmath{A}_{1}\cdot\bmath{\partial}_{1}\right)f_{1}\nonumber\\
&\qquad\quad=\bmath{\partial}_{1}\cdot\int_{r_{12}>\triangle}g_{2}^{\blacktriangle}\bmath{a}_{12}\text{d}_{2}+\bmath{\partial}_{1}\cdot\int_{r_{12}<\triangle}g_{2}^{\blacktriangledown}\bmath{a}_{12}\text{d}_{2},\label{Eq.1stBBGKY_patched}
\end{align}
and the second equation splits into two equations; one is equation \eqref{Eq.2nd_BBGKY_grained_Kandrup} at distances $r_{12}>r_\text{o}$ (assuming weak-coupling limit), and another is 
\begin{align}
&\left(\partial_{t}+\bmath{\varv}_{1}\cdot\nabla_{1}+\bmath{\varv}_{2}\cdot\nabla_{2}+\bmath{a}_{12}\cdot\bmath{\partial}_{12}\right)g^{\blacktriangledown}(1,2,t)\nonumber\\
&\qquad\qquad\qquad=\bmath{a}_{12}\cdot\bmath{\partial}_{12}f(1,t)f(2,t).\qquad (r_{12}<r_\text{o})\label{Eq.2ndBBGKY_in_ro}
\end{align}
The correlation functions $g^{\blacktriangledown}(1,2,t)$ and $g^{\blacktriangle}(1,2,t)$ can be simply decoupled since they are essentially local in configuration space and momentum space\footnote{One must be careful when considering the gravitational polarization effect since such effect couples the correlations in the spaces.}. In the outside of the Landau sphere, the correlation $g^{\blacktriangle}(1,2,t)$ reduces to the g-Landau equation. On the other hand, inside the sphere, employing the method of characteristics and the theory of \citep{Klimontovich_1982}, one obtains a basic form of correlation function, equation \eqref{Eq.g(1,2,t)_collision}, for a non-ideal star cluster. The correlation function , equation\eqref{Eq.g(1,2,t)_collision}, will be discussed for a dense core in section \ref{sec.ret} and a halo with strong inhomogeneity in section \ref{sec.inh}. 

It is to be noted that, to find the Boltzmann collision term based on the domain-patched DF, one must employ the Grad's derivation (section \ref{sec.WC_strong}), not the Bogoliubov's derivation (Appendix \ref{Appendix:Bogorigouv}) that has been one of standard methods to derive the collision term \citep{Uhlenbeck_1963,Klimontovich_1982,Landau_1987}. This is since the Bogoliubov's derivation needs a cylindrical spatial volume in which possible encounters may occur while the volume mismatches the Landau sphere. Assume the encounter of concern inside the Landau sphere is ideal, then the collision term for the binary DF $f^{\blacktriangledown}(1,2,t)$ reads
\begin{subequations}
	\begin{align}
	I_\text{cor}&=-\int \bmath{a}_{12}\cdot\bmath{\partial}_{1}f^{\blacktriangledown}(1,2,t)\text{d}2,\\
	&\approx \int \left[\left(\bmath{\varv}_{12}\cdot\nabla_{12}\right)f^{\blacktriangledown}(\bmath{r}_{12},\bmath{r}_{1},\bmath{\varv}_{1},\bmath{\varv}_{2},t)\right]_{r_{12}<\triangle}\text{d}2.\label{Eq.ideal_g(1,2,t)_in}
	\end{align} 
\end{subequations}
To apply the Grad's derivation for equation \eqref{Eq.ideal_g(1,2,t)_in}, one needs to employ the Gauss's lemma in $r_{12}$ space;
 \begin{align}
 I_\text{cor}=\oiint\int f^{\blacktriangledown}(\bmath{r}_{12},\bmath{r}_{1},\bmath{p}_{1},\bmath{p}_{2},t)\bmath{\varv}_{12}\cdot\text{d}\bmath{\sigma}_{12}\text{d}^{3}\bmath{p}_{2}.\qquad (r_{12}=r_\text{o})\label{Eq.ideal_g(1,2,t)_surf}
 \end{align} 
Equation \eqref{Eq.ideal_g(1,2,t)_surf} is exactly the same as the surface integral term of the first equation \eqref{Eq.1stBBGKY_wc_strong} of the BBGKY hierarchy for the truncated DF. Hence one obtains the Boltzmann collision term $I^{\blacktriangledown\text{(loc)}}_\text{Bol}$ following section \ref{sec.WC_strong}.  It is to be noted that equation \eqref{Eq.ideal_g(1,2,t)_surf} is true only for ideal close encounter that satisfies
\begin{align}
&f^{\blacktriangledown}(\bmath{r}_{12},\bmath{r}_{1},\bmath{p}_{1},\bmath{p}_{2},t)\nonumber\\
&\qquad=f(1,t)f(\bmath{r}_{1},\bmath{p}_{1},t)+g^{\blacktriangledown}(\bmath{r}_{12},\bmath{r}_{1},\bmath{p}_{1},\bmath{p}_{2},t), \qquad (r_{12}<\triangle)
\end{align}
and the effect of non-ideality will be discussed in sections \ref{sec.ret} and \ref{sec.inh}.

\subsubsection{The retardation effect of strong encounter in a dense core}\label{sec.ret}
Consider that strong encounters may be likely to occur in the Landau sphere of test star in the dense core whose density gradient is not significant. Due to the short correlation time ,$\tau \lesssim 1/N$, inside the Landau sphere, equation \eqref{Eq.g(1,2,t)_collision} at $r_{12}<r_\text{o}$  reduces to
\begin{subequations}
	\begin{align}
	&g^{\blacktriangledown}(1,2,t)=g_\text{Bol}^{\blacktriangledown}(1,2,t)+g^{\blacktriangledown}_\text{ret}(1,2,t),\label{Eq.whole_g}\\
	&g^{\blacktriangledown}_\text{Bol}(1,2,t)\equiv f(\bmath{r}_{1},\bmath{p}_{1}(t-\tau),t)f(\bmath{r}_{2},\bmath{p}_{2}(t-\tau),t)-f(1,t)f(2,t),\\
	&g^{\blacktriangledown}_\text{ret}(1,2,t)\equiv-\int_{0}^{\tau} \tau'\left[\partial_{t}+\bmath{\varv}_{1}\left(t-\tau'\right)\cdot\nabla_{1}+\bmath{\varv}_{2}\left(t-\tau'\right)\cdot\nabla_{2}\right]\nonumber\\
	&\qquad\qquad\quad\times \frac{\text{d}}{\text{d}\tau'}f\left(\bmath{r}_{1},\bmath{p}_{1}\left(t-\tau'\right),t\right)f\left(\bmath{r}_{2},\bmath{p}_{2}\left(t-\tau'\right),t\right)\text{d}\tau'.\label{Eq.cor_strong_ret}
	\end{align}
\end{subequations}
where only the first order of $\tau$ is taken to expand the DFs following the \citep{Klimontovich_1982} and the non-locality in the argument for configuration spaces will be discussed in section \ref{sec.inh}. In limit of $\tau\to\infty$, the function $g^{\blacktriangledown}_\text{Bol}$ is a correlation function associated with the Boltzmann collision term while the function $g^{\blacktriangledown}_\text{ret}$ is associated with the retardation effect. As discussed in \citep{Snider_1995}, the retardation effect originated from the product of DFs, $f(\bmath{r}_{1}(t-\tau),\bmath{p}_{1}(t-\tau),t-\tau)f(\bmath{r}_{2}(t-\tau),\bmath{p}_{2}(t-\tau),t-\tau)$, is of significance to hold the conservation of the total energy of test star and a field star during the two-body encounter, while as explained in \citep{Klimontovich_1982} the function $g^{\blacktriangledown}_\text{ret}(1,2,t)$ is of special importance in considering the effect of high density of background field stars on the two-body encounter of concern. The retardation effect, however, is weak enough to be neglected compared to the Boltzmann collision term
\begin{align}
\frac{g^{\blacktriangledown}_\text{ret}(1,2,t)}{g^{\blacktriangledown}_\text{Bol}(1,2,t)}\sim\max\left(\frac{\tau}{t_\text{rel}},\tau\frac{\varv_{1}+\varv_{2}}{2}\cdot\nabla_{1}\right)\lesssim \mathcal{O}\left(1/N^{1/2}\right),
\end{align}
where the approximation $\bmath{r}_{1}-\bmath{r}_{2}\approx \bmath{r}_{1}+\mathcal{O}(1/N)$ was employed due to the locality in space. Hence, the total correlation function may reduce to $g^{\blacktriangledown}_\text{Bol}(1,2,t)$. The discussion in the present section is true if the system of concern is assumed a weakly-inhomogeneous in the core. If the density of the system, however, is strongly inhomogeneous in the halo, one must consider the effects of non-locality in arguments for spatial coordinates in equation \eqref{Eq.whole_g} as shown in section \ref{sec.inh}.

\subsubsection{The spatial non-locality of strong encounter in a strongly inhomogeneous halo}\label{sec.inh}
For simplicity, first neglect the effect of retardation in the present section. To some extent obvious though, the form of equation \eqref{Eq.whole_g} is the same as
\begin{align}
g(1,2,t)=f(1,2,t)-f(1,t)f(2,t),\label{Eq.cor_strong_2}
\end{align}
To consider the effect of non-locality in space one may expand equation \eqref{Eq.cor_strong_2} in series of the locality parameter $\mid\frac{\bmath{r}_{12}}{2}\cdot\nabla_{1}\mid$ \citep[e.g.][]{Belyi_2002} as follows
\begin{subequations}
	\begin{align}
	g^{\blacktriangledown}(1,2,t)=&g^{\blacktriangledown}(\bmath{R},\bmath{r}_{12},\bmath{p}_{1},\bmath{p}_{2},t),\\
	=&\exp{\left[-\frac{\bmath{r}_{12}}{2}\cdot\nabla_{1}\right]}g^{\blacktriangledown}(\bmath{r}_{1},\bmath{r}_{12},\bmath{p}_{1},\bmath{p}_{2},t),\\
	=&\exp{\left[-\frac{\bmath{r}_{12}}{2}\cdot\nabla_{1}\right]}\nonumber\\
	&\times\left[f^{\blacktriangledown}(\bmath{r}_{1},\bmath{r}_{12},\bmath{p}_{1} ,\bmath{p}_{2},t)-f(1,t)f(\bmath{r}_{1},\bmath{p}_{2},t)\right],\label{Eq.cor_strong_general}
	\end{align}
\end{subequations}
where the following approximation for small $\bmath{r}_{12}$ is exploited
\begin{align}
\bmath{r}_{12}\cdot\nabla_{R}\approx\frac{\bmath{r}_{12}}{2}\cdot\nabla_{1}.
\end{align}
In zeroth order of $\mid\frac{\bmath{r}_{12}}{2}\cdot\nabla_{1}\mid$, encounters modeled by the function, equation \eqref{Eq.cor_strong_general}, may occur locally in space as if the stars behaved as neutral particles in rarefied gas \citep[e.g.][]{Balescu_1960,Balescu_1997}
\begin{subequations}
	\begin{align}
	g^{\blacktriangledown}(1,2,t)=&f(\bmath{r}_{1},\bmath{r}_{12},\bmath{p}_{1},\bmath{p}_{2},t)-f(\bmath{r}_{1},\bmath{p}_{1},t)f(\bmath{r}_{1},\bmath{p}_{2},t),\label{Eq.cor_strong_Balescu}\\
	=&\delta f(\bmath{r}_{1},\bmath{r}_{12},\bmath{p}_{1},\bmath{p}_{2},t),\label{Eq.cor_strong_Balescu_soln}\\
	\end{align}
\end{subequations}
where equation \eqref{Eq.Stoss} is employed at distances $r_{12}<r_\text{o}$. In the zeroth order, the product of DFs, $f(1,t)f(\bmath{r}_{1},\bmath{p}_{2},t)$, does not contribute to the corresponding kinetic equation as a collision term (termed as 'the unweighted average potential \citep{Balescu_1960}' in wave kinetic description) due to the factors associated with distances $\bmath{r}_{12}$ being missing; 
\begin{align}
-&\int \bmath{a}_{12}\cdot\bmath{\partial}_{1}f(1,t)f(\bmath{r}_{1},\bmath{p}_{2},t)\text{d}\bmath{r}_{12}\nonumber\\
&\quad=\int \bmath{\varv}_{12}\cdot\nabla_{12}f(1,t)f(\bmath{r}_{1},\bmath{p}_{2},t)\text{d}\bmath{r}_{12}=0.
\end{align}
This implies one must employ the truncated m.f. acceleration for the kinetic equation since the m.f. acceleration itself is ignorable at distances $r_{12}<r_\text{o}$. 

To consider the spatial inhomogeneity of the star cluster of concern, the next order is of importance. In the first order of the parameter $\mid\frac{\bmath{r}_{12}}{2}\cdot\nabla_{1}\mid$, the stars behave as if they were constituent particles of weakly non-ideal (or imperfect) gases \citep[e.g.][]{Green_1952,Klimontovich_1982,Snider_1995}
\begin{subequations}
	\begin{align}
	&g^{\blacktriangledown}(1,2,t)\nonumber\\
	&=\left(1-\frac{\bmath{r}_{12}}{2}\cdot\nabla_{1}\right)\delta f(\bmath{r}_{1},\bmath{r}_{12},\bmath{p}_{1},\bmath{p}_{2},t),\\
	&=\delta f(\bmath{r}_{1},\bmath{r}_{12},\bmath{p}_{1},\bmath{p}_{2},t)-\frac{\bmath{r}_{12}}{2}\cdot\nabla_{1}f(\bmath{r}_{1},\bmath{p}_{1}(-\infty),t)f(\bmath{r}_{1},\bmath{p}_{2}(-\infty),t)\nonumber\\
	&\qquad +\frac{\bmath{r}_{12}}{2}\cdot\nabla_{1}f(\bmath{r}_{1},\bmath{p}_{1},t)f(\bmath{r}_{1},\bmath{p}_{2},t).\label{Eq.cor_strong_Klimon}
	\end{align}
\end{subequations}
where the second term on the R.H.S in the last equation is associated with the spatial non-locality of background field stars and the collision term can be written as
\begin{align}
&I_\text{Bol}^{\blacktriangledown (\text{inh})}=\int \left[\bmath{a}_{12}\cdot\bmath{\partial}_{1}\frac{\bmath{r}_{12}}{2}\right]_{r_{12}<\triangle}\cdot\nabla_{1}\nonumber\\
&\qquad\qquad \times f(\bmath{r}_{1},\bmath{p}_{1}(-\infty),t)f(\bmath{r}_{1},\bmath{p}_{2}(-\infty),t)\text{d}_{2}.
\end{align}\label{Eq.I_nonl}
The third term on the R.H.S of equation \eqref{Eq.cor_strong_Klimon} is related to the m.f. acceleration. The m.f. acceleration in the first order of the parameter $\mid\frac{\bmath{r}_{12}}{2}\cdot\nabla_{1}\mid$ reads
\begin{subequations}
	\begin{align}
	\bmath{A}^{\blacktriangledown}_{1}=&\int_{r_{12}<r_\text{o}}\bmath{a}_{12}f(2,t)\text{d}_{2},\\
	\approx&\int_{r_{12}<r_\text{o}}\bmath{a}_{12}(1-\bmath{r}_{12}\cdot\nabla_{1})f(\bmath{r}_{1},\bmath{p}_{2},t)\text{d}_{2},
	\end{align}
\end{subequations}
and the term, $\frac{\bmath{r}_{12}}{2}\cdot\nabla_{1}f(\bmath{r}_{1},\bmath{p}_{1},t)f(\bmath{r}_{1},\bmath{p}_{2},t)$, in equation \eqref{Eq.cor_strong_Klimon} is associated with the m.f. acceleration, hence their total contribution may be written as follows
	\begin{align}
	A_\text{1.can}^{\blacktriangledown}\cdot\bmath{\partial}_{1} f(1,t)\equiv-\frac{\pi Gm\triangle^{2}}{3} \left[n(1,t)\nabla_{1}-\nabla_{1}n(1,t)\right]\cdot\bmath{\partial}_{1}f(1,t).\label{Eq.I_inh}
	\end{align}
and equation \eqref{Eq.I_inh} may be termed a 'corrected' m.f. acceleration of star 1 inside the Landau sphere. 
	
The expression of equation \eqref{Eq.I_inh} clearly shows that the spatial non-locality of close encounter is of significance when the system has a \emph{strong inhomogeneity} over the Landau radius scale. This implies the spatial non-locality may possibly contribute to the evolution of stellar clusters especially in the inner halo at the late stage of core collapse, which has not been discussed in the conventional collision kinetic theories. The OoM of the term $A_\text{1.can}^{\blacktriangledown}\cdot\bmath{\partial}_{1} f(1,t)$ for a weakly inhomogeneous system (described by the scaling of section \ref{subsec:scalings}) is approximately 
\begin{align}
\left|\frac{A_\text{1.can}^{\blacktriangledown}\cdot\bmath{\partial}_{1} f(1,t)}{\partial_{t} f(1,t)}\right|\sim G r_\text{o}^{2}n(\bmath{r}_{1},t)\frac{t_\text{rel}}{|\bmath{r}_{1}|}\frac{\text{d}\ln n(\bmath{r}_{1},t)}{\text{d}\ln |\bmath{r}_{1}|}\sim \mathcal{O}(1/N^{2}),
\end{align}
where $\frac{\text{d}\ln n(\bmath{r}_{1},t)}{\text{d}\ln |\bmath{r}_{1}|}$ is assumed to be finite and the center of the system is placed at the origin of coordinate space. On the other hand, in a strongly inhomogeneous system, following the numerical results for the core collapse of star clusters \citep{Cohn_1979,Takahashi_1995}, the magnitude of density-profile gradient at the inner halo (the self-similar region) may be scaled by
\begin{align}
\frac{\text{d}\ln n(\bmath{r}_{1},t)}{\text{d}\ln |\bmath{r}_{1}|}\approx-2,\label{Eq.grad_neg2}
\end{align}
and note that the density profile of the system approximately follows that of an isothermal sphere\footnote{In the result of \citep{Cohn_1979,Takahashi_1995}, the correct gradient in density is characterised by 
\begin{align}
\frac{\text{d}\ln n(\bmath{r}_{1},t)}{\text{d}\ln |\bmath{r}_{1}|}\approx-2.23.
\end{align}
However, considering the fact that the density profiles of various isolated self-gravitating systems are approximately modeled by Lane-Emden functions \citep[e.g.][]{Ito_2018} and the isothermal sphere has the most gentle gradient \citep[e.g.][]{Chandra_1939,Ito_2018}, employing equation \eqref{Eq.grad_neg2} is the most conservative way for the scaling of spatial inhomogeneity.}
\begin{align}
n(\bmath{r}_{1},t)\sim \frac{1}{r_{1}^{2}}.\label{Eq.norm_dens}
\end{align}
At an epoch that the possible maximum density reaches order of $\sim N^{2}$ on the Landau radius scales, the corrected m.f. term $A_\text{1.can}^{\blacktriangledown}\cdot\bmath{\partial}_{1} f(1,t)$ is a necessary term at central radii $r_{1}\lesssim \mathcal{O}\left(1/N^{2/3}\right)$;
\begin{align}
\left|\frac{A_\text{1.can}^{\blacktriangledown}\cdot\bmath{\partial}_{1} f(1,t)}{\partial_{t} f(1,t)}\right|\gtrsim \mathcal{O}(1).
\end{align}
One may consider the second order of $\left|\frac{\bmath{r}_{12}}{2}\cdot\nabla_{1}\right|$ since they are just of less significance by a factor of at least $\mathcal{O}(N^{1/3})$ compared to the first order meaning it is of importance at central radius of $\mathcal{O}(1/N^{3/4})$. However, due to the mathematical complication of the Boltzmann collision term itself and the higher orders of non-local terms and another mathematical divergent problem in strong encounter\footnote{Higher orders of density expansion must be typically dealt with the ring approximation \citep[e.g.][]{Ernst_1998}.}, holding only the first order would be a moderate extension of ideal strong encounters. Hence, it is to be noted that the applicability of weakly-nonideal encounter (the first order expansion of Boltzmann collision term) would be applicable for part of star cluster at central radii $r_{1}>1/N^{3/4}$ and to correctly handle the inhomogeneity one must hold all the higher orders of spatial non-local. 

\subsubsection{Kinetic equation for weakly- and fully non-local close encounters}\label{sec.weak-non}
Combining the results of sections \ref{sec.ret} and \ref{sec.inh}, the total of collision terms for strongly inhomogeneous systems associated with weakly-nonideal close encounters inside the Landau sphere reads
\begin{align}
&I_\text{nonid}=I_\text{Bol}^{\blacktriangledown\text{(loc)}}+I_\text{Bol}^{\blacktriangledown (\text{inh})} -A_\text{1.can}^{\blacktriangledown}\cdot\bmath{\partial}_{1} f(1,t)+I_\text{ret}^{\blacktriangledown},\label{Eq.correct_kinetic_eqn_strong}
\end{align}
where the retardation term associated with spatial inhomogeneity at distances $r_{12}<\triangle$ is
\begin{align}
I_\text{ret}^{\blacktriangledown}=&-\int \bmath{a}_{12}\cdot\bmath{\partial}_{1}\int_{0}^{\tau} \tau'\left[\frac{\bmath{\varv}_{1}+\bmath{\varv}_{2}}{2}\cdot\nabla_{1}\right]\nonumber\\
&\quad\times \frac{\text{d}}{\text{d}\tau'}f\left(\bmath{r}_{1},\bmath{p}_{1}\left(t-\tau'\right),t\right)f\left(\bmath{r}_{1},\bmath{p}_{2}\left(t-\tau'\right),t\right)\text{d}\tau'\text{d}2.\label{Eq.I_ret_inh}
\end{align}
Equation \eqref{Eq.I_ret_inh} vanishes for a weakly inhomogeneous system while it remains as a collision term for strongly inhomogeneous systems. The first two terms on the R.H.S of equation \eqref{Eq.correct_kinetic_eqn_strong} originate from pure two-body encounter and the effect of inhomogeneity at the beginning of the encounter while the last two terms are due to the effect of background stars inside the Landau sphere.

To find the whole expression of the kinetic equation, one must also consider the scaling of physical quantities outside the Landau sphere. Employing the scaling of the density, equation \eqref{Eq.norm_dens}, one may find for $r_{1}>>r_\text{o}$
\begin{subequations}
	\begin{align}
	&f(1,t)\sim 1/r_{1}^{2},\\
	&g(1,2,t)\sim \max\left(\frac{t_\text{cor}}{Nr_{12}^{2}r^{2}_{1}r_{2}^{2}},\frac{t_\text{cor}}{N^{2}r^{4}_{1}r_{2}^{2}} \right),\\
	&\bmath{A}_{1}\sim \frac{1}{r_{1}^{2}N},\label{Eq.scale_A_inh}\\
	&1/t_\text{cor}\sim \max\left(\bmath{A}_{1}\cdot\bmath{\partial}_{1},\bmath{A}_{2}\cdot\bmath{\partial}_{1},\bmath{\varv}_{1}\cdot\bmath{\nabla}_{1}+\bmath{\varv}_{2}\cdot\bmath{\nabla}_{2}\right),\\
	&\bmath{\varv}_{1}\sim\bmath{\varv}_{2}\sim 1.
	\end{align}\label{Eq.scale_inh}
\end{subequations}
One sees a complication of scaling due to the effect of strong inhomogeneity in equation \eqref{Eq.scale_inh}, only one concern in formulation, however, is if the m.f. acceleration is of significant or not. The effect of the m.f. acceleration is very weak on the system-size scales as seen in equation \eqref{Eq.scale_A_inh} while it is of significance on small scales; one may keep exploiting the g-Landau equation outside the Landau sphere (The weak-coupling approximation is employed yet for simplicity). As a result one obtains the following kinetic equation
\begin{align}
&\left(\partial_{t}+\bmath{\varv}_{1}\cdot\nabla_{1}+\left[\bmath{A}^{(2,2)}_{1}+A_\text{1.can}^{\blacktriangledown}\right]\cdot\bmath{\partial}_{1}\right)f(1,t)\nonumber\\
&\qquad\qquad\qquad\qquad=I_\text{Bol}^\text{(loc)}+I_\text{ret}^{\blacktriangledown}+I_\text{Bol}^{\blacktriangledown (\text{inh})}+I_\text{g-L}^{\blacktriangle}.\label{Eq.coptimal_collision}
\end{align}

Equation \eqref{Eq.coptimal_collision} is obviously different from typical kinetic equations of collision kinetic theory that have a form of
\begin{align}
&\left(\partial_{t}+\bmath{\varv}_{1}\cdot\nabla_{1}+\bmath{A}_{1}\cdot\bmath{\partial}_{1}\right)f(1,t)=I_\text{Bol}^{\text{(loc)}}\label{Eq.coptimal_collision_typical}.
\end{align}\label{conv_collision}
First, kinds of equation \eqref{Eq.coptimal_collision_typical} do not correctly show the relation between the m.f. acceleration and Boltzmann collision term. Also, the effect of the non-ideality has been neglected in collision kinetic theories while the order of the non-ideality terms can be the same as that of the m.f. acceleration inside the Landau sphere, $\bmath{A}_{1}^{\blacktriangledown}$. Even if considering the fully non-local collision term inside the Landau sphere, one obtains the following kinetic equation

	\begin{align}
	&\left(\partial_{t}+\bmath{\varv}_{1}\cdot\nabla_{1}+\bmath{A}^{(2,2)}_{1}\cdot\bmath{\partial}_{1}\right)f(1,t)=I_\text{Bol}^{\blacktriangledown\text{(inh)}}+I_\text{ret}^{\blacktriangledown\text{(inh)}}+I_\text{g-L}^{\blacktriangle},\label{Eq.correct_kinetic_eqn_strong_com}
	\end{align}
	where
	\begin{subequations}
		\begin{align}
	&I_\text{Bol}^{\blacktriangledown\text{(inh)}}\equiv\int\text{d}^{3}\bmath{p}_{2}\int_{0}^{\triangle}b\text{d}b \int_{0}^{2\pi}\text{d}\psi \varv_{1,2}[-f(\bmath{r}_{1},\bmath{p}_{1},t)f(\bmath{r}_{2},\bmath{p}_{2},t)\nonumber\\
	&\qquad+f\left(\bmath{r}_{1}(t-\tau),\bmath{p}_{1}(t-\tau),t\right)f\left(\bmath{r}_{2}(t-\tau),\bmath{p}_{2}(t-\tau),t\right)],\\
	&I_\text{ret}^{\blacktriangledown\text{(inh)}}=-\int \bmath{a}_{12}\cdot\bmath{\partial}_{1}\int_{0}^{\tau}\left[\frac{\bmath{\varv}_{1}+\bmath{\varv}_{2}}{2}\cdot\nabla_{1}\right]\nonumber\\
	&\quad\qquad\times f\left(1\left(t-\tau'\right),t\right)f\left(2\left(t-\tau'\right),t\right)\text{d}\tau'\text{d}2.
	\end{align}
\end{subequations}
where the retardation term associated with the spatial inhomogeneity must take non-approximated form since it is the same order as the non-locality of strong encounter. Hence, one does obtain the truncated acceleration, not the standard m.f. acceleration. 

\subsubsection{The moments of collision terms and the insignificance of m.f. acceleration inside the Landau sphere}\label{sec.mom}
It was obviously shown that the truncated m.f. acceleration must be employed in place of typical one; for the zeroth order of non-ideality (equation \eqref{Eq.cor_strong_Balescu_soln}), the first order (equation \eqref{Eq.correct_kinetic_eqn_strong_com}) and the maximum order (fully spatially non-local equation \eqref{Eq.coptimal_collision}). Only one concern left is if the m.f. acceleration can perfectly give place to the m.f. correction term $A_\text{1.can}^{\blacktriangledown}\cdot\bmath{\partial}_{1} f(1,t)$. This may be discussed by taking the moments of the two terms; $A_\text{1.can}^{\blacktriangledown}\cdot\bmath{\partial}_{1} f(1,t)$ and $\bmath{A}^{\blacktriangledown}_{1}$\footnote{The moments of the terms $I_\text{ret}^{\blacktriangledown}$ and $I_\text{Bol}^{\blacktriangledown (\text{inh})}$ were already discussed in \citep{Klimontovich_1982}, while the moments of $A_\text{1.can}^{\blacktriangledown}\cdot\bmath{\partial}_{1} f(1,t)$ were not discussed. Interestingly, \citep{Klimontovich_1982} has mentioned that the moments of the term $A_\text{1.can}^{\blacktriangledown}\cdot\bmath{\partial}_{1} f(1,t)$ vanished. This occurs if one takes the following unclear (at least to me) approximation
\begin{align}
f(1,t)f(2,t)\approx\left(1-\bmath{r}_{12}\cdot\nabla\right)f(1,t)f(\bmath{r}_{1},\bmath{p}_{2},t).
\end{align}
where $r_{12}<<1$.}. The first three moments of the m.f. acceleration $\bmath{A}^{\blacktriangledown}_{1}$ simply read
\begin{subequations}
	\begin{align}
	&\int \bmath{A}^{\blacktriangledown}_{1}\text{d}^{3}\bmath{p}_{1}=0,\\
	&\int \bmath{\varv}_{1} \bmath{A}^{\blacktriangledown}_{1}\text{d}^{3}\bmath{p}_{1}=\bmath{A}^{\blacktriangledown}_{1}n(1,t),\\
    &\int \bmath{\varv}_{1}^{2} \bmath{A}^{\blacktriangledown}_{1}\text{d}^{3}\bmath{p}_{1}=\bmath{A}^{\blacktriangledown}_{1}\cdot n(1,t)\bmath{u}(1,t).
	\end{align}\label{mom_m.f.}
\end{subequations}
On the other hand, the moments of the term $A_\text{1.can}^{\blacktriangledown}\cdot\bmath{\partial}_{1} f(1,t)$ may be written as
\begin{subequations}
	\begin{align}
	&\int A_\text{1.can}^{\blacktriangledown}\cdot\bmath{\partial}_{1} f(1,t)\text{d}^{3}\bmath{p}_{1}=0,\\
	&\int \bmath{\varv}_{1} A_\text{1.can}^{\blacktriangledown}\cdot\bmath{\partial}_{1} f(1,t)\text{d}^{3}\bmath{p}_{1}=0,\\
	&\int \bmath{\varv}_{1}^{2} A_\text{1.can}^{\blacktriangledown}\cdot\bmath{\partial}_{1} f(1,t)\text{d}^{3}\bmath{p}_{1}=\frac{\pi mG\triangle^{2}}{3}n(1,t)^{2}\nabla_{1}\cdot \bmath{u}(1,t).\label{mom_inh3}
	\end{align}\label{mom_inh}
\end{subequations}
The m.f. acceleration, $\bmath{A}^{\blacktriangledown}_{1}$, breaks the conservation of the total- momentum and energy in two-body encounter as part of momentum- and energy- fluxes, however, the term $A_\text{1.can}^{\blacktriangledown}\cdot\bmath{\partial}_{1} f(1,t)$ holds the conservation of momentum and energy if the system of concern should be incompressible. Also, since the momentum dependence of the corrected m.f. acceleration, $A_\text{1.can}^{\blacktriangledown}\cdot\bmath{\partial}_{1} f(1,t)$, is the same as the m.f.  acceleration $A_\text{1}^{\blacktriangledown}\cdot\bmath{\partial}_{1} f(1,t)$, the corrected m.f. acceleration does not contribute to the Boltzmann $H$ function, while the term, $A_\text{1.can}^{\blacktriangledown}\cdot\bmath{\partial}_{1} f(1,t)$, partially originates from the correlation function though.
 
Another relevant discussion for the secular evolution of star cluster would be the moment (gaseous) models \citep[e.g.][]{Louis_1991}. Typically, the moment models assume the dynamical equilibrium due to the slow relaxation process, meaning the (kinetic) pressure must be self-consistently supported by the pressure due to the self-gravitation. However, this description breaks down if one considers the effect of strong encounter inside the Landau sphere. The internal kinetic pressure must be balanced out by the non-locality of Boltzmann collision term \footnote{This is due the effect of pair-wise acceleration of star 1 being stronger than that of the m.f. acceleration inside the Landau sphere.}. Hence, the dynamical equilibrium on small space scales must have the form of 
\begin{align}
\partial_{r}p=\partial_{r}\int \frac{mG}{6r_{12}}f(\bmath{r},\bmath{p}_{1}(-\infty),t)f(\bmath{r},\bmath{p}_{2}(-\infty),t)\text{d}^{3}\bmath{p}_{1}\text{d}^{3}\bmath{p}_{2}\text{d}^{3}\bmath{r}_{12},\label{Eq.dyn_equilib}
\end{align}
where the system is assumed as an isotropic cluster for simplicity. 

It would be obvious that the effect of m.f. acceleration of star 1 is less significant inside the Landau sphere in collision kinetic theory; (i) the moments of the term $A_\text{1.can}^{\blacktriangledown}\cdot\bmath{\partial}_{1} f(1,t)$ does not function as a m.f. acceleration and (ii) the fundamental role of the m.f. acceleration of star 1 in dynamical equilibrium gives place to the effect of the pair-wise acceleration. 

\subsection{Remarks on the domain-patched DF and 'test-particle' method with truncated DF}\label{sec.patch_trunc_DF}
As explained in section \ref{sec:intro}, kinetic equations based on the 'test-particle' method can be also derived from typical BBGKY hierarchy; the \citep{Kandrup_1981}'s g-Landau equation is mathematically the same as the \citep{Gilbert_1971}'s equation without gravitational polarization; \citep{Kaufman_1960}'s 'test-particle' method can reproduce a standard FP equation, as explained in \citep{Montgomery_1964}.

It is obviously seen in section \ref{sec.strong_Klimon} that one can retrieve the g-Landau equation with the effect of strong encounter based on the truncated DF (equation \eqref{Eq.correct_kinetic_eqn}), if employing the domain-patched DF and neglecting the effect of non-ideality of encounter. The derivation due to the truncated DF can correctly execute an expansion of $N$-body DF in terms of the discreteness parameter, $1/N$, while it needs a strong physical assumption (no possibility of finding simultaneous strong encounters at a time). On the other hand, the domain-patched DF can deal with the effect of strong spatial inhomogeneity of a star cluster even inside the Landau sphere (but not for the infinite density) while it necessitates mathematically hard treatment of the scaling of OoM of physical quantities (the BBGKY hierarchy includes mathematical divergences due to pair-wise Newtonian acceleration $\bmath{a}_{12}$). In section \ref{sec.strong_Klimon}, the fundamental scaling employed for the second equation of the BBGKY hierarchy \emph{inside} the Landau sphere implicitly relies on a density expansion (Thermodynamic limit) \footnote{As a matter of fact, the theory of \citep{Klimontovich_1992} essentially depends on the density expansion neglecting the effects of polarization and triple collisions. It, however, is to be noted that the expansion can be correct only for 'weakly-dense(non-ideal)' cluster (in which the occupation number of stars  is less than one.); one may possibly introduce the \citep{Zubarev_1996}'s non-equilibrium operator method as done for a dense one-component plasma \cite{Kobryn_1996} to discuss the dynamics inside Landau sphere in a 'moderately dense' core of star clusters.}, aside from discussion of the actual applicability. In this sense, one may consider the domain-patched DF is also a self-consistent method; the BBGKY hierarchy was truncated at the second-equation level based on weak-coupling limit and density expansion inside- and outside the Landau sphere respectively. 

It may not be obvious to understand the importance in use of the truncated DF compared to that of the domain-patched DF. The domain-patched DF is proper to model strongly inhomogeneous systems if one can correctly deal with spatial inhomogeneity of the Boltzmann collision terms in equation \eqref{Eq.correct_kinetic_eqn_strong_com}; the relevant situation would be the late stage of core-collapse of star clusters without binaries. It, however, is of importance in stellar dynamics to model finite and more realistic star clusters based on the truncated DF. As discussed in \citep{Chandra_1943a} and \citep{Takase_1950}, the Landau radius transits from the possible randomness of DF (fluctuations in m.f. acceleration) to important astrophysical events associated with discreteness inside the Landau sphere e.g. formation of binaries and possibly triple encounters, which were neglected in the present work. The truncated DF has a 'potential' to model the evolution of star clusters after the core collapse correctly including the effects of loss of stars from the DF of stars. Also, some dense compact star clusters may have Landau spheres smaller than the size of stars; one may apply the truncated DF to discuss inelastic direct collisions (e.g. coalescence \citep{Quinlan_1987} and disruption \citep{Kandrup_1985}) as partially discussed in Appendix \ref{Appendix:extra}.

%% file: Section6_CKT.tex
\section{Star-cluster convergent kinetic theory}\label{sec:discussion}
Employment of the truncated DF (section \ref{sec.WC_strong}) or domain-patched DF (section \ref{sec.strong_Klimon}), in essence, induces a mathematical divergence on large-wavenumber scales due to the weak-coupling limit being inconsistent with the spatial locality of the DFs at short distances ($r_{12}\to r_\text{o}$). To hold the weak-coupling limit outside the Landau sphere, in section \ref{subsec:conv_CKT} a 'conventional' CKT \citep{Frieman_1963} is applied to the first two equations of standard BBGKY hierarchy. In section \ref{sec:convergent} the generalization of the conventional CKT is shown; a mathematical program to self-consistently find convergent kinetic equations of star clusters with a dispersion approximation. Lastly, in section \ref{section:full_wave} a convergent kinetic equation to deal with the infinite-density problem of collisional star cluster at the late stage of core-collapse is suggested.
 
\subsection{A star-cluster convergent kinetic equation based on the 'conventional' CKT}\label{subsec:conv_CKT}
In sections \ref{sec:complete_WC} and \ref{sec:strong}, the OoM of physical quantities  was throughly considered at BBGKY-hierarchy-level to solve the second equation for the correlation function $g(1,2,t)$ or binary DF $f(1,2,t)$. In order to systematically find convergent kinetic equations, one needs to overlook the scaling (or mathematical divergence) problem at the hierarchy-level. The second equation of standard BBGKY hierarchy, equation \eqref{Eq.BBGKY_orth}, reads
\begin{align}
&\left(\partial_{t}+\bmath{\varv}_{1}\cdot\nabla_{1}+\bmath{\varv}_{2}\cdot\nabla_{2}+\bmath{a}_{12}\cdot\bmath{\partial}_{12}+\bmath{A}_{1}\cdot\bmath{\partial}_{1}+\bmath{A}_{2}\cdot\bmath{\partial}_{2}\right)f(1,2,t)\nonumber\\
&=\left[\partial_{t}+\bmath{\varv}_{1}\cdot\nabla_{1}+\bmath{\varv}_{2}\cdot\nabla_{2}+\left(1+\frac{1}{N}\right)\left(\bmath{A}_{1}\cdot\bmath{\partial}_{1}+\bmath{A}_{2}\cdot\bmath{\partial}_{2}\right)\right]\nonumber\\
&\quad \times f(1,t)f(2,t),\label{Eq.2ndBBGKY_convergent_col}
\end{align}
where the effects of triple encounters and gravitational polarisation are ignored. Employing the kinetic theory of non-ideal systems (section \ref{subsec:nonideal}) and the method of characteristics, one can solve equation \eqref{Eq.2ndBBGKY_convergent_col} for the correlation function $g(1,2,t)$ in collision kinetic description; 
\begin{align}
&g(1,2,t)\nonumber\\
&=f(1(t-\tau),t-\tau)f(2(t-\tau),t-\tau)-f(1,t)f(2,t)\nonumber\\
&\quad+\int_{t-\tau}^{t}\frac{\text{D}}{\text{D}t'} f\left(1\left(t'\right),t')f(2\left(t'\right),t'\right)\hspace{3pt}\text{d}t',\label{Eq.g(1,2,t)_convergent_col}
\end{align}
where the Mayer-cluster expansion of equation \eqref{Eq.2ndBBGKY_convergent_col} is rendered and the total derivative is defined as follows
\begin{align}
&\frac{\text{D}}{\text{D}t'}=\left[\partial_{t}+\bmath{\varv}_{1}\cdot\nabla_{1}+\bmath{\varv}_{2}\cdot\nabla_{2}+\left(1+\frac{1}{N}\right)\left(\bmath{A}_{1}\cdot\bmath{\partial}_{1}+\bmath{A}_{2}\cdot\bmath{\partial}_{2}\right)\right]_{t=t'}.\label{Eq.tot_deriv}
\end{align}
Equation \eqref{Eq.g(1,2,t)_convergent_col} in wave kinetic description reads
\begin{align}
&g(1,2,t)\nonumber\\
&=-\int^{t}_{t-\tau}\left[(\tilde{\bmath{a}}_{12}\cdot\bmath{\partial}_{1}+\tilde{\bmath{a}}_{21}\cdot\bmath{\partial}_{2})f\left(1\left(t'\right),t'\right)f\left(2\left(t'\right),t'\right)\right]\text{d}t'.\label{Eq.g(1,2,t)_convergent_wave}
\end{align}
For equations \eqref{Eq.g(1,2,t)_convergent_col} and \eqref{Eq.g(1,2,t)_convergent_wave}, assume stars 1 and 2 follow the complete trajectory, equation \eqref{Eq.characteristics_general}. 

The \citep{Frieman_1963}'s method (a conventional plasma CKT) is, in essence, a matched-asymptote method to find an approximated global solution of the second equation of BBGKY hierarchy of concern. Contrasting the boundary layer problem \citep[e.g.][]{Bender_1999} with the present problem, the boundary layer corresponds with the Landau sphere and the outer region is the outside of the sphere. One can easily find out the method of characteristics is naive to the distance $r_{12}$ through the correlation time $\tau$ (section \ref{subsec:scalings}). The matching region is the distances $r_{12}$ much greater than the Landau radius $r_\text{o}$ but at most the order of the BG radius $a_\text{BG}$, on which one can employ the rectilinear approximation, equation\eqref{Eq.characteristics_rectilinear}. Hence the composite solution is
\begin{align}
g(1,2,t)=g_\text{Bol}(1,2,t)-g_\text{L}(1,2,t)+g_\text{g-L}(1,2,t),\label{Eq.composit_CF}
\end{align}
where each correlation function on the R.H.S. satisfies the following condition
\begin{enumerate}
	\item $g_\text{Bol}(1,2,t)$ is the most accurate at distances $r_{12}\lesssim r_\text{o}$ 
	\item[]\hspace{0.5cm} and correct up to $\sim a_\text{BG}$.
	\item $g_\text{L}(1,2,t)$ is the most accurate at distances $r_\text{o} <<r_{12}\lesssim a_\text{BG}$.
	\item $g_\text{g-L}(1,2,t)$ is the most accurate at distances 
	\item[]\hspace{0.5cm}$a_\text{BG}<<r_{12}\lesssim R$ and correct down to $\sim r_\text{o} $.
\end{enumerate}
The binary correlation function  $g_\text{Bol}(1,2,t)$ can be approximated to be local in configuration spaces also the retardation effect is neglected for simplicity. Following equation \eqref{Eq.cor_strong_Balescu_soln} for ideal encounter, one obtains
	\begin{align}
	g_\text{Bol}(1,2,t)=\delta f(\bmath{r}_{1},\bmath{r}_{12},\bmath{p}_{1},\bmath{p}_{2},t),\label{Eq.twoDF_local_str}
	\end{align}
where  the trajectory of test star is determined through the local Newtonian two-body interaction, equation \eqref{Eq.characteristics_v_jump}. 

The correlation function $g_\text{L}(1,2,t)$ can also be approximated to be local in space with the weak-coupling limit;
\begin{align}
g_\text{L}(1,2,t)=&-\int^{t}_{t-\tau}\bmath{a}_{12}\left(t'\right)\text{d}\tau'\cdot\bmath{\partial}_{12}f\left(\bmath{r}_{1},\bmath{p}_{1},t\right)f\left(\bmath{r}_{1},\bmath{p}_{2},t\right),\label{Eq.twoDF_local_L}
\end{align}
where the trajectory of test star follows rectilinear motion, equation \eqref{Eq.characteristics_rectilinear}, and the effect of non-ideality was neglected.

Lastly, since the correlation function $g_\text{g-L}(1,2,t)$ is to be non-local, only the weak-coupling limit can be applied to it; the function $g_\text{g-L}(1,2,t)$ is the same form as the correlation function for the g-Landau collision term in equation \eqref{Eq.Generalized_Landau}. Hence, the whole correlation function, equation \eqref{Eq.composit_CF}, reduces to
\begin{align}
&g(1,2,t)\nonumber\\
&=f(\bmath{r}_{1},\bmath{p}_{1}(t-\tau),t)f(\bmath{r}_{1},\bmath{p}_{2}(t-\tau),t)\nonumber\\
&\quad+\int^{t}_{t-\tau}\bmath{a}_{12}\left(t'\right)\text{d}\tau'\cdot\bmath{\partial}_{12}f\left(\bmath{r}_{1},\bmath{p}_{1},t\right)f\left(\bmath{r}_{1},\bmath{p}_{2},t\right)\nonumber\\
&\qquad-\int^{t}_{t-\tau}(\tilde{\bmath{a}}_{12}\cdot\bmath{\partial}_{1}+\tilde{\bmath{a}}_{21}\cdot\bmath{\partial}_{2})_{t=t'}f\left(1\left(t'\right),t'\right)f\left(2\left(t'\right),t'\right)\text{d}t',\label{strong_collision_similar_plasma}
\end{align}
where it is to be noted that the domain of the distance $r_{12}$ for the functions $g_\text{L}(1,2,t)$ and $g_\text{g-L}(1,2,t)$ spans from $r_{12}=0$ to $r_{12}=R$. After taking a proper integral of equation \eqref{strong_collision_similar_plasma}, one obtains the following mathematically non-divergent collision terms; 
\begin{align}
I_\text{con}=I_\text{g-L}+I_\text{Bol}^\text{(loc)}-I_\text{L}.\label{Eq.plasma_conv_eqn}
\end{align}
The collision terms on the R.H.S of equation \eqref{Eq.plasma_conv_eqn} follows the scalings associated with logarithmic divergence (Section \ref{subsec:Log_col})
\begin{align}
I_{\text{con}}&\sim \ln[k_{\text{max}}]+\ln[b_{\text{max}}]-\ln[k_{\text{max}}/k_{\text{min}}],\label{Eq.plasma_conv_scale}
\end{align}
where $b_{\text{max/min}}$ and $k_{\text{max/min}}$ are the maximum- and minimum- values of the impact parameter and the wave number respectively. The mathematical convergence of equation \eqref{Eq.plasma_conv_eqn} may be achieved if the lower $k_{\text{min}}$ is canceled out (leaving a proper constant value) by the larger $b_{\text{max}}$. As explained in section \ref{subsec:Coulomb_log}, to correctly find the logarithmic divergence in the Boltzmann collision term one must assume the dispersion approximation, equation \eqref{Eq.thermo_approx_b}, or must extend the range of distance to the range between $b_\text{min}=0$ and $b_\text{max}\to\infty$ to hold small-angle deflection approximation, equation \eqref{Eq_close_encounter2}, and small change in velocity of test star, equation \eqref{Eq_close_encounter3}. In this sense, one may also take $k_\text{min}\to 0$ while the treatment is a purely mathematical operation rather than it seriously takes the physical context.

Three fundamental problems in the 'conventional' CKT are that one can not find (i) higher orders of correction terms to equation \eqref{Eq.plasma_conv_eqn} and the forms of overlapped terms \footnote{To raise the order of correction terms, one must consider the OoM of the inner- and outer- solutions (correlation functions) then find the correlation function in the overlapped region. For example, the accuracy of the outer solution $g_\text{g-L}(1,2,t)$ and inner one $g_\text{Bol}(1,2,t)$ can be improved by taking the higher orders of weak-coupling approximation and non-ideality respectively. Then one must find a correct form of the function $g_\text{L}$ which is overlapped terms of the functions $g_\text{g-L}(1,2,t)$ and $g_\text{Bol}(1,2,t)$. This, however, is not a correct way to find the generalised equation with higher orders since one must incorrectly include the effect of strong encounter at distances $r_{12}<r_\text{o}$ into the correlation function $g_\text{g-L}(1,2,t)$.}, (ii) the relation between the collision terms and the m.f. potential and (iii) if the Landau collision term in equation \eqref{Eq.plasma_conv_eqn} can be either of collision- and wave- kinetic descriptions. The three problems can be resolved by employing \citep{Klimontovich_1982}'s theory and the results of section \ref{sec:strong}, as shown in section \ref{sec:convergent}.

\subsection{Generalised convergent kinetic equation}\label{sec:convergent}
To initiate a star-cluster CKT, the \citep{Klimontovich_1982}'s theory of non-ideal systems, the truncated m.f. acceleration of star 1 and truncated g-Landau collision term (sections \ref{sec:complete_WC} and \ref{sec:strong}) are thoroughly employed in the present section. Since the mathematical divergence problem in configuration spaces at BBGKY-hierarchy level was already discussed in section \ref{sec:complete_WC} and \ref{sec:strong}, the problem is neglected in the derivation of convergent kinetic equations. The goal of the derivation is to generalise the plasma-CKT-like collision term, equation \eqref{Eq.plasma_conv_eqn}, and \emph{self-consistently} establish a star-cluster CKT. 

To contrast the present method with the 'conventional' CKT (section \ref{subsec:conv_CKT}), consider the first two equations of standard BBGKY hierarchy; the correlation functions, equations \eqref{Eq.g(1,2,t)_convergent_col} and \eqref{Eq.g(1,2,t)_convergent_wave}, are the exact solutions of the second equation of the BBGKY hierarchy with the complete trajectory, equation \eqref{Eq.characteristics_general}, as done in section \ref{subsec:conv_CKT}. Now recall the main idea of domain-patched DF; to separate the complete trajectory into two trajectories. If separating the complete trajectory of the time-integral term in equation \eqref{Eq.g(1,2,t)_convergent_wave} at the Landau radius, one obtains 
\begin{align}
g(1,2,t)=&f(1(t-\tau),t-\tau)f(2(t-\tau),t-\tau)-f(1,t)f(2,t)\nonumber\\
&+\int_{t-\tau}^{t}\frac{\text{D}}{\text{D}t'}\left[f\left(1\left(t'\right),t')f(2\left(t'\right),t'\right)\right]_{r_{12}<\triangle}\text{d}t'\nonumber\\
&\quad+\int_{t-\tau}^{t}\frac{\text{D}}{\text{D}t'}\left[f\left(1\left(t'\right),t')f(2\left(t'\right),t'\right)\right]_{r_{12}>\triangle}\text{d}t',\label{Eq.f(1,2,t)_convergent_2}
\end{align}
where the initial times for correlations both outside and inside the Landau sphere may take the same expression $\tau$ for brevity\footnote{It is to be noted that the domain of time integral should be taken as
\begin{align}
\int^{t}_{t-\tau}\cdot\text{d}t'=\int^{t}_{t-\tau_{1}}\cdot\text{d}t'+\int^{t-\tau_{1}}_{t-\tau}\cdot\text{d}t',
\end{align}
where the initial correlation time $\tau_{1}$ inside the Landau sphere has the correlation-time scale during which test star undergoes close encounter. Since the time scale of correlation inside the Landau sphere is less significant than that outside the sphere by a factor of $N$, one obtains
\begin{align}
\int^{t}_{t-\tau}\cdot\text{d}t'\approx\int^{t}_{t-\tau_{1}}\cdot\text{d}t'+\int^{t}_{t-\tau}\cdot\text{d}t'. \label{Eq._approx_tau_int}
\end{align}
The approximation of equality in equation \eqref{Eq._approx_tau_int} works at any order of the smallness parameter since the both initial correlation times $\tau_{1}$ and $\tau$ can never be compatible and the values are taken to an infinity in senses discussed for the weakly-coupled- and truncated- DFs in sections \ref{sec:complete_WC} and \ref{sec:strong} respectively.}. Employing an integral-by-parts reduces equation \eqref{Eq.f(1,2,t)_convergent_2} into
\begin{align}
&g(1,2,t)\nonumber\\
&=f(1(t-\tau),t-\tau)f(2(t-\tau),t-\tau)-\left[f(1,t)f(2,t)\right]_{r_{12}<\triangle}\nonumber\\
&\quad+\int_{t-\tau}^{t}\frac{\text{D}}{\text{D}t'}\left[f\left(1\left(t'\right),t')f(2\left(t'\right),t'\right)\right]_{r_{12}<\triangle}\text{d}t'\nonumber\\
&\qquad-\left[f(1(t-\tau),t-\tau)f(2(t-\tau),t-\tau)\right]_{r_{12}>\triangle}\nonumber\\
&\qquad\quad-\int_{t-\tau}^{t}\left[\left(\tilde{\bmath{a}}_{12}\cdot\bmath{\partial}_{1}+\tilde{\bmath{a}}_{21}\cdot\bmath{\partial}_{2}\right)\right]^{r_{12}>\triangle}_{t=t'}\nonumber\\
&\qquad\qquad\qquad\times f\left(1\left(t'\right),t')f(2\left(t'\right),t'\right)\hspace{3pt}\text{d}t',\label{Eq.f(1,2,t)_convergent_3}\\
&=g_\text{coll}^{\blacktriangledown}(1,2,t)+g_\text{wave}^{\blacktriangle}(1,2,t),\label{Eq.f(1,2,t)_convergent_4}
\end{align}
where both of the functions $g_\text{coll}^{\blacktriangledown}(1,2,t)$ and $g_\text{wave}^{\blacktriangle}(1,2,t)$ have the complete trajectory, equation \eqref{Eq.characteristics_general}. Equation \eqref{Eq.f(1,2,t)_convergent_4} is in essence the domain-patched correlation function, meaning one needs to employ the truncated acceleration in kinetic equation.

\subsubsection{Convergent star-cluster kinetic equation with ideal-close encounter}\label{sec:star_cluster_CKT_ideal}
Since the non-ideality of strong encounter is not of significance in discussion of CKTs, assume a weakly-inhomogeneous star cluster (section \ref{sec.WC_strong}) in which the effect of non-ideality of strong encounter is negligible inside the Landau sphere. The zeroth order of correlation function, equation \eqref{Eq.f(1,2,t)_convergent_3}, reduces to
\begin{align}
&g(1,2,t)\nonumber\\
&=\left[f(\bmath{r}_{1},\bmath{p}_{1}(t-\tau),t)f(\bmath{r}_{1},\bmath{p}_{2}(t-\tau),t)\right]_{r_{12}<\triangle}\nonumber\\
&\quad-\int_{t-\tau}^{t}\left[\left(\tilde{\bmath{a}}_{12}\cdot\bmath{\partial}_{1}+\tilde{\bmath{a}}_{21}\cdot\bmath{\partial}_{2}\right)\right]^{r_{12}>\triangle}_{t=t'}\nonumber\\
&\qquad\qquad\times f\left(1\left(t'\right),t')f(2\left(t'\right),t'\right)\hspace{3pt}\text{d}t'.\label{Eq.g(1,2,t)_conv_zeroth}
\end{align}
Hence, one obtains the corresponding convergent kinetic equation
\begin{align}
&\left(\frac{\partial}{\partial t}+\bmath{\varv}_{1}\cdot\frac{\partial}{\partial \bmath{r}_{1}}+\bmath{A}_{1}^{(2,2)}\cdot\frac{\partial}{\partial \bmath{\varv}_{1}}\right)f(\bmath{r}_{1},\bmath{p}_{1},t)=I_\text{st-g-L}^{\blacktriangle}+I_\text{Bol}^{\blacktriangledown},\label{Eq.convergent_eqn_COL}
\end{align}
where the collision term $I_\text{st-g-L}^{(2,2)}$ is the g-Landau collision term but the motion of test star follows the completer trajectory, equation \eqref{Eq.characteristics_general}, hence the term does not mathematically diverge. It would be obvious that equation \eqref{Eq.correct_kinetic_eqn} based on the truncated DF is a weak-coupling limit outside the Landau sphere of the convergent equation \eqref{Eq.convergent_eqn_COL}. While equation \eqref{Eq.convergent_eqn_COL} is a convergent kinetic equation, all the collision terms in the equation include the effect of strong pair-wise interaction, $\bmath{a}_{12}$, in the trajectory of test star. Such equation is a complete description though, it faces a mathematical difficulty in finding the explicit form.

A question that one may have is 'what is the relation between equations \eqref{Eq.convergent_eqn_COL} and \eqref{Eq.plasma_conv_eqn}'. To answer the question, one may leave the fourth term on the R.H.S of equation \eqref{Eq.f(1,2,t)_convergent_3} at the zeroth order of non-ideality
\begin{align}
&g(1,2,t)\nonumber\\
&=\left[f(\bmath{r}_{1},\bmath{p}_{1}(t-\tau),t)f(\bmath{r}_{1},\bmath{p}_{2}(t-\tau),t)\right]\nonumber\\
&\quad-\left[f(\bmath{r}_{1},\bmath{p}_{1}(t-\tau),t)f(\bmath{r}_{1},\bmath{p}_{2}(t-\tau),t)\right]_{r_{12}>\triangle}\nonumber\\
&\qquad-\int_{t-\tau}^{t}\left[\left(\tilde{\bmath{a}}_{12}\cdot\bmath{\partial}_{1}+\tilde{\bmath{a}}_{21}\cdot\bmath{\partial}_{2}\right)\right]^{r_{12}>\triangle}_{t=t'}\nonumber\\
&\qquad\qquad\quad\times f\left(1\left(t'\right),t')f(2\left(t'\right),t'\right)\hspace{3pt}\text{d}t'.\label{Eq.g(1,2,t)_conv_zeroth_con}
\end{align}
The second- and third- terms on the R.H.S of equation \eqref{Eq.g(1,2,t)_conv_zeroth_con} are defined only at distances $r_{12}>\triangle=r_\text{o}$, meaning one can ideally expand the two terms in terms of the following operator for momenta $\bmath{p}_{1}$ and $\bmath{p}_{2}$
\begin{align}
\overrightarrow{\mathcal{L}_{p_{12}}}(t;t-\tau)\equiv-\int^{t}_{t-\tau}\left[\bmath{a}_{12}\left(t'\right)\right]_{r_{12}>\triangle}\text{d}t'\cdot\bmath{\partial}_{12}.\label{Eq.operator}
\end{align}
Hence, one may Taylor-expand equation \eqref{Eq.g(1,2,t)_conv_zeroth_con}
\begin{align}
&g(1,2,t)\nonumber\\
&=\left[f(\bmath{r}_{1},\bmath{p}_{1}(t-\tau),t)f(\bmath{r}_{1},\bmath{p}_{2}(t-\tau),t)\right]\nonumber\\
&\quad-e^{\overrightarrow{\mathcal{L}_{p_{12}}}(t;t-\tau)}f(\bmath{r}_{1},\bmath{p}_{1},t)f(\bmath{r}_{1},\bmath{p}_{2},t)\nonumber\\
&\quad-\int^{\tau}_{0}(\tilde{\bmath{a}}_{12}\cdot\bmath{\partial}_{1}+\tilde{\bmath{a}}_{21}\cdot\bmath{\partial}_{2})_{t=t-\tau'}e^{\overrightarrow{\mathcal{L}_{p_{12}}}(t;t-\tau')}\nonumber\\
&\qquad \times f(\bmath{r}_{1}(t-\tau'),\bmath{p}_{1},t-\tau')f(\bmath{r}_{2}(t-\tau'),\bmath{p}_{2},t-\tau')\text{d}\tau',\label{Eq.g(1,2,t)_conv_zeroth_con_W.C.}
\end{align}
where, on the R.H.S, the first term is defined on $0<r_{12}<R$ while the second- and third- terms are defined on $r_\text{o}<r_{12}<R$ . The basic idea to deal with equation \eqref{Eq.g(1,2,t)_conv_zeroth_con_W.C.} is still the same as the 'conventional' CKT; one would like to consider the first two terms can be canceled out only at long distances. Hence, one needs to keep only the lower order of weak-coupling approximation for equation \eqref{Eq.g(1,2,t)_conv_zeroth_con_W.C.}. Recalling the second term in equation \eqref{Eq.g(1,2,t)_conv_zeroth_con_W.C.} is more significant than the third term at most by a factor of $N$ at $r_{12}\approx R$, expand the second term up to the first order and the third term to the zeroth order on the R.H.S. of equation \eqref{Eq.g(1,2,t)_conv_zeroth_con_W.C.};
\begin{align}
&g(1,2,t)\nonumber\\
&=\left[f(\bmath{r}_{1},\bmath{p}_{1}(t-\tau),t)f(\bmath{r}_{1},\bmath{p}_{2}(t-\tau),t)\right]\nonumber\\
&\quad-\left(1-\overrightarrow{\mathcal{L}_{p_{12}}}(t;t-\tau)\right) f(\bmath{r}_{1},\bmath{p}_{1},t)f(\bmath{r}_{1},\bmath{p}_{2},t)\nonumber\\
&\qquad\quad-\int^{\tau}_{0}(\tilde{\bmath{a}}_{12}\cdot\bmath{\partial}_{1}+\tilde{\bmath{a}}_{21}\cdot\bmath{\partial}_{2})_{t=t-\tau'}\nonumber\\
&\qquad\qquad\qquad \times f(\bmath{r}_{1}(t-\tau'),\bmath{p}_{1},t)f(\bmath{r}_{2}(t-\tau'),\bmath{p}_{2},t)\text{d}\tau',\label{Eq.g(1,2,t)_conv_zeroth_con_W.C._z}
\end{align}
where the Non-Markovian effect in the third term on the R.H.S. may vanish due to its weakness. Hence, one obtains a star-cluster kinetic equation.
\begin{align}
&\left(\frac{\partial}{\partial t}+\bmath{\varv}_{1}\cdot\nabla_{1}+\bmath{A}_{1}^{(2,2)}\cdot\bmath{\partial}_{1}\right)f(\bmath{r}_{1},\bmath{p}_{1},t)=I_\text{g-L}^{\blacktriangle}+I_\text{Bol}^\text{(loc)}-I_\text{L}^{\blacktriangle}.\label{Eq.star_cluster_conv_eqn}
\end{align}
Equation \eqref{Eq.star_cluster_conv_eqn} is superior to equation \eqref{Eq.plasma_conv_eqn} in sense that one can (i) obtain a correct relation between the m.f. acceleration and the collision terms on scales smaller than the Landau radius (ii) find the general form of the correlation function, equation \eqref{Eq.convergent_eqn_COL}, (and the corresponding kinetic equation) in a mathematically well-defined manner. A unique property of equation \eqref{Eq.star_cluster_conv_eqn} is that the Landau collision term $I_\text{L}^{\blacktriangle}$ must be described by the wave kinetic description. This is since the Landau collision term is defined at distances $r_\text{o}<r_{12}< R$. As a matter of fact, if one should replace the Landau collision term in equation \eqref{Eq.star_cluster_conv_eqn} by the corresponding collision kinetic description, the equation is no longer a convergent kinetic equation since the logarithmic divergences of the Landau- and g-Landau- collision terms are not canceled out on large wavenumber scales. 

The accuracy of the convergent kinetic equation \eqref{Eq.star_cluster_conv_eqn} may be discussed simply by separating the distances $r_\text{o}<r_{12}< R$ at the BG radius $a_\text{B.G.}$. Assume the system of concern follows the scaling for the weakly-inhomogeneous cluster (section \ref{subsec:scalings}). Due to the weakness of the m.f. acceleration of stars at distances $r_\text{o}<r_{12}< a_\text{BG}$, one may expand the g-Landau collision term $I_\text{g-L}$ in terms of the following operator
\begin{align}
\overrightarrow{\mathcal{L}_{p_{A}}}(t;t-\tau)\equiv&-\int^{t}_{t-\tau}\left[\bmath{A}_{1}\left(t'\right)\right]_{r_{12}>\triangle}\text{d}t'\cdot\bmath{\partial}_{1}\nonumber\\
&\quad-\int^{t}_{t-\tau}\left[\bmath{A}_{2}\left(t'\right)\right]_{r_{12}>\triangle}\text{d}t'\cdot\bmath{\partial}_{2}.\label{Eq.operator_A}
\end{align}
Hence, the g-Landau collision term can approximately reduces to the Landau collision term
\begin{align}
I_\text{g-L}=I_\text{L}+\mathcal{O}\left(1/N^{1/2}\right),
\end{align}
On the same distances ($r_\text{o}<r_{12}< a_\text{BG}$), the Boltzmann collision term is at most order of $\mathcal{O}(\ln[N])$, meaning the total of collision terms in equation \eqref{Eq.star_cluster_conv_eqn} reduces to the Boltzmann collision term
\begin{align}
I=I_\text{Bol}^\text{(loc)}+\mathcal{O}\left(1/N^{1/2}\right), \qquad (r_\text{o}<r_{12}< a_\text{BG})
\end{align}

On the other hand, at distances $a_\text{BG}<r_{12}< R$, one can approximate the Boltzmann collision term in series of the operator, equation \eqref{Eq.operator}, employing the weak-coupling approximation (section \ref{subsec:Log_col});
\begin{align}
I_\text{Bol}=I_\text{L}+\mathcal{O}\left(1/N^{1/2}\right).
\end{align}
On the same domain $(a_\text{BG}<r_{12}< R)$, the g-Landau collision term is order of $1$. Hence, the collision terms in equation \eqref{Eq.star_cluster_conv_eqn} reduce to
\begin{align}
I=I_\text{g-L}+\mathcal{O}\left(1/N^{1/2}\right).  \qquad (a_\text{BG}<r_{12}< R)
\end{align}

At kinetic-equation level of order $\sim 1/N^{1/2}$, the transition between collision- and wave- kinetic descriptions may occur at the BG radius $a_\text{BG}$. This implies that improving equation \eqref{Eq.g(1,2,t)_conv_zeroth} inside the Landau sphere makes the resulting collision term more accurate on the interval of $r_\text{o}<r_{12}< a_\text{BG}$, while an improvement of the g-Landau collision term is also possible by taking the higher orders of weak-coupling approximation. It, however, is to be noted that the convergent kinetic equation \eqref{Eq.star_cluster_conv_eqn} may allow a 'coexistence' of the m.f. acceleration and the Boltzmann collision term at $r_\text{o}<r_{12}<a_\text{BG}$ due to the weak-coupling approximation; one needs to hold all orders of weak-coupling approximation to perfectly separate the collision- and wave kinetic description at the Landau radius $r_\text{o}$. Since equation \eqref{Eq.star_cluster_conv_eqn} is the result of the Boltzmann-collision description with ideal-encounter approximation, one may move on to incorporating the effect of non-ideality in the equation as shown in section \ref{sec:star_cluster_CKT_nonideal}.

\subsubsection{Convergent star-cluster kinetic equation with weakly non-ideal-close encounter}\label{sec:star_cluster_CKT_nonideal}
One can extend the method employed in section \ref{sec:star_cluster_CKT_ideal} to a convergent kinetic equation including the effect of non-ideal-close encounter. If keeping the effect of non-idealty in the correlation function, equation \eqref{Eq.f(1,2,t)_convergent_3}, up to the first order of non-ideality, one obtains
\begin{subequations}
	\begin{align}
	&g(1,2,t)\nonumber\\
	&=\delta f(\bmath{r}_{1},r_{12}<\triangle,\bmath{p}_{1},\bmath{p}_{2},t)\nonumber\\
	&\quad+\int_{t-\tau}^{t}\frac{\text{D}}{\text{D}t'}\left[f\left(1\left(t'\right),t')f(2\left(t'\right),t'\right)\right]_{r_{12}<\triangle}\text{d}t'\nonumber\\	
	&\qquad-\int_{t-\tau}^{t}\left[\left(\tilde{\bmath{a}}_{12}\cdot\bmath{\partial}_{1}+\tilde{\bmath{a}}_{21}\cdot\bmath{\partial}_{2}\right)\right]^{r_{12}>\triangle}_{t=t'}\nonumber\\
	&\qquad\qquad\quad \times f\left(1\left(t'\right),t')f(2\left(t'\right),t'\right)\hspace{3pt}\text{d}t',\label{Eq.g(1,2,t)_non_ideal}\\
	&\approx\left(1-\frac{r_{12}}{2}\cdot\nabla_{1}\right)\delta f(\bmath{r}_{1},r_{12}<\triangle,\bmath{p}_{1},\bmath{p}_{2},t)\nonumber\\
	&\quad-\int_{0}^{\tau}\left[\tau'\bmath{V}\left(t-\tau'\right)\cdot\nabla_{1}\right]_{r_{12}<\triangle}\nonumber\\
	&\qquad\qquad\times \frac{\text{d}}{\text{d}\tau'}f\left(\bmath{r}_{1},\bmath{p}_{1}\left(t-\tau'\right),t\right)f\left(\bmath{r}_{1},\bmath{p}_{2}\left(t-\tau'\right),t\right)\text{d}\tau'\nonumber\\
	&\qquad-\int_{t-\tau}^{t}\left[\left(\tilde{\bmath{a}}_{12}\cdot\bmath{\partial}_{1}+\tilde{\bmath{a}}_{21}\cdot\bmath{\partial}_{2}\right) f\left(1\left(t\right),t)f(2\left(t\right),t\right)\right]^{r_{12}>\triangle}_{t=t'}\hspace{3pt}\text{d}t',\label{Eq.g(1,2,t)_non_ideal_1st}
\end{align}
\end{subequations}

One may extend the range of the distance domain of the first two terms on the R.H.S of equation \eqref{Eq.g(1,2,t)_non_ideal_1st} into distances $0<r_{12}<R$ as follows
\begin{align}	
&g(1,2,t)\nonumber\\
&=\left(1-\frac{r_{12}}{2}\cdot\nabla_{1}\right)\delta f(\bmath{r}_{1},\bmath{r}_{12},\bmath{p}_{1},\bmath{p}_{2},t)\nonumber\\
&\quad-\int_{0}^{\tau}\left[\tau'\bmath{V}\left(t-\tau'\right)\cdot\nabla_{1}\right]\nonumber\\
&\qquad\qquad\times \frac{\text{d}}{\text{d}\tau'}f\left(\bmath{r}_{1},\bmath{p}_{1}\left(t-\tau'\right),t\right)f\left(\bmath{r}_{1},\bmath{p}_{2}\left(t-\tau'\right),t\right)\text{d}\tau'\nonumber\\
&\quad-\left(1-\frac{r_{12}}{2}\cdot\nabla_{1}\right)\delta f(\bmath{r}_{1},r_{12}>\triangle,\bmath{p}_{1},\bmath{p}_{2},t)\nonumber\\
&\quad+\int_{0}^{\tau}\left[\tau'\bmath{V}\left(t-\tau'\right)\cdot\nabla_{1}\right]_{r_{12}>\triangle}\nonumber\\
&\qquad\qquad\times \frac{\text{d}}{\text{d}\tau'}f\left(\bmath{r}_{1},\bmath{p}_{1}\left(t-\tau'\right),t\right)f\left(\bmath{r}_{1},\bmath{p}_{2}\left(t-\tau'\right),t\right)\text{d}\tau'\nonumber\\
&\quad-\int_{t-\tau}^{t}\left[\left(\tilde{\bmath{a}}_{12}\cdot\bmath{\partial}_{1}+\tilde{\bmath{a}}_{21}\cdot\bmath{\partial}_{2}\right)\right]^{r_{12}>\triangle}_{t=t'}\nonumber\\
&\qquad\qquad\quad\times f\left(1\left(t'\right),t')f(2\left(t'\right),t'\right)\hspace{3pt}\text{d}t',\label{Eq.g(1,2,t)_non_ideal_1st_conv}
\end{align}
Since third through fifth terms on R.H.S of equation \eqref{Eq.g(1,2,t)_non_ideal_1st_conv} are defined on $r_\text{o}<r_{12}<R$, one can Taylor-expand the equation in terms of the momenta operator, equation \eqref{Eq.operator}, recalling one may expand the third- and fourth terms up to the first order and the fifth term up to the zeroth order;
\begin{subequations}
\begin{align}
&g(1,2,t)\nonumber\\
&=\left(1-\frac{r_{12}}{2}\cdot\nabla_{1}\right)\delta f(\bmath{r}_{1},\bmath{r}_{12},\bmath{p}_{1},\bmath{p}_{2},t)\nonumber\\
&\quad-\int_{0}^{\tau}\left[\tau'\bmath{V}\cdot\nabla_{1}\right] \frac{\text{d}}{\text{d}\tau'}\nonumber\\
&\qquad\qquad\times f\left(\bmath{r}_{1},\bmath{p}_{1}\left(t-\tau'\right),t\right)f\left(\bmath{r}_{1},\bmath{p}_{2}\left(t-\tau'\right),t\right)\text{d}\tau'\nonumber\\
&\quad+\left(1-\frac{r_{12}}{2}\cdot\nabla_{1}\right)\left[\overrightarrow{\mathcal{L}_{p_{12}}}(t;t-\tau)f(1,t)f(\bmath{r}_{1},\bmath{p}_{2},t)\right]_{r_{12}>\triangle}\nonumber\\
&\quad+\int_{0}^{\tau}\left[\tau'\bmath{V}\cdot\nabla_{1}\right]_{r_{12}>\triangle}\frac{\text{d}}{\text{d}\tau'}\left(1-\overrightarrow{\mathcal{L}_{p_{12}}}(t;t-\tau)\right)\nonumber\\
&\qquad\qquad \times f\left(\bmath{r}_{1},\bmath{p}_{1},t\right)f\left(\bmath{r}_{1},\bmath{p}_{2},t\right)\text{d}\tau'\nonumber\\
&\quad-\int^{\tau}_{0}(\tilde{\bmath{a}}_{12}\cdot\bmath{\partial}_{1}+\tilde{\bmath{a}}_{21}\cdot\bmath{\partial}_{2})_{t=t-\tau'}\nonumber\\
&\qquad\qquad\times f(\bmath{r}_{1}(t-\tau'),\bmath{p}_{1},t)f(\bmath{r}_{2}(t-\tau'),\bmath{p}_{2},t)\text{d}\tau'.\label{Eq.g(1,2,t)_non_ideal_1st_conv_W.C.}
\end{align}
\end{subequations}
Hence one obtains the following convergent kinetic equation with weak non-ideality
\begin{align}
&\left(\frac{\partial}{\partial t}+\bmath{\varv}_{1}\cdot\nabla_{1}+\left[\bmath{A}_{1}^{(2,2)}+\bmath{A}_\text{1.can}^{\blacktriangledown}\right]\cdot\bmath{\partial}_{1}\right)f(\bmath{r}_{1},\bmath{p}_{1},t)\nonumber\\
&\qquad=I_\text{g-L}^{\blacktriangle}+\left(I_\text{Bol}^\text{(loc)}+I_\text{Bol}^{\text{(inh)}}+I_\text{ret}\right)-\left(I_\text{L}^{\blacktriangle}+I_\text{L(ret)}^{\blacktriangle}+I_\text{L(inh)}^{\blacktriangle}\right),\label{Eq.star_cluster_conv_eqn_non_ideal_2}
\end{align}
where the effects of the retardation and spatial inhomogeneity on the Landau collision terms are written as
\begin{subequations}
\begin{align}
&I_\text{L(inh)}^{\blacktriangle}=\int\bmath{a}_{12}\cdot\bmath{\partial}_{1}\frac{\bmath{r}_{12}}{2}\cdot\bmath{\nabla}_{1}\int_{0}^{\tau}\bmath{a}_{12}(t-\tau')\text{d}\tau'\cdot\bmath{\partial}_{12}\nonumber\\
&\qquad\qquad \times f\left(\bmath{r}_{1},\bmath{p}_{1},t\right)f\left(\bmath{r}_{1},\bmath{p}_{2},t\right)\text{d}_{2},\label{Eq.I_L_inh}\\
&I_\text{L(ret)}^{\blacktriangle}=\int\bmath{a}_{12}\cdot\bmath{\partial}_{1}\bmath{V}\cdot\nabla_{1}\int_{0}^{\tau}\tau'\bmath{a}_{12}(t-\tau')\text{d}\tau'\cdot\bmath{\partial}_{12}\nonumber\\
&\qquad\qquad \times f\left(\bmath{r}_{1},\bmath{p}_{1},t\right)f\left(\bmath{r}_{1},\bmath{p}_{2},t\right)\text{d}_{2}\label{Eq.I_L_ret}.
\end{align}
\end{subequations}

The accuracy of equation \eqref{Eq.star_cluster_conv_eqn_non_ideal_2} for strongly inhomogeneous systems (section \ref{sec.inh}) may also be discussed (of course somehow). As the distance $r_{12}$ in the collision terms becomes greater than the B.G radius, the Boltzmann collision terms with non-ideality $\left(I_\text{Bol}^\text{(loc)}+I_\text{Bol}^{\text{(inh)}}+I_\text{ret}\right)$ approximately turns into the Landau collision terms with non-ideality $\left(I_\text{L}^{\blacktriangle}+I_\text{L(ret)}^{\blacktriangle}+I_\text{L(inh)}^{\blacktriangle}\right)$ due to the weak-coupling limit. Then, the dominant collision term on the domain of $a_\text{B.G.}<r_{12}<R$ is still the g-Landau collision term;
\begin{align}
I\sim I_\text{g-L}^{\blacktriangle}+\mathcal{O}\left(\left[\overrightarrow{\mathcal{L}_{p_{12}}}(t;t-\tau)\right]^{2}I_\text{g-L}^{\blacktriangle}\right) \qquad (\text{for}\quad a_\text{B.G.}<r_{12}<R),\label{Eq.I_approx}
\end{align}
where the scalings of OoM for the operator, equation \eqref{Eq.operator}, and the non-ideality of Boltzmann collision term ($\sim\tau\bmath{V}\cdot\bmath{\nabla}_{1}$) are as follows
\begin{align}
&\overrightarrow{\mathcal{L}_{p_{12}}}(t;t-\tau)\sim\frac{1}{r_{12}N}, \label{Eq.scale_L_12}\\
&\tau\bmath{V}\cdot\bmath{\nabla}_{1}\sim\frac{r_{12}}{r_{1}}.\label{Eq.scale_ret}
\end{align}
The factor, equation \eqref{Eq.scale_ret}, may be significant in the limits of $r_{1}\to R$ and $r_{12}\to R$, however, the collision terms in equation \eqref{Eq.I_approx} themselves could be negligible at the outer halo ($r_{1}>>a_\text{BG}$) of the strongly inhomogeneous system due to the very low density. On the other hand, the Landau collision terms $\left(I_\text{L}^{\blacktriangle}+I_\text{L(ret)}^{\blacktriangle}+I_\text{L(inh)}^{\blacktriangle}\right)$ can be canceled out with the g-Landau collision term on small scales $r_\text{o}<r_{12}<a_\text{BG}$. This is since one may take the limit of $r_{12}\to r_\text{o}$ for the g-Landau collision term; meaning the first order of non-ideality may be taken
\begin{align}
I_\text{g-L}^{\blacktriangle}&\approx\int\text{d}_{2}\bmath{a}_{12}\cdot\bmath{\partial}_{1}\left(1-\frac{\bmath{r}_{12}}{2}\right)\nonumber\\
&\qquad\times\int_{0}^{\tau}\bmath{a}_{12}(1-\tau'\bmath{V}\cdot\nabla_{1})f(1,t)f(\bmath{r}_{1},\bmath{p}_{2},t)\text{d}\tau'.\nonumber\\
&=I_\text{L}^{\blacktriangle}+I_\text{L(ret)}^{\blacktriangle}+I_\text{L(inh)}^{\blacktriangle},
\end{align}
where the effect of the m.f. acceleration is entirely ignored. Comparing the scaling in equation \eqref{Eq.scale_ret} to 
that of the operator, equation \eqref{Eq.operator_A},
\begin{align}
\overrightarrow{\mathcal{L}_{p_{A}}}(t;t-\tau)\sim\frac{r_{12}}{r_{1}^{2}N}, \label{Eq.scale_L_A}
\end{align}
the effect of m.f. acceleration is always less important compared to the effect of non-ideality of strong encounters  at the distances $r_\text{o}<r_{12}<a_\text{B.G.}$ \emph{and} $r_\text{o}<<r_{1}<R$. (Recall that in the present work, the focus of inhomogeneity of core-halo structure of a singular isothermal sphere is on the scales larger than the Landau radius). Hence, on scales of distances $r_\text{o}<r_{12}<a_\text{B.G.}$, the convergent collision terms approximate to 
\begin{align}
	I\sim I_\text{Bol}^\text{(loc)}+I_\text{Bol}^{\text{(inh)}}+I_\text{ret}+\mathcal{O}\left(\left[\overrightarrow{\mathcal{L}_{p_A}}(t;t-\tau)\right] I_\text{L}\right), \label{Eq.I_approx_non}
\end{align}
where it is assumed that the order of the factor $\overrightarrow{\mathcal{L}_{p_{A}}}(t;t-\tau)$ is compatible or more than that of $(\tau\bmath{V}\cdot\nabla_{12})^{2}$.

\subsubsection{Comments on Convergent star-cluster kinetic equation}\label{section:star_cluster_CKT}
One would be able to systematically find the higher orders of star-cluster convergent kinetic equations based on the method employed in sections \ref{sec:star_cluster_CKT_ideal} and \ref{sec:star_cluster_CKT_nonideal}\footnote{It is to be noted that as explained in section \ref{sec:star_cluster_CKT_nonideal} the effect of m.f. acceleration of may be of importance at the second order of the non-ideality effect of close-encounter collision terms, implying the form of higher orders will be more complicated than the zeroth- and first- orders.}, while one can know the relation of the equations with the correlation function, equation \eqref{Eq.f(1,2,t)_convergent_3}, that is a formal solution to the second equation of standard BBGKY hierarchy. In this sense, the star-cluster CKT is a self-consistent method to find the convergent kinetic equations. 

The kernel of the star-cluster CKT is use of convergent relations, equations \eqref{Eq.g(1,2,t)_collision} and \eqref{Eq.g(1,2,t)_wave}. Recalling the theory of \citep{Klimontovich_1982}, the second term, $\left[f(1(t-\tau),t-\tau)f(2(t-\tau),t-\tau)\right]_{r_{12}>\triangle}$, on the R.H.S of equation \eqref{Eq.f(1,2,t)_convergent_3} has a duality of collision- and wave- kinetic descriptions. This is since the term itself is a collision kinetic description while its zero-th order of the ideal-encounter- and weak-coupling- approximations takes the wave-kinetic description of the Landau collision term at kinetic-equation level. The duality term has a more general form than the 'idealized interaction term \citep{Aono_1968}', which does not include the effects of close encounters and m.f. acceleration (spatial non-locality) in the trajectory of test star.

The star-cluster CKT is more general than 'conventional' plasma CKTs in sense that one can \emph{self-consistently} obtain a more accurate kinetic equation based on the methods in sections \ref{sec:star_cluster_CKT_ideal} and \ref{sec:star_cluster_CKT_nonideal}. This is since the star cluster CKT is merely a series expansion of the correlation function, equation \eqref{Eq.f(1,2,t)_convergent_4}, based on the weak-coupling- and non-ideal- approximations, rather than a direct application of the method of matched-asymptote. A problem in the star-cluster CKT is that the effective collision terms, equations \eqref{Eq.I_approx} and \eqref{Eq.I_approx_non}, at distances $r_\text{o}<r_{12}<R$ is written in collision kinetic description against the fact that the truncated m.f. acceleration is defined on the same distances. However, considering the nature of the Taylor-expansion, the collision kinetic description is very accurate only at limit of $r_{12}\to r_\text{o}$. Hence, one needs to hold the truncated acceleration of star for the star-cluster CKT. This implies that one may need a fully wave convergent kinetic equation (See section \ref{section:full_wave}).

Lastly, we may be able to find the convergent kinetic equation with completely weak-coupling limit;
\begin{align}
&\left(\frac{\partial}{\partial t}+\bmath{\varv}_{1}\cdot\nabla_{1}+A_{1}^{(2,2)}\cdot\bmath{\partial}_{1}\right)f(\bmath{r}_{1},\bmath{p}_{1},t)=I_\text{g-L}^{\blacktriangle}+I_\text{Bol}^{\blacktriangle}-I_\text{L}^{\blacktriangle},\label{Eq.convergent_eqn_Boltzmann_WC}
\end{align}
where the Boltzmann collision terms is also defined on scales $r_{12}>r_\text{o}$. One can find out that even equation \eqref{Eq.convergent_eqn_Boltzmann_WC} is not a convergent kinetic equation if the Landau collision term is rewritten in collision kinetic description.

\subsection{Needs of fully wave kinetic equations and an open question}\label{section:full_wave}
As shown in sections \ref{sec.WC_strong} and \ref{sec.strong_Klimon}, due to the employment of collision kinetic description inside the Landau sphere, typical m.f. acceleration can not be applied in modeling of the evolution of star clusters. However, if binaries should not be formed in the core of systems (without the primordial) for mathematical simplicity, the system undergoes the gravithermal instability. Then, one may admit of a mathematical infinite density at theoretical level. This encourages one to hold a typical self-consistent m.f. acceleration since truncated m.f. acceleration does not cover small scale dynamics. Hence, one faces a conjecture of how one can include a fine structure smaller than Landau-distance scale, meaning the equation may be written in a fully wave kinetic description. In the description, the effect of occasional two-body close encounter may be understood as "a transient fluctuation of larger amplitude as a star passes by relatively closely" phrased by \cite{Spitzer_1988}. The mathematical formulation of the fully wave kinetic equation can be done simply based on \citep{Klimontovich_1982}'s theory.

One possibility to keep the standard m.f. acceleration of stars in a kinetic equation is to employ a wave kinetic theory even for short distance (Appendix \ref{Appendix:wave_stron_Landau}). Equation \eqref{Eq.convergent_eqn_COL} may be rewritten employing equation \eqref{Eq.twoDF_strong_weak} without exploiting the Boltzmann collision term  (See equation \eqref{Eq.I_st_L} in Appendix \ref{Appendix:wave_stron_Landau}) as follows
\begin{align}
&\left(\frac{\partial}{\partial t}+\bmath{\varv}_{1}\cdot\frac{\partial}{\partial \bmath{r}_{1}}+\bmath{A}_{1}\cdot\frac{\partial}{\partial \bmath{\varv}_{1}}\right)f(\bmath{r}_{1},\bmath{p}_{1},t)\nonumber\\
&=\bmath{\partial}_{1}\cdot\int\text{d}_{2} \bmath{a}_{12}\int_{0}^{\tau}\text{d}\tau'\left[\tilde{\bmath{a}}_{12}\cdot\bmath{\partial}_{12}\right]_{t-\tau'}\nonumber\\
&\quad\times f(1(t-\tau'),t-\tau')f(2(t-\tau'),t-\tau'),\label{Eq.convergent_eqn_wave}
\end{align}
where the spatial non-locality is held to keep the form of typical m.f. acceleration and the trajectory of test star follows equation \eqref{Eq.characteristics_general}. One should be aware that equation \eqref{Eq.convergent_eqn_wave} is, of course, the formal solution to second equation of BBGKY hierarchy, equation \eqref{Eq.2ndBBGKY_convergent_col}.

If one chooses the wave kinetic description for both of the correlation functions $g_\text{Bol}(1,2,t)$ and $g_\text{L}(1,2,t)$, the correlation function, equation \eqref{Eq.g(1,2,t)_conv_zeroth_con}, for the star-cluster CKT reduces to
\begin{align}
&g(1,2,t)\nonumber\\
&=-\int^{\tau}_{0}\bmath{a}_{12}\cdot\bmath{\partial}_{12}f\left(\bmath{r}_{1},\bmath{p}_{1}\left(t-\tau'\right),t\right)f\left(\bmath{r}_{1},\bmath{p}_{2}\left(t-\tau'\right),t\right)\text{d}\tau'\nonumber\\
&\quad+\int^{t}_{t-\tau}\left[\bmath{a}_{12}\left(t'\right)\right]_{r_{12}>\triangle}\text{d}\tau'\cdot\bmath{\partial}_{12}\left[f\left(\bmath{r}_{1},\bmath{p}_{1},t\right)f\left(\bmath{r}_{1},\bmath{p}_{2},t\right)\right]_{r_{12}>\triangle}\nonumber\\
&\quad-\int^{t}_{t-\tau}\left[(\tilde{\bmath{a}}_{12}\cdot\bmath{\partial}_{1}+\tilde{\bmath{a}}_{21}\cdot\bmath{\partial}_{2})f\left(1\left(t'\right),t'\right)f\left(2\left(t'\right),t'\right)\right]_{r_{12}>\triangle}\text{d}t',\label{strong_collision_wave_description}
\end{align}
where the trajectories of first, second and third terms follow the Local Newtonian interaction, equation \eqref{Eq.characteristics_two_body}, rectilinear motion, equation \eqref{Eq.characteristics_rectilinear}, and motion due to m.f. acceleration, equation \eqref{Eq.characteristics_m.f.}, respectively. The convergent kinetic equation to be obtained from equation \eqref{strong_collision_wave_description} is 
\begin{align}
&\left(\frac{\partial}{\partial t}+\bmath{\varv}_{1}\cdot\nabla_{1}+A_{1}^{\blacktriangle}\cdot\bmath{\partial}_{1}\right)f(\bmath{r}_{1},\bmath{p}_{1},t)=I_\text{g-L}^{\blacktriangle}+I_\text{st-L}^{(\text{loc})}-I_\text{L}^{\blacktriangle}.\label{Eq.convergent_eqn_Landau}
\end{align}
Equation \eqref{Eq.convergent_eqn_Landau} merely implies that the m.f. acceleration must be truncated even if one employs the fully wave kinetic description for the convergent CKT due to the spatial locality of the strong Landau collision term, equation \eqref{Eq.I_st_L}. Hence, one must employ the following fully non-local strong-Landau collision term\footnote{Of course in the same way, even the Landau collision term, $I_\text{L}^{\blacktriangle}$, in equation \eqref{Eq.convergent_eqn_Landau} must be fully non-local in space.} inside the Landau sphere to hold the form of typical m.f. acceleration
\begin{align}
I_\text{st-L}=&\int\text{d}_{2}\bmath{a}_{12}\cdot\bmath{\partial}_{1}\int^{\tau}_{0}\text{d}\tau'\bmath{a}_{12}\left(t-\tau'\right)\cdot\bmath{\partial}_{12}\left(t-\tau'\right)\nonumber\\
&\times f\left(\bmath{r}_{1}\left(t-\tau'\right),\bmath{p}_{1}\left(t-\tau'\right),t\right)f\left(\bmath{r}_{2}\left(t-\tau'\right),\bmath{p}_{2}\left(t-\tau'\right),t\right),\label{Eq.I_st_L_non_local}
\end{align}
where the effect of retardation was neglected and the trajectories of test- and field stars follows equation \eqref{Eq.characteristics_two_body}.

To the best of my knowledge, the explicit form of the strong-Landau collision term, equation \eqref{Eq.I_st_L_non_local}, has not been found hence one needs another mathematical breakthrough; the derivation of explicit form of equation \eqref{Eq.I_st_L_non_local} should be left as an open question in the present paper.

%% file: Section7_conclusion.tex
\section{conclusion}\label{sec:conclusion}
The truncated m.f. acceleration (potential) is of significance in statistical dynamics of dense star clusters when discussions of the relation is necessitated between small- and large- scale relaxation processes in kinetic formulation, which occurs inevitably for the secular evolution of the clusters. In the present paper, based on the truncated m.f. acceleration, a star-cluster CKT was initiated to self-consistently derive fundamental kinetic equations to model the secular evolution of dense star clusters of 'point' stars interacting via pair-wise Newtonian interaction, beginning with the $N$-body Liouville equation of stars. In the theory, collision- and wave- kinetic theories were combined to find mathematically non-divergent kinetic equations of star clusters including the effects of 'discreteness' of the clusters and strong two-body encounters, while neglecting the effects of gravitational polarization and relative velocity between two stars and some realistic astrophysical events (triple encounters, primordial binaries, stellar evolution...).

In section \ref{sec:truncated_BBGKY}, the \emph{weakly-coupled} DF, truncated DF and \emph{domain-patched} DF were introduced to model the evolution of star clusters and the corresponding BBGKY hierarchies were derived. The lower limit of 'discreteness' fluctuations in m.f. acceleration was defined based on the weakly-coupled DF (without close encounters) and the truncated DF (with close encounters). It was especially shown that the truncated DF could hold the conservation of total- number and energy of stars if one employs the weakly-coupled DF or 'test-particle' method. The domain-patched DF assumes the systems do not have the lower limit of the 'discrete' fluctuation. Also, the conversion relation of correlation function between collision- and wave- kinetic descriptions was explained based on the \citep{Klimontovich_1982}'s theory. 

In section \ref{sec:complete_WC}, beginning with the BBGKY hierarchy for the weakly-coupled DF (assuming no stars can approach each other closer than the Landau radius), the g-Landau equation with 'completely'-weak-coupling approximation was derived. The mathematical formulation based on the weakly-coupled DF is corresponding to a kinetic formulation of the classical works \citep{Chandra_1943a,Takase_1950}. For the simple relation between the system size and Landau radius, equation \eqref{Eq.simple_R_ro}, the effect of truncated phase-space volume elements in the g-Landau collision integral term weakens typical Coulomb logarithm, $\ln[N]$, for relatively small-number star cluster ($N=10^{5}$) by 14.0 $\%$ and for relatively large-number clusters ($N=10^{7}$) by 10.0 $\%$. Another effect of discreteness appears in the Poisson equation where the truncated volume elements simply corresponds with a coarse-graining of the density of stars by isotropising the density at the Landau radius. The formulation correctly shows the relation between uncorrelated DF and correlation function inside the Landau sphere, which has been neglected in both conventional collision- and wave- kinetic theories. 

In section \ref{sec:strong} employing the truncated DF and 'test-particle' method (any third star does not come into play in close two-body encounter of concern), a kinetic equation for \emph{weakly-inhomogeneous} star clusters was derived in which no mathematical divergence occurs in distance spaces. The kinetic equation can even include close, strong encounters holding the OoM of collision terms at kinetic-equation level up to $\mathcal{O}(1)$ (for $\partial_{t}f(1,t)\sim 1$).  The importance of use of the truncated DFs is that one can correctly formulate the lower limit of occurrence of fluctuation in m.f. acceleration in the same way as the weakly-coupled DF, while one needs to neglect the existence of stars inside the Landau sphere. If one assumes that the star cluster is a continuum even inside the Landau sphere, then uses of \emph{domain-patched} DF and \citep{Klimontovich_1982}'s theory of non-ideal systems are of importance. Especially, at the late stage of star-cluster evolution, the non-locality in space may be of importance due to the strong inhomogeneity of the inner halo structure. While the physical point views between the truncated DF and domain-patched DF are different, the kinetic equation based on the domain-patched DF holds the same mathematical form at the first order as that based on the truncated DF and 'test-particle' method.

In section \ref{sec:discussion}, a convergent kinetic theory was shown based on the results of sections \ref{sec:truncated_BBGKY}, \ref{sec:complete_WC} and \ref{sec:strong}. First, a 'conventional' convergent kinetic equation of star clusters was derived based on the \citep{Frieman_1963}'s work. To find the relation among the equations derived based on the truncated DF, domain-patched DF and conventional CKT, a formally convergent kinetic equation \eqref{Eq.convergent_eqn_COL} was derived from the first two equations of standard BBGKY hierarchy. Then, employing \citep{Klimontovich_1982}'s theory and expanding equation \eqref{Eq.convergent_eqn_COL} in terms of weak-coupling- and weakly-ideal- approximations, the star-cluster convergent kinetic equations \eqref{Eq.star_cluster_conv_eqn} and \eqref{Eq.star_cluster_conv_eqn_non_ideal_2} were derived. The important differences of star-cluster CKT from plasma CKTs are that; (i) the star-cluster CKT is a self-consistent method to find convergent kinetic equations from the standard BBGKY hierarchy while 'conventional' plasma CKTs are not, (ii) the duality terms (the Landau collision term and the non-ideal terms) must be in wave kinetic description, (iii) one may resort to a fully wave kinetic description to hold typical m.f. acceleration of stars in convergent kinetic equation while it necessitates typical Landau collision term to include the spatial non-locality and strong encounters. 

In later papers, the following generalization and application will be done. The equations derived in the present work are, of course, not for an immediate full numerical application though, some analytical/numerical applications with approximations are not hard to accomplish. First place, the explicit expressions for the same kind of equations have been obtained, at least orbit-averaged level \citep{Ipser_1980, Polyachenko_1982,Heyvaerts_2010,Chavanis_2012}. Hence, if one accepts some simple assumptions e.g. Maxwellian distribution or fixed field star DF, one may challenge finding the self-consistent form or the value of Coulomb logarithm, accordingly the relaxation time of star clusters. Also for classical gravothemal-instability problem \citep{Antonov_1985,Binney_2011}, one may employ the star-cluster CKT as done in collision kinetic description \citep[e.g][]{Inagaki_1980,Ipser_1980} and wave one \citep{Wren_2018}. Aside from the application of the main equations, some theoretical extension works are necessary due to the strong assumptions done in the present work. The 'actual' Landau distance is essentially naive to the relative speed of test star to field one, hence the distance must be correctly handled without velocity dispersion approximation. This necessitate even reapplying \citep{Klimontovich_1982}'s theory (together with \citep{Grad_1958}'s method) to velocity space, following the basic result of \citep{Takase_1950,Chang_1992,Shoub_1992,Ipser_1983}. Hence, a future goal is to extend the present work to a 'perfectly' convergent kinetic theory that has no mathematical divergences in distance-, wavenumber- and momentum- spaces. Also, in the present paper the weak correlation condition was employed while employing the \citep{Zubarev_1996}'s non-equilibrium operator method to handle the dynamics inside Landau spheres in the core of clusters without binaries would be of special-mathematical interest.
 

%% file: Appendix.tex
\begin{appendices}
\renewcommand{\theequation}{\Alph{section}.\arabic{equation}}
\section{'test-particle' method and \emph{mean-field} stars in N-body DF and two-body DF for a star clusters}\label{sec.many_to_two}
In the present Appendix, it is shown that the \emph{mean-field} star approximations introduced in \citep{Kandrup_1981} based on ($N-1$)-body DF, and the 'test-particle' method introduced in \citep{Kaufman_1960} based on the two-body DF do have the same role to obtain g-Landau equation, unless one find the exact form of the ($N-1$)-body DF. In Appendix \ref{sec:rev.Kandrup} the Kandrup's g-Landau equation is revisited by assuming the explicit form of $(N-1)$-body DF and in Appendix \ref{subsec:Kaufman} the \citep{Kaufman_1960}'s method is compared to the \citep{Kandrup_1981}'s method; the 'test-particle' method employed in section \ref{sec:strong} is explained. 

\subsection{An explicit expression for the generalized Landau equation for stellar systems}\label{sec:rev.Kandrup}
In the present section, an explicit form of the g-Landau equation for many-body encounters is derived beginning with the $N$-body Liouville equation by assuming the form of $(N-1)$-body DF. The derivation here is simple since one does not need exploiting the time-dependent operator method.
\subsubsection{N-body DF description for many body encounters}\label{sec:FNmany}
As done in \citep{Kandrup_1981}, first arrange the Liouville equation \eqref{Eq.Liouville} into 'test' star (star 1) and $(N-1)$- 'field' stars (stars $2, \cdots, N$)
\begin{align}
&\left(\partial_{t}+\bmath{\varv}_{1}\cdot\nabla_{1}+\bmath{a}_{1}\cdot \bmath{\partial}_{1}+\sum_{i=2}^{N}\left[\bmath{\varv}_{i}\cdot\nabla_{i}+\bmath{a}_{i}\cdot \bmath{\partial}_{i}\right]\right)F_{N}(1,\cdots,N,t)\nonumber\\
&\equiv\frac{\text{D}}{\text{D}t}F_{N}(1,\cdots,N,t)=0.\label{Eq.Liouville_many}
\end{align}
Integration of the L.H.S of equation \eqref{Eq.Liouville_many} over the two entire phase-space volume elements $\text{d}_{2}\cdots\text{d}_{N}$ and $\text{d}_{1}$ respectively (\emph{without} truncating the spatial volume) results in the following two equations
\begin{subequations}
	\begin{align}
	&\left(\partial_{t}+\bmath{\varv}_{1}\cdot\nabla_{1}\right)F_{1}(1,t)+\sum_{j=2}^{N}\int\bmath{a}_{1j}\cdot \bmath{\partial}_{1}F_{N}(1,\cdots,N,t)\text{d}_{2}\cdots\text{d}_{N}\nonumber\\
	&\qquad=0,\label{Eq.Liouville_many1}\\
	&\left(\partial_{t}+\sum_{i=2}^{N}\left[\bmath{\varv}_{i}\cdot\nabla_{i}+\sum_{j=2\neq i}^{N}\bmath{a}_{ij}\cdot \bmath{\partial}_{i}\right]\right)F_{N-1}(2,\cdots,N,t),\nonumber\\
	&\qquad\quad+\sum_{j=2}^{N}\int\bmath{a}_{j1}\cdot \bmath{\partial}_{j}F_{N}(1,\cdots,N,t)\text{d}_{1} =0.\label{Eq.Liouville_many2}
	\end{align}
\end{subequations}
Equations \eqref{Eq.Liouville_many}-\eqref{Eq.Liouville_many2} are described in terms of only the three DFs $(F_{1},$ $F_{N-1}$ and $F_{N})$, meaning the equations are \emph{possibly} solved for the DFs, and especially one's concern is how the Newtonian interaction due to the field stars on the test star affects the test star's DF. In order to consider the relation between the test-star DF $F_{1}(1,t)$ and the field-stars' DF $F_{N-1}(2, \cdots, N,t)$, define the explicit form of $N$-body DF as 
\begin{align}
F_{N}(1,\cdots,N,t)=F_{1}(1,t)F_{N-1}(2, \cdots, N,t)+g_{N}(1, \cdots,  N, t).\label{Eq.many_g}
\end{align}
Insert the equation \eqref{Eq.many_g} into equations \eqref{Eq.Liouville_many}-\eqref{Eq.Liouville_many2}
\begin{subequations}
	\begin{align}
	\frac{\text{D}}{\text{D}t}g_{N}(1,\cdots,N,t)=&-\partial_{t}(F_{1}F_{N-1})-F_{N-1}(\bmath{\varv}_{1}\cdot\nabla_{1}+\bmath{a}_{1}\cdot\bmath{\partial}_{1})F_{1}\nonumber\\
	&-F_{1}\sum_{i=2}^{N}(\bmath{\varv}_{i}\cdot\nabla_{i}+\bmath{a}_{i}\cdot\bmath{\partial}_{i})F_{N-1},\label{Eq.time_evoln_g_N}
	\end{align}
	\begin{align}
	\left(\partial_{t}+\bmath{\varv}_{1}\cdot\nabla_{1}+\int\bmath{a}_{1}F_{N-1}\text{d}_{2}\cdots\text{d}_{N}\cdot\bmath{\partial}_{1}\right)F_{1}\nonumber\\
	+\int\bmath{a}_{1}\cdot\bmath{\partial}_{1}g_{N}\text{d}_{2}\cdots\text{d}_{N}=0,\label{Eq.time_evoln_F1_many}
	\end{align}
	\begin{align}
	&\left(\partial_{t}+\sum_{i=2}^{N}\left[\bmath{\varv}_{i}\cdot\nabla_{i}+\sum_{j=2(\neq i)}^{N}\bmath{a}_{ij}\cdot\partial_{i}\right]\right)F_{N-1}\nonumber\\
	&\qquad+\sum_{i=2}^{N}\left(\int\bmath{a}_{i1}F_{1}\text{d}_{1}\cdot\bmath{\partial}_{i}F_{N-1}+\int\bmath{a}_{i1}\cdot\bmath{\partial}_{i}g_{N}\text{d}_{1}\right)=0.\label{Eq.time_evoln_F_N-1}
	\end{align}
\end{subequations}
Find the relation of the correlation $g_{N}$ with the DFs $F_{N-1}$ and $F_{1}$ from equations \eqref{Eq.time_evoln_g_N}-\eqref{Eq.time_evoln_F_N-1};
\begin{align}
&\frac{\text{D}}{\text{D}t}g_{N}=\sum_{i=2}^{N}\left([\bmath{A}_{i}-\bmath{a}_{i1}]\cdot\partial_{i}+[\bmath{A}_{1}-\bmath{a}_{1i}]\cdot\partial_{1}\right)F_{1}F_{N-1}\nonumber\\
&\qquad\quad+F_{1}\sum_{i=2}^{N}\int\bmath{a}_{i1}\cdot\bmath{\partial}_{i}g_{N}\text{d}_{1}+F_{N-1}\int\bmath{a}_{1}\cdot\bmath{\partial}_{1}g_{N}\text{d}_{2}\cdots\text{d}_{N}.\label{Eq.time_evoln_g_many}
\end{align}
Equations \eqref{Eq.time_evoln_F1_many} and \eqref{Eq.time_evoln_g_many} form exactly the same equation derived based on the time-dependent operator method \citep[equation 33 of][]{Kandrup_1981} in which the last two terms on the R.H.S of equation \eqref{Eq.time_evoln_g_many} vanish due to the standard-time scale.

\subsubsection{Two-body DF description for many-body encounters}\label{sec:kinetic_eqn_many}
\cite{Kandrup_1981} assumed the field stars $2$ through $N$ behave as if they were collisionless stars; no encounters among field stars can occur during test star undergoes a two-body encounter with one of field stars. This assumption encourages one to prescribe the explicit form of $N$-body joint-probability function $F_{N}(1,2,\cdots,N)$ in terms of correlation functions and DFs as
\begin{align}
F_{N}(1,\cdots,N,t)\equiv\prod_{j=1}^{N}F_{1}(j,t)+\frac{1}{N}\sum_{k=2}^{N}G_{2}(1,k,t)\prod_{m=2\neq k}^{N}F_{1}(m,t),\label{Eq.many_FN}
\end{align}
where the correlation functions are defined as follows
\begin{align}
G_{2}(1,k,t)\equiv\frac{1}{N}g_{2}(1,k,t). \qquad(k=2,\cdots,N)
\end{align}
Equation \eqref{Eq.Liouville_many2} can reduce to a kinetic-like equation for each $k=2,3,\cdots N$
\begin{align}
&\left(\partial_{t}+\bmath{\varv}_{k}\cdot\nabla_{k}+\left[\sum_{j=2\neq k}^{N}\bmath{a}_{kj}+\int\bmath{a}_{k1}F_{1}\text{d}_{1}\right]\cdot\bmath{\partial}_{k}\right)F_{1}(k,t)\nonumber\\
&\qquad=-\frac{1}{N}\bmath{\partial}_{2}\cdot\int \bmath{a}_{k1}G_{2}(1,k,t)\text{d}_{1}\approx 0.\label{Eq.mf_eqn_each}
\end{align}
where the last equality is due to the weak-coupling limit. Equation \eqref{Eq.mf_eqn_each} shows a similarity to the collisionless Boltzmann equation though, only the accelerations of $k$-th star due to star 1 (test star) are smoothed out. 

The most important difference of the ($N-1$)-DF, equation \eqref{Eq.many_FN}, from that of standard BBGKY hierarchy is that it is \emph{asymmetric} about permutation between the test- and m.f.- star while symmetric between two m.f. stars, meaning one needs to employ the set of equations derived in Appendix \ref{sec:FNmany} rather than resorting to the standard BBGKY hierarchy. The L.H.S of equation \eqref{Eq.time_evoln_g_many} is
\begin{align}
&\frac{\text{D}}{\text{D}t}\left[\sum_{k=2}^{N}G_{2}(1,k,t)\prod_{l=2\neq k}^{N}F_{1}(l,t)\right]\nonumber\\
&=\sum_{k=2}^{N}\left[ \frac{\text{D}^{(k)}}{\text{D}^{(k)}t} G_{2}(1,k,t)\right]\prod_{l=2\neq k}^{N}F_{1}(l,t)\nonumber\\
&\qquad+\sum_{k=2}^{N}G_{2}(1,k,t)\left(\partial_{t}+\sum_{i=2}^{N},\left[\bmath{\varv}_{i}\cdot\nabla_{i}+\bmath{a}_{i}\cdot \bmath{\partial}_{i}\right]\right)\prod_{l=2\neq k}^{N}F_{1}(l,t)\label{Eq.time_evoln_G_asym_LHS_0}\\
&=\sum_{k=2}^{N}\left[ \frac{\text{D}^{(k)}}{\text{D}^{(k)}t} G_{2}(1,k,t)\right]\prod_{l=2\neq k}^{N}F_{1}(l,t)\nonumber\\
&\qquad+\sum_{k=2}^{N}G_{2}(1,k,t)\left(\sum_{i=2}^{N}\left(\bmath{a}_{i1}-\bmath{A}_{i}\right)\cdot \bmath{\partial}_{i}\right)\prod_{l=2\neq k}^{N}F_{1}(l,t),\label{Eq.time_evoln_G_asym_LHS}
\end{align}
where, for the second equality, equation \eqref{Eq.mf_eqn_each} was employed and the total derivative in equation \eqref{Eq.time_evoln_G_asym_LHS_0} is defined as
\begin{align}
\frac{\text{D}^{(k)}}{\text{D}^{(k)}t}=\partial_{t}+\bmath{\varv}_{1}\cdot\nabla_{1}+\bmath{\varv}_{k}\cdot\nabla_{k}+\bmath{a}_{1}\cdot \bmath{\partial}_{1}+\bmath{a}_{k}\cdot \bmath{\partial}_{k}.\label{Eq.tot_D}
\end{align}
As a result
\begin{align}
&\sum_{k=2}^{N}\left[ \frac{\text{D}^{(k)}}{\text{D}^{(k)}t} G_{2}(1,k,t)\right]\prod_{l=2\neq k}^{N}F_{1}(l,t)\nonumber\\
&\quad+\sum_{k=2}^{N}G_{2}(1,k,t)\left(\sum_{i=2}^{N}\left(\bmath{a}_{i1}-\bmath{A}_{i}\right)\cdot \bmath{\partial}_{i}\right)\prod_{l=2\neq k}^{N}F_{1}(l,t)\nonumber\\
&=\sum_{i=2}^{N}\left([\bmath{A}_{i}-\bmath{a}_{i1}]\cdot\partial_{i}-[\bmath{A}_{1}-\bmath{a}_{1}]\cdot\partial_{1}\right)F_{1}(1)F_{1}(i)\prod_{l=2\neq i}^{N}F_{1}(l,t).\label{Eq.time_evoln_g_asym_RHS}
\end{align}
If possible, one can isolate only the terms associated with $G_{2}(1,2)$, then
\begin{align}
&\left[ \frac{\text{D}^{(k)}}{\text{D}^{(k)}t} G_{2}(1,k,t)\right]+\left(\sum_{i=2}^{N}\left(\bmath{a}_{i1}-\bmath{A}_{i}\right)\cdot \bmath{\partial}_{i}\ln\left[F_{1}(i,t)\right]\right)G_{2}(1,k,t)\nonumber\\
&\qquad=\left([\bmath{A}_{k}-\bmath{a}_{k1}]\cdot\partial_{k}-[\bmath{A}_{1}-\bmath{a}_{1k}]\cdot\partial_{1}\right)F_{1}(1)F_{1}(k).\label{Eq.time_evoln_g_asym_RHS_sum}
\end{align}
Equation \eqref{Eq.time_evoln_g_asym_RHS_sum} describes the explicit form of the g-Landau equation for many-body encounter in two-body-DF description, where the effect of many-body interaction on test star appears as 'raw' pair-wise accelerations of stars in the total derivative, equation \eqref{Eq.tot_D}.

\subsubsection{Two-body DF description for many-body encounters}\label{sec:GLandau_many}
The second term on the L.H.S. in equation \eqref{Eq.time_evoln_g_asym_RHS_sum} did not appear in \citep{Kandrup_1981}. In order to deal with a complication of many-body interaction, Following \citep{Kandrup_1981} one needs to assume the interaction can be replaced by the mean filed 
\begin{align}
\sum^{N}_{j=2\neq i}\bmath{a}_{ij}\approx\sum^{N}_{j=2\neq i}\int\bmath{a}_{ij}F_{1}(j,t)\text{d}_{j}.
\end{align}
Accordingly, equation \eqref{Eq.mf_eqn_each} reduces to
\begin{align}
&\left(\partial_{t}+\bmath{\varv}_{k}\cdot\nabla_{k}+N\bmath{A}_{k}\cdot\bmath{\partial}_{k}\right)F_{1}(k,t)=0.\label{Eq.vlasov_mf_stars}
\end{align}
The assumption that field stars follow the Vlasov equations \eqref{Eq.vlasov_mf_stars} results in
\begin{align}
&\left[ \frac{\text{D}^{(k)}}{\text{D}^{(k)}t} G_{2}(1,k,t)\right]+\left(\sum_{i=2}^{N}\left(\bmath{a}_{i}-\bmath{A}_{i}\right)\cdot \bmath{\partial}_{i}\ln\left[F_{1}(i,t)\right]\right)G_{2}(1,k,t)\nonumber\\
&\qquad=\left([\bmath{A}_{k}-\bmath{a}_{k1}]\cdot\partial_{k}-[\bmath{A}_{1}-\bmath{a}_{1k}]\cdot\partial_{1}\right)F_{1}(1)F_{1}(k),\label{Eq.time_evoln_g_mf}
\end{align}
and if the approximation $\bmath{a}_{i}\approx\bmath{A}_{i}$ holds, then one obtains 
\begin{align}
&\left[ \frac{\text{D}^{(k)}}{\text{D}^{(k)}t} G_{2}(1,k,t)\right]\nonumber\\
&\qquad=\left([\bmath{A}_{k}-\bmath{a}_{k1}]\cdot\partial_{k}-[\bmath{A}_{1}-\bmath{a}_{1k}]\cdot\partial_{1}\right)F_{1}(1)F_{1}(k).\label{Eq.g_Landau_eqn}
\end{align}
The total derivative, equation \eqref{Eq.tot_D}, is also to be smoothed out
\begin{subequations}
	\begin{align}
	\frac{\text{D}^{(k)}}{\text{D}^{(k)}t}&=\partial_{t}+\bmath{\varv}_{1}\cdot\nabla_{1}+\bmath{\varv}_{k}\cdot\nabla_{k}+\bmath{a}_{1k}\cdot\partial_{1k}\nonumber\label{Eq.tot_many}\\
	&\quad+\sum^{N}_{j=2}\bmath{a}_{1j}\cdot\bmath{\partial}_{1}+\sum^{N}_{j=2\neq k}\bmath{a}_{kj}\bmath{\partial}_{k},\\
	&\approx\partial_{t}+\bmath{\varv}_{1}\cdot\nabla_{1}+\bmath{\varv}_{k}\cdot\nabla_{k}+\bmath{A}_{1j}\cdot\bmath{\partial}_{1}+\bmath{A}_{kj}\bmath{\partial}_{k},
	\label{Eq.tot_derv}
	\end{align}
\end{subequations}
where the term $\bmath{a}_{1k}\cdot\bmath{\partial}_{1k}$ disappears due to the weak-coupling limit. Equations \eqref{Eq.g_Landau_eqn} and \eqref{Eq.tot_derv} form the evolution of correlation function for the g-Landau equation. 

\subsubsection{Generalized Landau equation derived by Kandrup}\label{Appendix:Kandrup}
The present section reviews the strong approximations done in \citep{Kandrup_1981}. The last two terms in equation \eqref{Eq.time_evoln_g_many} corresponds with the projection operator used in \citep{Kandrup_1981}, where the concept of \emph{standard time} scale was introduced and those terms vanished from the equation. Essentially, the reason why the last two terms can vanish can be explained based on the weakly-coupling approximation; the first term on the R.H.S of equation \eqref{Eq.time_evoln_g_many}  is order of $\mathcal{O}(1/N)$, meaning the last two terms is recursively small up to order of $\mathcal{O}(1/N^{2})$ if one can ignore the convergence problem. Hence employing the method of characteristic, equation \eqref{Eq.time_evoln_F1_many} reduces to
\begin{align}
&\left(\partial_{t}+\bmath{\varv}_{1}\cdot\nabla_{1}+\int\bmath{a}_{1}F_{N-1}\text{d}_{2}\cdots\text{d}_{N}\cdot\bmath{\partial}_{1}\right)F_{1}\nonumber\\
&\quad+\int\bmath{a}_{1}\cdot\bmath{\partial}_{1}\int\left(\sum_{i=2}^{N}\left[\tilde{\bmath{a}}_{i1}\cdot\partial_{i}-\tilde{\bmath{a}}_{1}\cdot\partial_{1}\right]F_{1}F_{N-1}\right)\text{d}\tau\text{d}_{2}\cdots\text{d}_{N}\nonumber\\
&=0.
\end{align}
In addition to the weak-coupling approximation, mean-filed-particle approximation is used
\begin{align}
F_{N-1}\approx\prod_{k=2}^{N} F_{1}(k,t),\label{Eq.F_N_1}
\end{align}
meaning any interaction among the $(N-1)$ field stars does not affect the DF of test star. Due to the mean-filed-particle approximation and weak-coupling approximation the $(N-1)$- m.f. stars follows
\begin{align}
&\left(\partial_{t}+\sum_{i=2}^{N}\bmath{\varv}_{i}\cdot\nabla_{i}+\sum_{i=2}^{N}\int\bmath{a}_{i1}F_{1}\text{d}_{1}\cdot\bmath{\partial}_{i}\right)F_{N-1}=0,\label{Eq.mf_eqn}
\end{align}
where  the interaction among field particles are ignored. As a result, the g-Landau equation in terms of $s$-tuple DF is
\begin{align}
&\left(\partial_{t}+\bmath{\varv}_{1}\cdot\nabla_{1}+\int\bmath{a}_{1}F_{N-1}\text{d}_{2}\cdots\text{d}_{N}\cdot\bmath{\partial}_{1}\right)F_{1}\nonumber\\
&\quad+\int\bmath{a}_{1}\cdot\bmath{\partial}_{1}\int\left(\sum_{i=2}^{N}\left[\tilde{\bmath{a}}_{i1}\cdot\partial_{i}-\tilde{\bmath{a}}_{1}\cdot\partial_{1}\right]F_{1}F_{N-1}\right)\text{d}\tau\text{d}_{2}\cdots\text{d}_{N}\nonumber\\
&=0.
\end{align}
In the work of \citep{Kandrup_1981}, some strong assumptions are done such as
\begin{align}
&F_{1}(1(t-\tau),t-\tau)F_{N-1}(2(t-\tau),\cdots,N(t-\tau),t-\tau)\nonumber\\
&\qquad\approx F_{1}(1(t),t) F_{N-1}(2(t),\cdots,N(t),t),
\end{align}
together with the approximation
\begin{align}
\left[\tilde{\bmath{a}}_{i1}\cdot\partial_{i}-\tilde{\bmath{a}}_{1}\cdot\partial_{1}\right]_{t-\tau}\approx\tilde{\bmath{a}}_{i1}(t-\tau)\cdot\partial_{i}-\tilde{\bmath{a}}_{1}(t-\tau)\cdot\partial_{1}.
\end{align}
Fixing the time of characteristics to the time $t$, the corresponding g-Landau equation reads
\begin{align}
&\left(\partial_{t}+\bmath{\varv}_{1}\cdot\nabla_{1}+\int\bmath{A}_{1}\cdot\bmath{\partial}_{1}\right)F_{1}\nonumber\\
&\quad-\int\bmath{a}_{12}\cdot\bmath{\partial}_{1}\int\tilde{\bmath{a}}_{12}(t-\tau)\cdot\partial_{12}F_{1}(1,t)F_{1}(2,t)\text{d}\tau\text{d}_{2}=0.
\end{align}
As clearly shown above, one needs a number of approximations in \citep{Kandrup_1981}'s method. Especially in equation \eqref{Eq.tot_many}, without m.f. star approximation, one must deal with $(N-1)$-body problems. Hence one may be better to find a way to avoid the strong approximation done above as shown in Appendix \ref{subsec:Kaufman}. 

\subsection{'test-particle' method: test star interacting with collisionless field stars}\label{subsec:Kaufman}
The method employed in \citep{Kandrup_1981} is of physical importance but not mathematically straightforward unless one finds a correct form of $(N-1)$-body DF. Another method to find the effect of many-body encounters was challenged in plasma physics \citep{Kaufman_1960} where implicitly \emph{m.f. star approximation} was shown. In the present section, \citep{Kaufman_1960}'s work is extended to the g-Landau equation. 

The advantage of \citep{Kaufman_1960}'s method is that all stars may be labeled, or \emph{apparently} non-identical. Assume test star (star 1) can be described by a typical $s$-body DF, equation \eqref{Eq.SbodyJointDF}, and the BBGKY hierarchy, equation \eqref{Eq.BBGKY_orth}, while $s$-body DFs for field stars 2, 3 $\cdots$ are defined as 
\begin{subequations}
	\begin{align}
	U_{s}(2,\cdots, s+1, t)=&\int{F_{s+1}(1,\cdots s+1)}\quad\text{d}_1,\nonumber\\
	=&\int F_N(1,\cdots,N, t)\quad\text{d}_{1}\text{d}_{s+2}\cdots\text{d}_{N},\label{Eq.SbodyJointDF_U2}\\
	U_{s}(3,\cdots, s+2, t)=&\int{F_{s+2}(1,\cdots s+2)}\quad\text{d}_{1}\text{d}_2,\nonumber\\
	=&\int F_N(1,\cdots,N, t)\quad\text{d}_{1}\text{d}_2\text{d}_{s+3}\cdots\text{d}_{N}.\label{Eq.SbodyJointDF_U3}\\
	\qquad    \vdots \nonumber
	\end{align} 
\end{subequations} 
The BBGKY hierarchy for field star 2\footnote{To find the hierarchy, simply take the integral of the $(s+1)$-th equation of the BBGKY hierarchy over $\int\cdot \text{d}_{1}$} is
\begin{align}
&\partial_{t}U_{s}+\sum_{i=2}^{s+1}\left[\bmath{\partial}_{i}\cdot\int F_{s+1}\bmath{a}_{i,1}\text{d}_{1}\right]\nonumber\\
&\qquad+\sum_{i=2}^{s+1}\left[\bmath{\varv}_{i}\cdot\nabla_{i}+\sum_{\quad{j=2(\neq i)}}^{s+1}\bmath{a}_{ij}\cdot\bmath{\partial}_{i}\right]U_{s}\nonumber\\
&\qquad\quad +(N-s-1)\sum_{i=2}^{s+1}\left[\bmath{\partial}_{i}\cdot\int U_{s+1}\bmath{a}_{i,s+2}\text{d}_{s+2}\right]=0.\label{Eq.BBGKY_test}
\end{align}
The corresponding $s$-tuple DFs may be defined as follows
\begin{subequations}
	\begin{align}
	&F_{1}(1,t)=f(1,t),\\
	&(N-1)F_{2}(1,2,t)=f(1,t)u(2,t)+g(1,2,t),\\
	&(N-1)(N-2)F_{3}(1,2,3,t)\nonumber\\
	&\qquad=f_{1}(2,t)u_{1}(3,t)u_{1}(4,t)+f_{1}(1)\tilde{g}(2,3,t)+u_{1}(2)g(1,3,t)\nonumber\\
	&\qquad\quad+u_{1}(3)g(1,2,t),\\
	&(N-1)U_{1}(1,t)=u_{1}(1,t),\\
	&(N-1)(N-2)U_{2}(2,3,t)=u_{1}(2,t)u_{1}(3,t)+\tilde{g}(2,3,t),\\
	&(N-1)(N-2)U_{3}(2,3,4,t)\nonumber\\
	&\qquad=u_{1}(2,t)u_{1}(3,t)u_{1}(4,t)+u_{1}(2)\tilde{g}(3,4,t)+u_{1}(3)\tilde{g}(1,4,t)\nonumber\\
	&\qquad\quad+u_{1}(4)\tilde{g}(2,3,t).
	\end{align}
\end{subequations}

 If one assumes that any of two field stars are not correlated and that the weakly-coupling limit is still applicable to the system of concern, then the first two equation of the BBGKY hierarchy for test star and field star 2 are described by 
\begin{subequations}
	\begin{align}
	&\left(\partial_{t}+\bmath{\varv}_{1}\cdot\nabla_{1}+\int u(2,t)\bmath{a}_{1,2}\text{d}_{2}\cdot\bmath{\partial}\right)f(1,t)\nonumber\\
	&\qquad=\bmath{\partial}_{1}\cdot\int g(1,2,t)\bmath{a}_{12}\text{d}_{2},\label{Eq.1stBBGKY_wc_t}\\
	&\left(\partial_{t}+\bmath{\varv}_{1}\cdot\nabla_{1}+\bmath{\varv}_{2}\cdot\nabla_{2}-\int u(3,t)\left[\bmath{a}_{13}\cdot\bmath{\partial}_{1}+\bmath{a}_{23}\cdot\bmath{\partial}_{2}\right]\text{d}_{3}\right)g(1,2,t)\nonumber\\
	&\qquad=-\left[\bmath{a}_{12}\cdot\bmath{\partial}_{1}-\int f(1,t)\bmath{a}_{2,1}\text{d}_{1}\cdot\bmath{\partial}_{2}-\int \frac{u(2,t)}{N}\bmath{a}_{1,2}\text{d}_{2}\cdot\bmath{\partial}_{1}\right]\nonumber\\
	&\qquad\qquad\times(f(1,t)u(2,t)),\label{Eq.2nd_BBGKY_wc_t}\\
	&\left(\partial_{t}+\bmath{\varv}_{2}\cdot\nabla_{2}+\int f(1,t)\bmath{a}_{2,1}\text{d}_{1}\cdot\bmath{\partial}_{2}+\int u(3,t)\bmath{a}_{2,3}\text{d}_{3}\cdot\bmath{\partial}_{2}\right)u(2,t)\nonumber\\
	&\qquad=-\bmath{\partial}_{2}\cdot\int \bmath{a}_{23}g(2,3,t)\text{d}_{3}.\label{Eq.1stBBGKY_wc_f}
	\end{align}
\end{subequations}
Equations \eqref{Eq.1stBBGKY_wc_t}-\eqref{Eq.1stBBGKY_wc_f} are in essence the same as the standard BBGKY hierarchy if one assumes that the system is invariant under a permutation between the states of two stars. Equation \eqref{Eq.1stBBGKY_wc_f} is of importance for comparison to the result of many-body encounter in Appendix \ref{sec:GLandau_many} since the function $g(2,3,t)$ represents the correlation of field stars described by the DF $F_{N-1}(2,\cdots,N)$. In \citep{Kaufman_1960}, the factor $\int \bmath{a}_{23}g(2,3,t)\text{d}_{3}$ was considered null since the acceleration  $\bmath{a}_{23}$ is odd about the distance $r_{23}$ while the correlation function is even due to the Debye shielding
\begin{align}
g(2,3,t)\propto u(\bmath{p}_{2})u(\bmath{p}_{3})\frac{e^{-r_{23}/\lambda_{D}}}{r_{23}},
\end{align} 
where the underlying assumption is the plasmas of field particles is in thermal equilibrium and the correlation function is local i.e. the zeroth order of the center of mass $\bmath{R}$ is of concern $\left(\bmath{R}\approx \bmath{r}_{2}\right)$. \citep{Kaufman_1960} did not clearly explain the time-dependence of $u(\bmath{p}_{2})$ and $ u(\bmath{p}_{3})$ though, one may understand the field particles does not change during the correlation time $\tau$ in a similar way to the discussion in \citep{Kandrup_1981}; one must 'control' the time scale of the effective correlation time, the standard time scale, to assume the m.f. stars follows collisionless Boltzmann equation. This is corresponding to defining the DF $F_{N-1}(2,\cdots,N)$  as equation \eqref{Eq.F_N_1} by assuming the DF $F_{N-1}(2, \cdots, N,t)$ is initially perfectly uncorrelated and can hold the form at the standard times scales. 

Up to now, field stars are considered as different species though, in order to retrieve the g-Landau equation without deleting the terms associated with the function $g(i,j,t)$ for $2\leq i, j \leq N$, rewrite the DFs as follows
\begin{subequations}
	\begin{align}
	&f(1,t)\equiv NF_{1}(1,t),\\
	&f(2,t)=u(2,t),\\
	&\int \bmath{a}_{21}F(1)\text{d}_{1}=\int \bmath{a}_{21}f(1)\text{d}_{1}-\int \bmath{a}_{21}u(1)\text{d}_{2}.
	\end{align}
\end{subequations}
As a result equations \eqref{Eq.1stBBGKY_wc_t} and \eqref{Eq.2nd_BBGKY_wc_t} form the generalized Landau equation while equation \eqref{Eq.1stBBGKY_wc_f} is a typical Vlasov equation for the field star:
\begin{align}
&\left(\partial_{t}+\bmath{\varv}_{2}\cdot\nabla_{2}+\int f(1,t)\bmath{a}_{21}\text{d}_{1}\cdot\bmath{\partial}_{2}\right)f(2,t)\approx 0.\label{Eq.1stBBGKY_wc_f_2}
\end{align}

In summary of the present Appendix, it is obviously convenient in formulation of the g-Landau equation to resort to the \citep{Kaufman_1960}'s 'test-particle' method rather than the \citep{Kandrup_1981}'s method, unless one can find the correct form of the $(N-1)$-DF. Also, \citep{Kaufman_1960}'s method is essentially the same as the standard BBGKY hierarchy except for labeling each star. Hence, one of the simplest ways to employ the 'test-particle' method is that one first resorts to the standard BBGKY hierarchy and then to assume that the DF for star 2 follows a different kinetic equation from the equation that the DF for star 1 follows at the first two equations of the hierarchy, as employed in section \ref{sec:strong}.

\section{An explanation for the order of magnitude of the effective distance of Newtonian interaction potentials}\label{Appendix:interaction}
In the Appendix, the scaling of the OoM of the effective interaction range of accelerations of stars due to Newtonian potentials are explained following the ranges below;
\begin{enumerate}
	\item m.f. (many-body) interaction \qquad  $a_\text{BG}$$<r_{12}<$ $R$
	\item weak two-body interaction \quad  $ r_\text{o}$ $<r_{12}<$ $a_\text{BG}$
\end{enumerate}
where the average distance of stars is neglected since it is not of essence in the present work. To find the discussion for the Landau radius $r_\text{o}$, refer to section \ref{subsection:scaling_strong}.
 
\subsection{The threshold between (i) and (ii)}
The transition between ranges (i) and (ii) is the radius of encounter \citep{Ogorodnikov_1965}, at which the order of the irregular force is compatible with that of m.f. potential force. In range (i), star 1 can be accelerated by the total of Newtonian interaction forces due to the rest of stars as follows
\begin{align}
\bmath{a}_{1}=-\sum_{k=2}^{N}\frac{Gm}{r_{1k}^{3}} (\bmath{r}_{1}-\bmath{r}_{k}).
\end{align}
As assumed in \citep{Kandrup_1981} and Appendix \ref{sec.many_to_two}, the acceleration due to many-body encounters (with $(N-1)$-stars) may be roughly replaced by the acceleration due to the smooth self-consistent m.f. acceleration of star 1;
\begin{align}
\bmath{A}_{1}=-\left(1-\frac{1}{N}\right)\int \frac{Gm}{r_{12}^{3}}(\bmath{r}_{1}-\bmath{r}_{2})f(2,t)\text{d}_{2}.
\end{align}
Some stars, however, can occasionally enter range (ii), then the main cause of acceleration of star 1 is due to the pair-wise Newtonian potential, equation \eqref{Eq.phi}, from star 2
\begin{align}
\bmath{a}_{12}=-\frac{Gm}{r_{12}^{3}} (\bmath{r}_{1}-\bmath{r}_{2}).
\end{align}

Employing the scaling $G\sim\mathcal{O}(1/N)$ for fixed finite stellar masses $m\sim\mathcal{O}(1)$ and fixed momenta $\bmath{p}_{1}\sim\bmath{p}_{2}\sim\mathcal{O}(1)$ as explained in section \ref{subsec:scalings}, the two accelerations are scaled as
\begin{subequations}
	\begin{align}
	&\bmath{a}_{12}\sim \frac{1}{N}\frac{1}{r_{12}^{2}},\label{Eq.scale_a_12}\\
	&\bmath{a}_{1}\sim R\sim\mathcal{O}(1).\label{Eq.scale_A_1}
	\end{align}
\end{subequations}
By equating the two accelerations, equations \eqref{Eq.scale_a_12} and \eqref{Eq.scale_A_1}, the threshold between m.f. (many-body) and two-body interaction forces is obtained
\begin{align}
&r_{12}\sim \frac{1}{N^{1/2}}\sim a_\text{BG}.
\end{align}

\subsection{The size of a cluster in (i)}
Assume the size of a star cluster corresponds with the Jean length. The celebrated Jeans instability \citep{Jeans_1902} of a self-gravitating system may be discussed even at kinetic-equation level for collisionless \citep[e.g.][]{Binney_2011} and collisional \citep[e.g.][]{Trigger_2004} self-gravitating systems assuming the dynamical stability condition as follows
\begin{align}
\bmath{\varv}_{1}\cdot\nabla_{1}+\bmath{A}_{1}\cdot\bmath{\partial}_{1}\sim \frac{\bmath{\varv}_{1}}{R}- \frac{Gmnr_{12}}{\varv_1}=0.
\end{align}
where $R$ is the size of the stellar system and $n$ the average density of the system. Due to the unscreened gravitational potential, the interaction range $r_{12}$ or the wavelength of fluctuation in m.f. potential can reach the system radius $R$ and may bring the system into an unstable state. The Jeans length occurs when the distance $r_{12}$ is compatible with the system radius $R$
\begin{align}
R\sim \sqrt{\frac{\varv^{2}_{1}}{Gmn}}\sim\mathcal{O}(1),
\end{align}
where scalings $G\sim\mathcal{O}(1/N)$ and $n\sim\mathcal{O}(N)$ are taken for fixed stellar mass $m\sim\mathcal{O}(1)$ and dispersion $<\varv>\sim\mathcal{O}(1)$.

\section{Derivation of BBGKY hierarchy for truncated distribution function}\label{Appendix_BBGHY}
In \citep{Cercignani_1988}, the derivation of the BBGKY hierarchy for the hard-sphere DFs was made in a mathematically strict manner, by employing the Gauss's lemma and integration-by-parts, while counting the correct patterns of combinations for the Gauss's lemma is confusing and the BBGKY hierarchy for the truncated DF is not shown explicitly. In the present section, the latter hierarchy is derived by exploiting integration-by-parts and a general Heaviside function
\begin{align}
\Theta(r_{ij}-\triangle)&=
\begin{cases}
1            & \text{if}\quad r_{ij}\geq \triangle,\\
0             & \text{otherwise}. \label{Eq.Heaviside}
\end{cases}\\
&\equiv\theta_{(i,j)},
\end{align}
together with the following mathematical identity
\begin{align}
\nabla_{i}\theta_{(i,j)}=\frac{\bmath{r}_{ij}}{r_{ij}}\delta(r_{ij}-\triangle).\label{Eq:delta_func}
\end{align}
Use of the Heaviside function $\Theta(r_{ij}-\triangle)$ may admit of violating a mathematical strictness in distribution theory; the product of two genralised functions may not be well-defined in the sense of distribution \citep[e.g.][]{Griffel_2002}, since the $N$-body distribution function $F_{N}(1,\cdots,t)$\citep{Cercignani_1988} and the function $\Theta(r_{ij}-\triangle)$ are both generalised functions, while one will find its convenience of exploiting the Heaviside function to derive the \citep{Cercignani_1972}'s hierarchy below. 

First define the following term
\begin{align}
I_{s}&\equiv\int_{\Omega_{s+1,N}}\text{d}_{s+1}\cdots\text{d}_{N} S(1,\cdots, N, t),\label{Eq.Is_def}
\end{align}
where $S(1,\cdots, N, t)$ is any function of arguments $\{1,\cdots, N, t\}$. Following the domain, equation \eqref{Int_Omega}, of integration for the truncated DF, one may explicitly express the term as follows
\begin{align}
I_{s}&=\int\text{d}_{s+1} \hspace{2pt} \theta_{(s+1,1)}\hspace{3pt}\cdots  \theta_{(s+1,s)}\nonumber\\
     &\times\int\text{d}_{s+2} \hspace{2pt}  \theta_{(s+2,1)}\hspace{5pt}\cdots  \theta_{(s+2,s)}\hspace{2pt}\theta_{(s+2,s+1)}\nonumber\\
&\hspace{60pt}\vdots\hspace{40pt}\vdots\hspace{30pt}\vdots\hspace{22pt}\ddots\nonumber\\  
     &\times\int\text{d}_{N-1}\hspace{2pt}  \theta_{(N-1,1)}\cdots  \theta_{(N-1,s)}\theta_{(N-1,s+1)} \cdots\theta_{(N-1,N-2)}\nonumber\\
     &\times\int\text{d}_{N} \hspace{13pt}  \theta_{(N,1)}\hspace{2pt}\cdots \theta_{(N,s)}\hspace{7pt}\theta_{(N,s+1)}\hspace{5pt} \cdots\hspace{5pt}\theta_{(N,N-2)} \theta_{(N,N-1)}\nonumber\\
&\times S(1,\cdots, N, t).\label{Eq.Is}
\end{align}
 
\subsection{Truncated integral over the terms $\sum_{i=1}^{N}\bmath{\varv}_{i}\cdot\nabla_{i}F_{N}(1,\cdots,N,t)$}
For the function $S(1,\cdots, N, t)=\sum_{i=1}^{N}\bmath{\varv}_{i}\cdot\nabla_{i}F_{N}(1,\cdots,N,t)$, the pattern of subscripts of the distance $r_{ij}$ in equation \eqref{Eq.Is} is simple; the number $1\leq i\leq s$ appears only as the first letter in subscript. Hence one may separate the summation in the function $S(1,\cdots,N,t)$ into case 1: $1\leq i\leq s$ and case 2: $s+1\leq i\leq N$.

\subsubsection{Case 1: $1\leq i\leq s$}
The goal of the present Appendix is to reduce the term $I_\text{s}$ associated with the terms $\bmath{\varv}_{i}\cdot\nabla_{i}F_{s}^{\triangle}(1,\cdots,s,t)$ by repeating integral-by-parts method. For the numbers $1\leq i\leq s$, define the following term
\begin{align}
I_{s}^{(1:s)}&\equiv\sum_{i=1}^{s}\int_{\Omega_{s+1,N}}\text{d}_{s+1}\cdots\text{d}_{N} \bmath{\varv}_{i}\cdot\nabla_{i}F_{N}(1,\cdots N,t).
\end{align}
Employing equation \eqref{Eq:delta_func}, one obtains
\begin{align}
&I_{s}^{(1:s)}\nonumber\\
&=\sum_{i=1}^{s}\bmath{\varv}_{i}\cdot\nabla_{i}F_{s}^{\triangle}(1,\cdots,s,t) \nonumber\\
&\quad-\sum_{i=1}^{s}\bmath{\varv}_{i}\cdot\sum_{j=s+1}^{N}\int\text{d}_{N}\int\text{d}_{N-1}\cdots\int\text{d}_{j}\cdots\int\text{d}_{s+2}\int\text{d}_{s+1} \nonumber\\
&\quad\times \theta_{(s+1,1)}\cdots  \theta_{(s+1,i)}\cdots \theta_{(s+1,s)}\nonumber\\
&\quad\times \theta_{(s+2,1)}\cdots  \theta_{(s+2,i)}\cdots \theta_{(s+2,s)}\hspace{2pt} \theta_{(s+2,s+1)}\nonumber\\
&\hspace{80pt}\vdots\hspace{65pt}\ddots\nonumber\\  
&\quad\times\theta_{(j,1)}\hspace{5pt}\cdots \hspace{5pt}\frac{\bmath{r}_{ij}}{r_{ij}}\delta_{(j,i)}\hspace{4pt}\cdots\hspace{2pt}\theta_{(j,s)} \hspace{20pt} \cdots  \hspace{10pt}\theta_{(j,j-1)}\nonumber\\
&\hspace{80pt}\vdots \hspace{100pt}\ddots\nonumber\\  
&\quad\times \theta_{(N-1,1)}\cdots  \theta_{(N-1,i)} \cdots\theta_{(N-1, s)}\hspace{10pt}\cdots\hspace{15pt}\theta_{(N-1,N-2)}\nonumber\\
&\quad\times \theta_{(N,1)}\hspace{5pt}\cdots \hspace{5pt} \theta_{(N,i)}\hspace{5pt} \cdots\theta_{(N, s)}\hspace{20pt}\cdots\hspace{40pt}\theta_{(N,N-1)}\nonumber\\
&\quad\times F_{N}(1,\cdots, N, t),\label{Eq.Is_nab_i}
\end{align}
where  $\delta_{(j,i)}\equiv\delta(r_{ij}-\triangle)$. Due to the delta function $\delta_{(j,i)}$, one can convert the volume integral into the surface integral
\begin{align}
&\int\text{d}_{j} \theta_{(j,1)}\cdots  \frac{\bmath{r}_{ij}}{r_{ij}}\delta_{(j,i)}\cdots\theta_{(j,j-1)}=\int \text{d}^{3}\bmath{p}_{j} \oint \text{d}\bmath{\sigma}_{ij}, \label{eq.delta1}
\end{align}
where the $\text{d}\bmath{\sigma}_{ij}$ is the surface element of a sphere of radius $\triangle$ with a radial unit vector $\frac{\bmath{r}_{ij}}{r_{ij}}$ around the position $\bmath{r}_{j}$. Employing equation \eqref{eq.delta1}, one obtains
\begin{align}
I_{s}^{(1:s)}&=\sum_{i=1}^{s}\left(\bmath{\varv}_{i}\cdot\nabla_{i}F_{s}^{\triangle}-\sum_{j=s+1}^{N}\int \text{d}^{3}\bmath{p}_{j} \oint \bmath{\varv}_{i}\cdot\text{d}\bmath{\sigma}_{ij}F^{\triangle}_{s+1}(1,\cdots, s+1, t)\right).\label{Eq.Is_nab_ii}
\end{align}

\subsubsection{Case 2: $s+1\leq i\leq N$}
Define the term $I_\text{s}$associated with the numbers $s+1\leq i\leq N$;
\begin{align}
I_{s}^{(s+1:N)}&\equiv\sum_{i=s+1}^{N}\int_{\Omega_{s+1,N}}\text{d}_{s+1}\cdots\text{d}_{N} \bmath{\varv}_{i}\cdot\nabla_{i}F_{N}(1,\cdots N,t).
\end{align}
To reduce the term $I_{s}^{(s+1:N)}$, one must modify equation \eqref{Eq.Is_nab_i} as follows 
\begin{align}
&I_{s}^{(s+1:N)}\nonumber\\
&=-\sum_{i=s+1}^{N}\sum_{j=1}^{s}\int \text{d}^{3}\bmath{p}_{i} \oint \bmath{\varv}_{i}\cdot\text{d}\bmath{\sigma}_{ij}F^{\triangle}_{s+1}(1,\cdots, s+1, t)\nonumber\\
&\quad+\sum_{i=s+1}^{N}\int\bmath{\varv}_{i}\cdot\nabla_{i}\left(\prod_{i=s+1}^{N}\prod_{j=1}^{i-1}\theta_{(j,i)} F_{N}(1,\cdots N,t)\right) \text{d}_{s+1}\cdots\text{d}_{N}\nonumber\\
&\quad-\sum_{i=s+1}^{N}\bmath{\varv}_{i}\cdot\sum_{j=s+1}^{N}\nonumber\\
&\quad\times\int\text{d}_{N}\int\text{d}_{N-1}\cdots\int\text{d}_{i}\cdots\int\text{d}_{j}\cdots\int\text{d}_{s+2}\int\text{d}_{s+1}  \nonumber\\
&\quad\times \theta_{(s+1,1)}\cdots \theta_{(s+1,s)}\nonumber\\
&\quad\times \theta_{(s+2,1)}\cdots  \theta_{(s+2,s)}\theta_{(s+2,s+1)}\nonumber\\
&\hspace{20pt}\vdots\hspace{45pt}\vdots\hspace{25pt}\vdots\hspace{20pt}\ddots\nonumber\\ 
&\quad\times\theta_{(i+1,1)}\cdots\theta_{(i+1,s)}\theta_{(i+1,s+1)} \cdots \theta_{(i+1,i)}\nonumber\\
&\hspace{20pt}\vdots\hspace{45pt}\vdots\hspace{25pt}\vdots\hspace{40pt}\vdots\hspace{20pt}\ddots\nonumber\\ 
&\quad\times\theta_{(j,1)}\cdots\hspace{5pt}\theta_{(j,s)}\hspace{5pt}\theta_{(j,s+1)} \cdots \hspace{5pt}\frac{\bmath{r}_{ij}}{r_{ij}}\delta_{(j,i)} \cdots \theta_{(j,j-1)}\nonumber\\
&\hspace{20pt}\vdots\hspace{45pt}\vdots\hspace{25pt}\vdots\hspace{40pt}\vdots\hspace{40pt}\vdots\hspace{15pt}\ddots\nonumber\\ 
&\quad\times \theta_{(N,1)} \cdots\theta_{(N, s)}\theta_{(N,s+1)} \cdots \theta_{(N,i)}  \cdots \theta_{(N,j-1)} \cdots\theta_{(N,N-1)}\nonumber\\
&\quad\times F_{N}(1,\cdots, N, t).\label{Eq.Is_nab_j}
\end{align}
where the first summation of terms is obtained in the same way as done for equation \eqref{Eq.Is_nab_ii}, but this time the functions of the displacement vector $\bmath{r}_{j}$ (associated with the latter subscript $j$ in the distance $r_{ij}$) was differentiated. The second summation of terms on the R.H.S. in equation \eqref{Eq.Is_nab_j} vanishes if one assumes the function  $F_{i}(1,\cdots,i,t)$ approaches rapidly enough to zero at the surface of the integrals. Since the delta function in the third summation of terms links two volume integrals to a surface integral, one obtains
\begin{align}
I_{s}^{(s+1:N)}&=\sum_{i=1}^{s}\sum_{j=s+1}^{N}\int \text{d}^{3}\bmath{p}_{j} \oint \bmath{\varv}_{j}\cdot\text{d}\bmath{\sigma}_{ij}F^{\triangle}_{s+1}(1,\cdots, s+1, t)\nonumber\\
&\quad+\sum_{i=s+1}^{N}\sum_{j=s+1}^{N}\int \text{d}_{i}\int \text{d}^{3}\bmath{p}_{j} \oint \bmath{\varv}_{j}\cdot\text{d}\bmath{\sigma}_{ij}\nonumber\\
&\qquad\qquad\qquad\times F^{\triangle}_{s+2}(1,\cdots, s+2, t),\label{Eq.I_s_s+1}
\end{align}
where the following relation is employed
\begin{align}
&\int\text{d}_{i}\int \text{d}_{j} \hspace{5pt}\theta_{(j,1)}\hspace{5pt}\cdots \hspace{5pt}\frac{\bmath{r}_{ij}}{r_{ij}}\delta_{(j,i)}\hspace{4pt}\cdots  \hspace{10pt}\theta_{(j,j-1)}\nonumber\\
&\quad=\int \text{d}_{i}\int \text{d}^{3}\bmath{p}_{j} \oint \bmath{\varv}_{j}\cdot\text{d}\bmath{\sigma}_{ij}.
\end{align}
Combining the results above, equation \eqref{Eq.I_s_s+1}, with the result of case 1 ($1\leq i \leq s$) and considering the dummy integral variables, one obtains
\begin{align}
I_{s}=&\sum_{i=1}^{s}\bmath{\varv}_{i}\cdot\nabla_{i}F_{s}^{\triangle}+\sum_{i=1}^{s}(N-s)\left[\int\text{d}^{3}\varv_{s+1}\oiint F_{s+1}^{\triangle} \bmath{\varv}_{i,s+1}\cdot\text{d}\bmath{\sigma}_{i,s+1}\right]\nonumber\\
&+\frac{(N-s)(N-s-1)}{2}\int\text{d}^{3}\varv_{s+2}\int\text{d}_{s+1}\nonumber\\
&\quad\times\oiint F_{s+2}^{\triangle}  \bmath{\varv}_{s+1,s+2}\cdot\text{d}\bmath{\sigma}_{s+1,s+2},
\end{align} 
where $\bmath{\varv}_{ij}=\bmath{\varv}_{i}-\bmath{\varv}_{j}$. Only the configuration space in the truncated DF must be deprived, hence the rest of treatment for the other terms in the Liouville equation is the same as for the standard BBGKY hierarchy \citep[e.g.][]{Lifschitz_1981,Saslaw_1985,McQuarrie_2000,Liboff_2003}, which results in equation \eqref{Eq.BBGKY_truncated} in terms of $s$-tuple DFs.

\section{An extra term in the BBGKY hierarchy for the truncated DF}\label{Appendix:extra}
As proved in \citep{Cercignani_1988}, the following term appearing in the BBGKY hierarchy for the truncated DF,
\begin{align}
I_\text{ex}=\frac{1}{2}\int\text{d}^{3}\bmath{p}_{s+2}\int\text{d}_{s+1}\oiint f_{s+2}  \bmath{p}_{s+1,s+2}\cdot\text{d}\bmath{\sigma}_{s+1,s+2},\label{Eq.I_ex}
\end{align} 
vanishes if one considers elastic encounter. The collision term $I_\text{ex}$, however, does not vanish if one considers the effect of non-ideal encounters and inelastic direct physical collision (loss of kinetic energy due to tidal effects, coalescence and stellar mass evaporation) between two finite-size stars \citep[e.g.][]{Quinlan_1987} where equation \eqref{Eq.I_ex} takes part of the Smoluchowski-coagulation collision term if the system has a continuous mass distribution.

For brevity $s=1$ is taken and basic treatment for strong encounter follows the idea discussed in section \ref{sec.WC_strong}. Consider that two field stars undergo a two-body \emph{elastic} encounter at the surface of Landau sphere and the trajectory is determined by the two-body Newtonian interaction, equation \eqref{Eq.characteristics_two_body}, between star $2$ and star $3$. If the gain of state $f(2,t)f(3,t)$ is considered, then the three-body DF $f_{3}(1,2,t)$ reads
\begin{align}
&f_{3}(1,2,3,t)\approx f_{1}(1,t)f_{1}(2(t-\tau),t-\tau)f_{2}(3(t-\tau),t-\tau),\label{Eq.constant}
\end{align}
where the effect of background stars are neglected and collision kinetic description is employed. Hence, on the hemisphere of Landau sphere for stars leaving the sphere, the integral term $I_\text{ex}$ reduces to
\begin{subequations}
	\begin{align}
	&\frac{1}{2}\oiint^{(+)} \int \text{d}^{3}\bmath{p}_{3}\int\text{d}_{2}\bmath{p}_{2,3}\cdot\text{d}\bmath{\sigma}_{2,3} f_{3}(1,2,3,t)\\
	&=\frac{1}{2}\oiint^{(+)} \int \text{d}^{3}\bmath{p}_{3}\int\text{d}_{2}\bmath{p}_{2,3}\cdot\text{d}\bmath{\sigma}_{2,3}\nonumber\\
	&\qquad\times f_{1}(1,t)f_{1}(2(t-\tau),t-\tau)f_{2}(3(t-\tau),t-\tau),\label{Eq.I_ex_p1}\\
	&=-\frac{1}{2}\oiint^{(+)} \int \text{d}^{3}\bmath{p}_{3}(t-\tau)\int\text{d}^{3}\bmath{r}_{2}\text{d}^{3}\bmath{p}_{2}(t-\tau)\text{d}\bmath{\sigma}_{2,3}(t-\tau)\nonumber\\ 
	&\qquad \cdot\bmath{p}_{2,3}(t-\tau)f_{1}(1,t)f_{1}(2(t-\tau),t-\tau)f_{2}(3(t-\tau),t-\tau),\label{Eq.I_ex_p2}
	\end{align}
\end{subequations}
where equation \eqref{Eq.I_ex_p1} is obtained employing equation \eqref{Eq.constant}. Also equation \eqref{Eq.I_ex_p2} is derived based on the properties of elastic encounter
\begin{subequations}
	\begin{align}
	&\bmath{\varv}_{s+1,s+2}(t)\cdot\text{d}\bmath{\sigma}_{s+1,s+2}(t)=-\bmath{\varv}_{s+1,s+2}(t-\tau)\cdot\text{d}\bmath{\sigma}_{s+1,s+2}(t-\tau),\\
	&\text{d}^{3}\bmath{p}_{2}\text{d}^{3}\bmath{p}_{3}=\text{d}^{3}\bmath{p}_{2}(t-\tau)\text{d}^{3}\bmath{p}_{3}(t-\tau).
	\end{align} 
\end{subequations}
For brevity, if one takes the first order of retardation effect neglecting the non-locality in configuration space, equation \eqref{Eq.I_ex_p2} reduces to 
\begin{align}
&\frac{1}{2}\oiint^{(+)} \int \text{d}^{3}\bmath{p}_{3}\int\text{d}_{2}\bmath{p}_{2,3}\cdot\text{d}\bmath{\sigma}_{2,3} f_{3}(1,2,3,t)\\
&=\left(1-\tau\partial_{t}\right)\frac{1}{2}\oiint^{(-)} \int \text{d}^{3}\bmath{p}_{3}\int\text{d}^{3}\bmath{r}_{2}\text{d}^{3}\bmath{p}_{2}\text{d}\bmath{\sigma}_{2,3}\nonumber\\ 
&\qquad \cdot\bmath{p}_{2,3}f_{1}(1,t)f_{1}(2,t)f_{2}(3,t),\label{Eq.I_ex_p3}
\end{align}
where the integral of the hemisphere $(+)$ (for stars leaving the Landau sphere) was changed to the $(-)$ (for stars entering the sphere) due to the negative sign in equation \eqref{Eq.I_ex_p2} where the dummy variables (such as $\varv_{2}(t-\tau)$) can be reverted to the original variables (such as $\varv_{2}$). The first term on the R.H.S of equation \eqref{Eq.I_ex_p3} is the same as the corresponding collision term for loss of stars from the state
\begin{align}
f(1,2,t)=f(1,t)f(2,t)f(3,t),
\end{align}
where the effect of correlation is neglected in wave kinetic description. Hence, for a short correlation time duration, the collision term, equation \eqref{Eq.I_ex}, reduces to
\begin{align}
&I_\text{ex}=-\tau\partial_{t}\frac{1}{2}\oiint^{(-)} \int \text{d}^{3}\bmath{p}_{3}\int\text{d}_{2}\text{d}\bmath{\sigma}_{2,3}\cdot\bmath{p}_{2,3}f(1,t)f(2,t)f(3,t).\label{Eq.I_ex_ret}
\end{align} 
If one neglects the effect of the retardation (and spatial non-locality), the collision term $I_\text{ex}$ always vanishes as shown in \citep{Cercignani_1988} for any $s$-tuple DF. In the present paper, if one employs the weakly-coupled DF or 'test-particle' method, the term $I_\text{ex}$ vanishes since any pair of field stars can not exist on the surface of the Landau sphere at the same time.

\section{The Bogoliubov's derivation of the Boltzmann collision term for long-range interaction potential}\label{Appendix:Bogorigouv}
In the present section, the Bogoliubov's derivation of the Boltzmann collision term is arranged for Newtonian interaction potential (long-range two-body interaction) employing the anti-normalisation condition, equation \eqref{Eq.Anti-norm}. One can refer to \citep{Uhlenbeck_1963,Liboff_2003} for neutral gases (short-range two-body interaction) and to \citep{Klimontovich_1982} for non-ideal gases/plasmas. Their methods, however, are not applicable to a system of particles interacting with long-range interaction; even the most relevant work \citep{Klimontovich_1982} needs the effect of retardation to find the Boltzmann collision term. In the present Appendix, the Botlzmann collision term is derived without employing the effect of retardation.

Integrate the second equation of the standard BBGKY hierarchy 
\begin{align}
&\left(\partial_{t}+\bmath{\varv}_{1}\cdot\nabla_{1}+\bmath{\varv}_{2}\cdot\nabla_{2}+\bmath{a}_{12}\cdot\bmath{\partial}_{12}\right)g(1,2,t)\nonumber\\
&\qquad=-\bmath{a}_{12}\cdot\bmath{\partial}_{12}f(1,t)f(2,t),\label{Eq.2ndBBGKY_strong}
\end{align}
over $\int\cdot\hspace{3pt}\text{d}2$, then anti-normalisation condition, equation \eqref{Eq.Anti-norm}, turns the equation into 
\begin{align}
&\int \left(\bmath{\varv}_{1}\cdot\nabla_{1}+\bmath{\varv}_{2}\cdot\nabla_{2}+\bmath{a}_{12}\cdot\bmath{\partial}_{12}\right)g(1,2,t)\text{d}2=-\bmath{A}_{1}\bmath{\partial}_{1}f(1,t).\label{Eq.2ndBBGKY_strong_norm}
\end{align}
The collision integral due to the correlation function $g(1,2,t)$ reduces to
\begin{align}
I_\text{cor}&=-\int \bmath{a}_{12}\cdot\bmath{\partial}_{1}g(1,2,t)\text{d}2,\nonumber\\
            &=\int \left(\bmath{p}_{12}\cdot\nabla_{12}+\bmath{p}_{R}\cdot\nabla_{R}\right)g(1,2,t)\text{d}2+\bmath{A}_{1}\bmath{\partial}_{1}f(1,t),\nonumber\\
            &\approx \int \left(\bmath{\varv}_{12}\cdot\nabla_{12}\right)g(\bmath{r}_{12},\bmath{r}_{1},\bmath{p}_{1},\bmath{p}_{2},t)\text{d}2,\nonumber\\
\end{align} 
where the last equation is obtained by assuming the encounter is local. The method of characteristic gives the solution to equation \eqref{Eq.2ndBBGKY_strong}
\begin{subequations}
\begin{align}
g(1,2,t)&=f(1(t-\tau),t-\tau)f(2(t-\tau),t-\tau),\\
        &=f(\bmath{r}_{1},\bmath{P}_{1}(-\infty),t)f(\bmath{r}_{1},\bmath{P}_{2}(-\infty),t),
\end{align}
\end{subequations}
where no effect of the non-ideality is considered and the correlation time $\tau$ is taken as infinity. The last equation implicitly depends on the $r_{12}$ due to the conservation of momentum and the conservation of total energy
\begin{align}
\frac{P_{1}^{2}(-\infty)}{2m}+\frac{P_{2}^{2}(-\infty)}{2m}=\frac{p_{1}^{2}}{2m}+\frac{p_{2}^{2}}{2m}+\phi(r_{12}).
\end{align}
Hence, the coordinate of collision integral can be converted into a cylindrical one along the relative velocity $\bmath{\varv}_{12}$
\begin{align}
I_\text{cor}=&\int\text{d}\bmath{p}_{2}\int_{0}^{2\pi}\text{d}\psi\int_{0}^{\infty}b\text{d}b \int \varv_{12}\nonumber\\
&\qquad\times\frac{\partial}{\partial z_{12}}f(\bmath{r}_{1},\bmath{P}_{1}(-\infty),t)f(\bmath{r}_{1},\bmath{P}_{2}(-\infty),t).
\end{align}
In the same way as the Grad's analysis, assume two cases of encounters; the gain of and loss from the star state $(\bmath{r}_{1},\bmath{p},t)$, which are to be separated by taking the limit of the distance $z_{12}$ to $+\infty$ or $-\infty$;
\begin{align}
&\bmath{p}_{1}\equiv\bmath{P}_{1}(r_{12},\bmath{p}_{1},\bmath{p}_{2},t=-\infty)_{z_{12}=\infty},\\
&\bmath{p}_{2}\equiv\bmath{P}_{2}(r_{12},\bmath{p}_{1},\bmath{p}_{2},t=-\infty)_{z_{12}=\infty},\\
&\bmath{p}'_{1}\equiv\bmath{P}_{1}(r_{12},\bmath{p}_{1},\bmath{p}_{2},t=-\infty)_{z_{12}=-\infty},\\
&\bmath{p}'_{2}\equiv\bmath{P}_{2}(r_{12},\bmath{p}_{1},\bmath{p}_{2},t=-\infty)_{z_{12}=-\infty}.
\end{align}
Employing the boundary conditions for the initial momenta for integration of the collision integral with respect to the distance $z_{12}$, one obtains the Boltzmann collision term $I_\text{Bol}^\text{(loc)}$. 

\section{Approximation of the conversion relation between collision- and wave- kinetic descriptions}\label{Appendix.Ext.non_ideality}
In the present section, the following two approximations of the conversion relation between wave kinetic description, equation \eqref{Eq.g(1,2,t)_wave}, and collision one, equation \eqref{Eq.g(1,2,t)_collision}, are shown for weakly inhomogeneous star cluster in the two cases (i) close-strong two-body encounter at distances $r_{12}\lesssim r_\text{o}$ and (ii) weak-distant two-body encounter: weak-coupling limit with rectilinear trajectory at distances $r_{12}\sim a_\text{BG}$. For simplicity, the effects of the m.f. acceleration of stars, gravitational polarization and triple encounters are neglected.

\subsection{Wave kinetic description of close-strong encounters}\label{Appendix:wave_stron_Landau}
One can find an alternative description of close-strong two-body encounter (Boltzmann collision term) based on equation \eqref{Eq.twoDF_strong_weak}. Employing the same method as done in \citep{Klimontovich_1982} or section \ref{subsec:nonideal}, one can expand equation \eqref{Eq.twoDF_strong_weak} in series of the correlation time $\tau$ and hold only the zeroth order;
\begin{subequations}
	\begin{align}
	&f(\bmath{r}_{1},\bmath{p}_{1}(t-\tau),t)f(\bmath{r}_{1},\bmath{p}_{2}(t-\tau),t),\nonumber\\
	&=-\int^{\tau}_{0}\bmath{a}_{12}\cdot\bmath{\partial}_{12}f\left(\bmath{r}_{1},\bmath{p}_{1}\left(t-\tau'\right),t\right)f\left(\bmath{r}_{1},\bmath{p}_{2}\left(t-\tau'\right),t\right)\text{d}\tau'\nonumber\\
	&\qquad+f(\bmath{r}_{1},\bmath{p}_{1},t)f(\bmath{r}_{1},\bmath{p}_{2},t), \label{Eq.twoDF_strong}
	\end{align}
\end{subequations} 
where the L.H.S is a typical binary DF associated with the Boltzmann collision term while the R.H.S shows two different binary DFs; the first term is associated with strong encounter in wave kinetic description and the second term 'the unweighted average potential \citep{Balescu_1960}'. One may realise the relation, equation \eqref{Eq.twoDF_strong}, still holds the characteristics of a local Newtonian two-body encounter, equation \eqref{Eq.characteristics_v_jump}, meaning the trajectory of test star is ideal (local in time and space). The ideal 'strong' Landau collision term reads
\begin{align}
I_\text{st-L}^{(\text{loc})}=&\int\text{d}_{2}\bmath{a}_{12}\cdot\bmath{\partial}_{1}\bmath{a}_{12}\cdot\int^{\tau}_{0}\bmath{\partial}_{12}\nonumber\\
&\qquad\times f\left(\bmath{r}_{1},\bmath{p}_{1}\left(t-\tau'\right),t\right)f\left(\bmath{r}_{1},\bmath{p}_{2}\left(t-\tau'\right),t\right)\text{d}\tau'.\label{Eq.I_st_L}
\end{align}
If extending the range of distance $r_{12}$ up to order of the BG radius $a_\text{BG}$ and taking the weak-coupling limit for the both sides of equation \eqref{Eq.twoDF_strong}, then one obtains the binary DFs associated with a typical Landau collision term 
\begin{align}
&f(\bmath{r}_{1},\bmath{p}_{1}(t-\tau),t)f(\bmath{r}_{1},\bmath{p}_{2}(t-\tau),t),\nonumber\\
&\qquad=\left(1-\int^{t}_{t-\tau}\bmath{a}_{12}\text{d}t'\cdot\bmath{\partial}_{12}\right)f(\bmath{r}_{1},\bmath{p}_{1},t)f(\bmath{r}_{1},\bmath{p}_{2},t).\label{Eq.Landau_wave}
\end{align}
The relation, equation \eqref{Eq.Landau_wave}, is of significance in the star-cluster CKT in section \ref{sec:discussion}. 

\subsection{The Landau collision term as the first orders of the weak-non-idaelity- and weak-coupling- approximation}\label{Appendix:I_L_nonideal}
Another asymptotic limit of distance $r_{12}$ for equation \eqref{Eq.g(1,2,t)_collision} is the BG radius $a_\text{BG}$ since the effect of the m.f. acceleration dominate the motion of test star at distances greater than the radius $a_\text{BG}$. The OoM of non-ideality and weak-coupling approximation at distances $r_{12}\approx a_\text{BG}\sim N^{-1/2}$ are to be the same order in weakly-inhomogeneous systems; one may expand the second equation of the BBGKY hierarchy in series of the radius $a_\text{BG}\sim1/N^{1/2}$. Recalling in \citep{Klimontovich_1982}'s theory that the spatial locality must be applied to correlation function $g(1,2,t)$, not two-body DF $f(1,2,t)$, one can find the following description for the correlation function
\begin{subequations}
	\begin{align}
	&g(1,2,t)\nonumber\\
	=&f(1(t-\tau),t-\tau)(2(t-\tau),t-\tau)-f(1,t)f(2,t)\nonumber\\
	&+\int^{t}_{t-\tau} \left(\partial_{t}+\bmath{\varv}_{1}\cdot\nabla_{1}+\bmath{\varv}_{2}\cdot\nabla_{2}\right)_{t=t'}f\left(1\left(t'\right),t'\right)f\left(2\left(t'\right),t'\right)\text{d}t',\\
	\approx&\left(1-\tau\partial_{t}-\int\bmath{V}\text{d}t'\cdot\nabla_{1}-\int\bmath{a}_{12}\text{d}t'\cdot\bmath{\partial}_{12} -\frac{\bmath{r}_{12}}{2}\cdot\nabla_{1}\right)\nonumber\\
	&\qquad\times f(1,t)f(\bmath{r}_{1},\bmath{p}_{2},t)-\left(1-\frac{\bmath{r}_{12}}{2}\cdot\nabla_{1}\right)f(1,t)f(\bmath{r}_{1},\bmath{p}_{2},t)\nonumber\\
	&\quad+\int \left(\partial_{t}+\bmath{V}\cdot\nabla_{1}\right)\text{d}t'f(1,t)f(\bmath{r}_{1},\bmath{p}_{2},t),\label{Eq.Landau_ret_col}\\
	=&-\int\bmath{a}_{12}\cdot\bmath{\partial}_{12}\text{d}\tau f(1,t)f(\bmath{r}_{1},\bmath{p}_{2},t), \label{Eq.Landau_ret_wave}
	\end{align}
\end{subequations} 
where equation \eqref{Eq.Landau_ret_col} is expanded up to the first orders of short correlation time $\tau$ and weak-coupling approximation. A unique property of the correlation function at distances $r_{12}\sim a_\text{BG}$ is that its mathematical form is the same as the form of correlation functions at the zero-th order, equation \eqref{Eq.Landau_wave}. This implies that \emph{not} taking the weak-coupling approximation of the correlation function (collision term) is an essential process to hold the effect of non-ideality in collision kinetic description at the first order of non-ideality.

\end{appendices}